\documentclass[11pt,a4paper]{article}

\usepackage{mlmodern}
\usepackage{amsfonts,amssymb,amsthm,ellipsis}
\usepackage{upgreek,mathrsfs}

\usepackage{twemojis} 
\usepackage{bold-extra}

\usepackage[normalem]{ulem} 

\usepackage{silence}
\usepackage{microtype}

\usepackage[dvipsnames]{xcolor}
\definecolor{maroon}{RGB}{167,74,74}
\definecolor{magreen}{RGB}{59,125,37}
\definecolor{mablue}{RGB}{59,136,195}
\definecolor{mayellow}{RGB}{242,147,24}
\colorlet{mabrown}{maroon!50!magreen}
\colorlet{maorange}{maroon!50!mayellow}
\colorlet{macyan}{magreen!50!mablue}

\colorlet{red}{maroon}
\colorlet{green}{magreen}
\colorlet{blue}{mablue}
\colorlet{yellow}{mayellow}
\colorlet{brown}{mabrown}
\colorlet{orange}{maorange}
\colorlet{cyan}{macyan}

\definecolor{rred}{RGB}{167,33,74}
\definecolor{bblue}{RGB}{29,136,255}
\definecolor{ppurple}{RGB}{113,84,165}
\definecolor{ppink}{RGB}{255,55,219}

\RequirePackage{hyperref}
\hypersetup{colorlinks,
    linkcolor=mablue,
    citecolor=magreen,
    urlcolor=maroon,
    anchorcolor=mablue,
    filecolor=mablue,
    menucolor=mablue,
    pdftitle=dS-Ising
    }

\RequirePackage[top=10mm,bottom=12mm,left=30mm,right=30mm,head=12mm,includeheadfoot]{geometry}
\makeatletter
\g@addto@macro\bfseries{\boldmath}
\makeatother

\makeatletter
\renewcommand\section{\@startsection{section}{1}{\z@}%
  {-3.5ex \@plus -1.3ex \@minus -.7ex}%
  {2.3ex \@plus.4ex \@minus .4ex}%
  {\large\bfseries}}
\renewcommand\subsection{\@startsection{subsection}{2}{\z@}%
    {-2.3ex\@plus -1ex \@minus -.5ex}%
    {1.2ex \@plus .3ex \@minus .3ex}%
    {\normalsize\bfseries}}
\renewcommand\subsubsection{\@startsection{subsubsection}{3}{\z@}%
    {-2.3ex\@plus -1ex \@minus -.5ex}%
    {1ex \@plus .2ex \@minus .2ex}%
    {\normalsize\bfseries}}
\renewcommand\paragraph{\@startsection{paragraph}{4}{\z@}%
    {1.75ex \@plus1ex \@minus.2ex}%
    {-1em}%
    {\normalsize\bfseries}}
\renewcommand\subparagraph{\@startsection{subparagraph}{5}{\parindent}%
    {1.75ex \@plus1ex \@minus .2ex}%
    {-1em}%
    {\normalsize\bfseries}}
\makeatother

\RequirePackage{tocloft}

\RequirePackage[nottoc,notlot,notlof]{tocbibind}

\usepackage{comment}

\usepackage{import}
\usepackage{enumitem}
\usepackage{csquotes}

\usepackage[new]{old-arrows}

\newcommand{\nn}{\nonumber \\}

\renewcommand{\leq}{\leqslant}

\usepackage{physics}
\usepackage{stackengine}

\newcommand{\parti}{\mathcal{Z}} 
\newcommand{\pd}{\partial} 

\newcommand{\half}{\tfrac{1}{2}} 

\newcommand{\w}{\mathbin{\scalebox{0.8}{$\wedge$}}} 
\newcommand{\ex}[1]{\mathrm{e}^{#1}} 
\newcommand{\ii}{i} 

\let\C\undefined

\newcommand{\R}{\mathbb{R}} 
\newcommand{\C}{\mathbb{C}} 
\newcommand{\Z}{\mathbb{Z}} 
\renewcommand{\S}{S} 

\newcommand{\eqv}{\Longleftrightarrow} 

\newcommand{\SU}{\mathrm{SU}} 
\renewcommand{\O}{\mathrm{O}} 
\newcommand{\SO}{\mathrm{SO}} 
\newcommand{\SL}{\mathrm{SL}} 
\newcommand{\alg}[1]{\mathfrak{#1}} 
\renewcommand{\sl}{\alg{sl}} 
\newcommand{\so}{\alg{so}} 
\renewcommand{\u}{\alg{u}} 

\renewcommand{\t}[1]{{\text{#1}}} 
\newcommand{\xmapsto}[1]{\overset{#1}{\mapsto}} 

\makeatletter
\renewcommand{\xmapsto}[2][]{\ext@arrow 0599{\mapstofill@}{#1}{#2}}
\def\mapstofill@{\arrowfill@{\mapstochar\relbar}\relbar\rightarrow}
\makeatother

\def \cD {\mathcal{D}}

\def \cI {\mathcal{I}}
\def \cJ {\mathcal{J}}

\def \cO {\mathcal{O}}
\def \cP {\mathcal{P}}

\def \cR {\mathcal{R}}

\DeclareSymbolFont{bbold}{U}{bbold}{m}{n}
\DeclareSymbolFontAlphabet{\mathbbold}{bbold}

\def \bb1 {{\mathbb{1}}}

\def \sfT {\mathsf{T}}

\usepackage{tikz} 

\usetikzlibrary{decorations.markings,arrows.meta,svg.path}

\tikzset{line/.style={line width=0.25mm},
curve/.style={line,smooth,tension=1},
->-/.style={decoration={
  markings,
  mark=at position #1 with {\arrow[>=stealth]{>}}},postaction={decorate}},
-<-/.style={decoration={
  markings,
  mark=at position #1 with {\arrow[>=stealth]{<}}},postaction={decorate}},
}

\tikzset{bg/.style={opacity=.5}}

\usepackage{tikz-cd}
\usepackage{multirow}
\RequirePackage{mdframed}
\usepackage{empheq}

\definecolor{pansypurple}{rgb}{0.47, 0.09, 0.29}
\definecolor{patriarch}{rgb}{0.5, 0.0, 0.5}
\definecolor{carmine}{rgb}{0.59, 0.0, 0.09}
\definecolor{blueflag}{rgb}{0.2, 0.2, 0.6}
\usepackage{tikz-3dplot}

\def\fdiffd{\mathcal{D}}
\DeclareDocumentCommand\fdifferential{ o g d() }{ 
    \IfNoValueTF{#2}{
        \IfNoValueTF{#3}
        {\fdiffd\IfNoValueTF{#1}{}{^{#1}}}
        {\mathinner{\fdiffd\IfNoValueTF{#1}{}{^{#1}}\argopen(#3\argclose)}}
    }
    {\mathinner{\fdiffd\IfNoValueTF{#1}{}{^{#1}}#2} \IfNoValueTF{#3}{}{(#3)}}
}

\DeclareDocumentCommand\variation{ o g d() }{ 
    \IfNoValueTF{#2}{
        \IfNoValueTF{#3}
        {\updelta \IfNoValueTF{#1}{}{^{#1}}}
        {\mathinner{\updelta \IfNoValueTF{#1}{}{^{#1}}\argopen(#3\argclose)}}
    }
    {\mathinner{\updelta \IfNoValueTF{#1}{}{^{#1}}#2} \IfNoValueTF{#3}{}{(#3)}}
}

\DeclareDocumentCommand\wedgecommutator{ l m m }{\braces#1{\lbrack}{\rbrack}{#2\w #3}} 

\DeclareMathOperator{\polylogarithm}{Li}
\DeclareDocumentCommand\Li{}{\opbraces{\polylogarithm}}
	
\DeclareMathOperator{\pfaffian}{Pfaff}
\DeclareDocumentCommand\pf{}{\opbraces{\pfaffian}}

\RequirePackage{interval}
\intervalconfig{soft open fences}
\newcommand{\ropen}[2]{\interval[open right]{#1}{#2}}
\newcommand{\lopen}[2]{\interval[open left]{#1}{#2}}
\newcommand{\open}[2]{\interval[open]{#1}{#2}}
\newcommand{\closed}[2]{\interval{#1}{#2}}

\usepackage[noabbrev]{cleveref}

\crefname{subsection}{subsection}{subsections}
\crefname{equation}{}{}

\numberwithin{equation}{section}

\usepackage[style=numeric-comp, eprint=true, natbib, backend=biber, maxbibnames=10, giveninits=true, isbn=false, url=false, doi=false, date=year, sorting=none, useprefix=true]{biblatex}

\usepackage{hyperref}

\DeclareFieldFormat*{title}{\emph{#1}}

\DeclareFieldFormat*{journaltitle}{#1}

\DeclareFieldFormat{pages}{#1}

\DeclareFieldFormat{labelalpha}{\textsc{#1}}

\renewbibmacro{in:}{}

\renewbibmacro*{date}{%
  \iffieldundef{year}
    {}
    {\printtext[parens]{\printdate}}}

\renewbibmacro*{issue+date}{%
    \printfield{issue}%
    \setunit*{\addspace}%
    \usebibmacro{date}%
    \newunit}

\renewbibmacro*{publisher+location+date}{%
    \printlist{location}%
    \iflistundef{publisher}
        {\setunit*{\addcomma\space}}
        {\setunit*{\addcolon\space}}%
    \printlist{publisher}%
    \setunit*{\addspace}%
    \usebibmacro{date}%
    \newunit}

\renewbibmacro*{institution+location+date}{%
    \printlist{location}%
    \iflistundef{institution}
        {\setunit*{\addcomma\space}}
        {\setunit*{\addcolon\space}}%
    \printlist{institution}%
    \setunit*{\addspace}%
    \usebibmacro{date}%
    \newunit}

\renewbibmacro*{organization+location+date}{%
    \printlist{location}%
    \iflistundef{organization}
        {\setunit*{\addcomma\space}}
        {\setunit*{\addcolon\space}}%
    \printlist{organization}%
    \setunit*{\addspace}%
    \usebibmacro{date}%
    \newunit}

\renewbibmacro*{location+date}{%
    \printlist{location}%
    \setunit*{\addspace}%
    \usebibmacro{date}%
    \newunit}

\AtEveryBibitem{\clearfield{urldate}}
\AtEveryBibitem{\clearfield{month}}
\AtEveryBibitem{\clearfield{day}}
\AtEveryBibitem{\clearfield{issn}}
\AtEveryBibitem{\clearfield{version}}

\DeclareFieldFormat{doilink}{%
	\iffieldundef{doi}{%
		\iffieldundef{url}{%
			\iffieldundef{isbn}{%
				\iffieldundef{issn}{%
					#1%
				}{%
					\href{http://books.google.com/books?vid=ISSN\thefield{issn}}{#1}%
				}%
			}{%
				\href{http://books.google.com/books?vid=ISBN\thefield{isbn}}{#1}%
			}%
		}{%
			\href{\thefield{url}}{#1}%
		}%
	}{%
		\href{http://dx.doi.org/\thefield{doi}}{#1}%
	}%
}

\DeclareBibliographyDriver{article}{%
	\usebibmacro{bibindex}%
	\usebibmacro{begentry}%
	\usebibmacro{author/translator+others}%
	\setunit{\labelnamepunct}\newblock
	\usebibmacro{title}\addcomma\space%
	\newunit
	\printlist{language}%
	\newunit\newblock
	\usebibmacro{byauthor}%
	\newunit\newblock
	\usebibmacro{bytranslator+others}%
	\newunit\newblock
	\printfield{version}%
	\newunit\newblock
	\printtext[doilink]{%
		\usebibmacro{journal+issuetitle}%
		\newunit
		\usebibmacro{byeditor+others}%
		\newunit
		\usebibmacro{note+pages}%
	}%
	\newunit\newblock
	\iftoggle{bbx:isbn}
	{\printfield{issn}}
	{}%
	\newunit\newblock
	\usebibmacro{doi+eprint+url}%
	\newunit\newblock
	\usebibmacro{addendum+pubstate}%
	\setunit{\bibpagerefpunct}\newblock
	\usebibmacro{pageref}%
	\usebibmacro{finentry}}

\usepackage{slashed}

\let\u\relax
\newcommand{\u}{u} 

\let\epsilon\roundepsilon
\newcommand{\epsilon}{\varepsilon}

\let\tilde\thintilde
\newcommand{\tilde}{\widetilde}

\usepackage{bbm}
\usepackage{subcaption}

\newcommand{\unit}{\mathbbm{1}}

\newcommand{\mat}[1]{\begin{pmatrix} #1 \end{pmatrix}}

\newcommand{\ds}{\ensuremath{\t{dS}}}

\usepackage{slashed}
\newcommand{\dsl}{\slashed{\nabla}}

\let\u\relax
\newcommand{\u}{u}
\newcommand{\uL}{u^{\scriptscriptstyle \t{L}}}

\newcommand{\lds}{\ell}

\newcommand{\Gm}{G_\mu}
\newcommand{\Gs}{G_\sigma}
\newcommand{\dGs}{\dot{G}_\sigma}
\newcommand{\ddGs}{\ddot{G}_\sigma}
\newcommand{\dGm}{\dot{G}_\mu}
\newcommand{\ddGm}{\ddot{G}_\mu}
\newcommand{\sds}{\tilde{\Delta}_\sigma}
\newcommand{\sdm}{\tilde{\Delta}_\mu}

\begin{document}

\begin{center}
    {\Large \textsc{Ising the way into de Sitter}} \\
	\bigskip
	Giovanni Galati 
	and Stathis Vitouladitis
	\\	
	\bigskip
	\footnotesize{Physique Théorique et Mathématique, Université Libre de Bruxelles \\ \&  International Solvay Institutes, CP 231, 1050 Brussels, BE \\
	\bigskip
    \href{mailto:giovanni.galati@ulb.be}{\small \sf giovanni.galati@ulb.be} \qquad
    \href{mailto:stathis.vitouladitis@ulb.be}{\small \sf stathis.vitouladitis@ulb.be}
    }
\end{center}

\begin{abstract}
\noindent
    We study the two-dimensional Ising model deformed by the relevant thermal operator and placed on de Sitter (dS) spacetime. Despite being strongly interacting in its original formulation, the theory is exactly solvable on account of fermionisation.
    We compute exact cosmological correlators and compare them with conformal perturbation theory. In the Euclidean formulation of the model, we first compute the renormalised sphere partition function and the exact two-point functions of the thermal operator and descendant-like operators. We analytically continue the two-point functions to Lorentzian dS$_2$. Their late-time behaviour is governed by de~Sitter representation theory and includes oscillations associated with principal-series scaling dimensions. We then analyse two-point functions of the spin and disorder operators, which are non-local in the fermionic variables, and derive non-perturbative constraints on their late-time scaling dimensions. In both cases, we compare the exact answers to conformal perturbation theory (CPT) and we show that divergent secular terms generically spoil the perturbative series at late times. The de Sitter Ising model shows explicitly how late-time perturbative pathologies are resummed in non-perturbative cosmological observables and provides a minimal solvable laboratory for quantum field theory dynamics in de Sitter space.
\end{abstract}

\tableofcontents

\section{Introduction}

In statistical mechanics, the Ising model is the prototypical example of an exactly solvable model. It is a very simple model, consisting of spins on a lattice interacting only with their nearest neighbours, yet it contains a remarkable amount of physics. Both on the lattice and as a continuum quantum field theory (QFT), the Ising model has consistently served as a source of exact results and predictions. Here we make the case that it can serve the same purpose in cosmology. The Ising model placed on an expanding universe is arguably the simplest setup displaying several thorny features of quantum cosmology, while remaining under non-perturbative control.

The standard cosmological picture suggests that, in its very early moments, our universe underwent a period of inflation \cite{Starobinsky:1980te,Guth:1980zm,Linde:1981mu,Albrecht:1982wi}, while observational data support that it is currently approaching another phase of accelerated expansion \cite{PlanckCollab,Suzuki,SDSS}. In both periods, the geometry of spacetime is well approximated by de~Sitter (dS) spacetime: the maximally symmetric solution to Einstein's equations with positive cosmological constant. Assuming that quantum gravity effects \cite{Anninos:2012qw,Witten:2001kn} can be set aside for the moment,%
\footnote{Though, certainly these effects are utterly important and we should eventually switch them back on. We comment on such effects in \cref{sec:outlook}.}
to give a convincing account of cosmology one is faced with the question of understanding QFT in dS spacetime. Compared with QFT in the other maximally symmetric spacetimes, Minkowski and anti-de~Sitter (AdS) space, QFT in de~Sitter remains much less well understood.

Building on early developments (see e.g. \cite{Nachtmann:1967nss,Chernikov:1968zm,Spindel:1976,Bunch:1978yq,Mottola:1984ar,Allen:1985ux,Higuchi:1986wu,Higuchi:1991tn,Higuchi:1991tk,Higuchi:1991tm}), recent years have seen substantial effort to clarify some of the foundational questions of QFT in dS. These include the structure of the associated Hilbert space, correlation functions, symmetries and representations, and unitarity; see for example \cite{Anninos:2014lwa,Gorbenko:2019rza,Sleight:2020obc,Hogervorst:2021uvp,Anninos:2020hfj,DiPietro:2021sjt,Letsios:2022tsq,Letsios:2023qzq,Penedones:2023uqc,Loparco:2023rug,Letsios:2026ypo}. However, the dearth of exactly solvable models to guide this effort has made progress slower and fuzzier than desired.

There is a clear need for such models in de~Sitter, so that observables can be computed exactly and used to test proposals and make predictions. The list of known solvable models in de~Sitter is rather short.%
\footnote{From this list we exclude free theories and conformal field theories as they do not see the interesting features of \ds.}
It consists of recent studies of the two-dimensional Schwinger model \cite{Anninos:2024fty}%
\footnote{See also \cite{Jayewardena:1988td} for an older study and \cite{Smith:2026dae} for a perturbative test of the results of \cite{Anninos:2024fty}.}
and generalisations thereof \cite{Aguilera-Damia:2026dbk,ThirringScwhinger}, and \(\O(N)\) vector models at large \(N\) \cite{LopezNacir:2016gzi,DiPietro:2023inn}. The purpose of this paper is to provide what is perhaps the simplest example of this kind: the two-dimensional Ising model. As will be made clear in the following sections, this model combines exact solvability with a range of non-trivial cosmological phenomena.

The two-dimensional Ising model in the continuum can be defined as the Ising conformal field theory (CFT) deformed by its relevant thermal scalar operator $\varepsilon(x)$ of dimension $\Delta = 1$. Its action can be schematically written as
\begin{equation}
    S[\tau] = S_\t{CFT} + \tau \int_{\ds_2} \dd[2]{x}\, \sqrt{\abs{g}} \;\varepsilon(x)~.
\end{equation}
This is a strongly coupled massive theory. First, the CFT action, $S_\t{CFT}$, does not have a weakly coupled description in terms of the fundamental CFT fields. Second, the relevant deformation, $\tau$, need not be small. This theory is the scaling limit of the famous Ising lattice model.%
\footnote{While traditionally the Ising lattice model is defined on a square lattice, it has also been studied on spherical lattices \cite{Diego:1993wq,Hoelbling:1995hi,Holm:1995wa,Hoelbling:1996gv,Brower:2024otr}.}
The solvability of this model rests on the fermionisation map, which relates the theory to a free massive Majorana fermion together with a discrete gauge field. However, not all observables of the Ising model can be easily computed by fermionisation. Some of the local operators in the Ising model are non-local in terms of the fermionic field. This is, for instance the case of the spin operator $\sigma(x)$, which is the continuum avatar of the lattice spin variable. The presence of such operators is one of the reasons why the study of this theory is interesting.

With this model in hand, we can study several aspects of de Sitter dynamics. A first natural question concerns correlation functions in the Euclidean formulation of the theory, namely on the two-dimensional sphere. After analytic continuation to Lorentzian \(\ds_2\), these become correlation functions of local operators computed in a de~Sitter-invariant state, the Euclidean or Bunch--Davies vacuum \cite{Chernikov:1968zm,Spindel:1976,Bunch:1978yq,HartleHawking}.%
\footnote{The Bunch--Davies vacuum is not the only de Sitter-invariant vacuum one can consider. Indeed, free theories in de Sitter have a one-complex-parameter family of vacuum states called $\alpha$-vacua \cite{Chernikov:1968zm}. However, the Bunch--Davies vacuum is the only one for which two-point functions satisfy the Hadamard condition. See \cite{Mottola:1984ar,Allen:1985ux,Miller:2025jbz} for some studies in this direction. In the presence of global symmetries there are also other zero-particle Hadamard states that are, however, generically not invariant under the full de~Sitter group \cite{Kirsten:1993ug,Letsios:2026ypo}. An exception to this statement is provided by theories with mixed anomalies between zero-form and $(d-1)$-form global symmetries. Such theories possess multiple degenerate vacua on any compact spacetime and, in de Sitter space, admit Hadamard and de Sitter-invariant vacuum states \cite{Aguilera-Damia:2026dbk}.}
Among the most interesting observables for cosmology and inflation are the so-called \emph{cosmological correlators}, namely correlation functions of local operators evaluated at future infinity. De~Sitter isometries constrain these correlators to transform as conformal correlation functions. This has motivated recent proposals to define such observables non-perturbatively through a bootstrap philosophy \cite{Hogervorst:2021uvp}.\footnote{We refer to \cite{Baumann:2022jpr} and references therein for an overview of this subject.} The exact solvability of the Ising model lets us evaluate such correlators at finite coupling and track their conformal behaviour. 

Finally, de Sitter space comes with an intrinsic length scale, its radius $\lds$. Any massive QFT on this background therefore carries the dimensionless combination $\nu \coloneqq  m\lds$. Here $\nu\propto \tau\lds$. This allows us to track how correlation functions evolve as a function of this parameter. In the regime $\nu \ll 1$, the de~Sitter Ising model can be analysed in its original formulation using conformal perturbation theory (CPT). Although technically challenging, this procedure should in principle provide perturbative estimates for the exact correlation functions. These estimates can then be compared with the exact results, thus allowing one to test the validity of perturbation theory. Remarkably, perturbation theory is generically (but not always) spoiled at late times by the appearance of secular terms \cite{Ford:1985,Antoniadis:1986,TSAMIS19951,Senatore:2009cf,Polyakov:2012uc,Anninos:2014lwa,Gorbenko:2019rza,Akhmedov:2019cfd,Green:2020dynamicalrg}. Nevertheless, as also noted in \cite{Anninos:2024fty}, the exact result obtained by resumming the perturbative series cures these divergences. The resulting cosmological correlators remain finite at late times, with either real or complex scaling dimensions. As we will see, all of these phenomena appear in the de~Sitter Ising model.

\paragraph{Structure of the paper.}

The rest of the paper is organised as follows. In
\cref{sec:review-dS-QFT}, we review some basic facts about
two-dimensional de~Sitter spacetime and the unitary irreducible
representations (UIRs) of its isometry group. In \cref{sec:Isingmodel}, we
introduce the Ising model. We describe its UV fixed point, the
fermionisation map to the free Majorana theory, and the massive deformation that will be studied throughout the paper. 

We then move on to observables. In \cref{sec:sphere partition function}, we compute the exact sphere partition function of the theory and match it with the conformal perturbation theory computation in the Ising frame, to two-loop order. In \cref{sec: energy 2ptf}, we begin our study of correlation functions with the two-point function of the thermal operator \(\varepsilon(x)\). Since this operator is a local composite of the fermion field, its correlators can be computed exactly in the Majorana presentation of the model. We derive its late-time limit and verify the conformal behaviour of the corresponding cosmological correlator. We also compute two-point functions of descendant-like operators, obtained by applying differential operators to the thermal operator. These correlators display qualitatively new features, including late-time oscillations. We close the section with a stress test of conformal perturbation theory against the exact results at two-loop order.

In \cref{sec:spin 2pt}, we turn to the spin two-point function. As
emphasised above, the spin operator is non-local with respect to the
fermion field, and its two-point function therefore cannot be obtained by
Wick contractions. Instead, we use the differential equations for this
correlator that follow from the symmetries of the model, first derived in
\cite{Doyon:2004fv}. These equations allow us to derive non-perturbative
constraints on the late-time scaling dimensions of the spin and disorder operators. We test these constraints by computing the two-point functions numerically for different values of the dimensionless parameter \(\nu\). We again compare the exact results with conformal perturbation theory, now at one loop. We end in \cref{sec:outlook} with open problems and future directions. Technical details and loop computations are collected in \cref{app:dirac-op,app:sphere-parti,app:eps-eps-app,app:antipodal}.

\section{QFT in \texorpdfstring{$\ds_2$}{dS2}}\label{sec:review-dS-QFT}

In this section we review basic facts about QFTs in rigid two-dimensional de~Sitter spacetime. We refer to \cite{Spradlin:2001pw,Anninos:2012qw,Galante:2023uyf} for a more comprehensive discussion.

\begin{figure}[thb]
    \centering
    \begin{tikzpicture}[scale=3,line cap=round,line join=round]
        \coordinate (SW) at (0,0);
        \coordinate (SE) at (2,0);
        \coordinate (NE) at (2,2);
        \coordinate (NW) at (0,2);

        \fill[blue!18] (NW) -- (NE) -- (SW) -- cycle;

        \draw[thick] (SW) rectangle (NE);
        \draw[very thin,black!55] (NW) -- (SE);
        \draw[thick,red!70] (SW) -- (NE);

        \node[above] at (1,2) {$\mathcal I^+$};
        \node[below] at (1,0) {$\mathcal I^-$};
        \node[left] at (0,1) {\rotatebox{90}{north pole}};
        \node[right] at (2,1) {\rotatebox{270}{south pole}};
        \node[red!70,above,rotate=45] at (0.58,0.58) {null horizon};
        \node[blue!70] at (1,1.58) {planar patch};
    \end{tikzpicture}
    \caption{Penrose diagram of global $\ds_2$. Time runs upwards, from the past conformal boundary, \(\cI^-\), to the future one, \(\cI^+\). The vertical sides are the worldlines of the north and south poles of the spatial circle. The shaded triangle is the expanding planar patch covered by $(\eta,x)$; its future boundary reaches $\mathcal I^+$ at $\eta\rightarrow0$. The red diagonal is the null horizon of the patch.}
    \label{fig:ds2-planar-patch}
\end{figure}
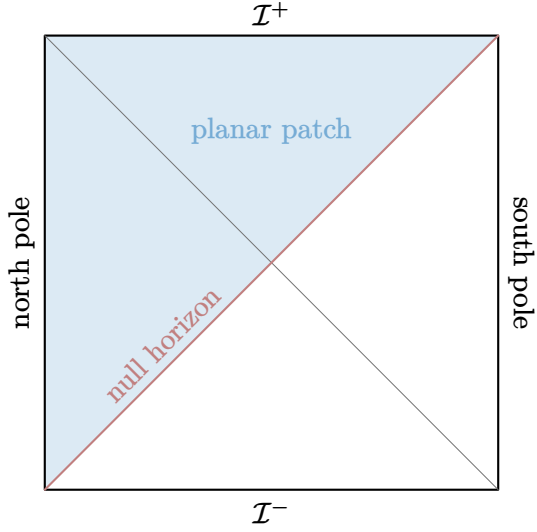

A convenient definition of $\ds_2$ is as a hyperboloid, embedded in $\R^{1,2}$ as:
\begin{equation}
    \eta_{AB}\,X^A\,X^B = \lds^2~, \qquad \eta = \operatorname{diag}(-1,1,1)~, 
\end{equation}
where $\lds$ is the Hubble radius setting the size of $\ds_2$. In global coordinates, the metric reads
\begin{equation}\label{eq:global-dS}
    \dd{s}^2 = \lds^2\qty(-\dd{t}^2+\cosh^2t\,\dd{\varphi}^2)~, \qquad  t \in \mathbb{R}~, \quad \varphi \sim \varphi + 2\pi~.
\end{equation}
Equivalently, introducing a compact time coordinate, $T=2\arctan(\ex{t})-\pi$, gives the metric
\begin{equation}
    \dd{s}^2 = \lds^2 \qty(\frac{-\dd{T}^2+\dd{\varphi}^2}{\sin^2 T})~, \qquad -\pi< T < 0~. 
\end{equation}
It is often useful to work in \emph{planar coordinates} in which the metric takes the form
\begin{equation}
    \dd{s}^2 = \lds^2\qty(\frac{-\dd{\eta}^2+\dd{x}^2}{\eta^2})~, \qquad \eta \in \open{-\infty}{0}~, \quad x \in \R~.
\end{equation}
These coordinates cover only half of the dS geometry. The coordinate $\eta$ plays the role of conformal time, with the late-time limit corresponding to $\eta \to 0$. The Penrose diagram of global $\ds_2$ and the subregion covered by the planar coordinates is depicted in \cref{fig:ds2-planar-patch}.

The Hilbert space of a unitary QFT with a dS-invariant vacuum is organised into unitary irreducible representations (UIRs) of the algebra of isometries of de~Sitter spacetime \cite{Nachtmann:1967nss}. Equivalently, it is organised in UIRs of the isometry group,%
\footnote{Strictly speaking, we mean the component of the isometry group connected to the identity. Other components can be obtained by a combination of spatial reflections and time reversal.}
or a cover thereof if the theory contains fermions. See \cite{Hirai:1962,Ottoson:1968,Schwarz:1971} for the original classification of the relevant UIRs and \cite{Basile:2016aen,Sun:2021thf,Sengor:2022kji,Enayati:2022hed,Schaub:2024rnl,Hinterbichler:2026xqf} for comprehensive accounts on de Sitter representation theory. 

In two dimensions, the isometry group of $\ds_2$ is isomorphic to the Euclidean conformal group in one dimension $\SO(2,1)$. Its double cover, relevant for fermionic theories, is $\SL(2,\R)$. We will be therefore concerned with $\SL(2,\R)$ representations. Denote the $\sl(2,\R)$ generators as $L_0$ and $L_\pm$, satisfying 
\begin{align}
    \comm{L_\pm}{L_0} = \pm L_\pm~, \qquad \comm{L_+}{L_-} = 2 L_0~.
\end{align}
The quadratic Casimir of $\sl(2,\R)$ is 
\begin{align}
    C = L_0^2 - \frac{1}{2}\qty\big(L_+L_-+L_-L_+)~.
\end{align}
We label the eigenvalues of $C$ as $\Delta(\Delta-1)$, where the complex number $\Delta$ is known as the \emph{conformal weight} and specifies the corresponding UIR. States $\ket{\Delta,n}$ in a UIR are eigenvectors of the maximal commuting subalgebra of $\sl(2,\mathbb{R})$ with eigenvalues
\begin{align}
   C \ket{\Delta,n} &= \Delta(\Delta-1) \ket{\Delta,n}~, \\
   L_0 \ket{\Delta,n} &= -n \ket{\Delta,n}~,  \\ 
   L_{\pm} \ket{\Delta,n} &= -(n\pm \Delta) \ket{\Delta,n\pm 1}~,
\end{align}
so $n$ is an $\so(2)$ quantum number. Exponentiating requires that $n\in\Z$ for $\SO(2,1)$ representations, and $n\in\frac{1}{2}\Z$ for $\SL(2,\R)$ representations. Representations with $n\in\Z$ are known as bosonic representations, while those with $n\in\Z+1/2$ are known as fermionic.%
\footnote{More generally, there are representations of $\sl(2,\R)$ that exponentiate only to representations of the universal cover $\tilde{\SL}(2,\R)$ of $\SO(2,1)$. These have $n\in\Z+\mu$ with $\mu\in \lopen{-\frac12}{\frac12}+\ii \R$ and are anyonic representations. See \cite{Kitaev:2017hnr,Hinterbichler:2026xqf} for details.}
Not every value of $\Delta$ gives rise to a UIR. In two dimensions, the UIRs fall into three families according to the possible values of $\Delta$:
\begin{itemize}
    \item The \emph{principal series}, $\cP_\Delta$, for which $\Delta \in \frac{1}{2}+\ii\,\mathbb{R}$. In this branch $n$ can either be an integer or a half-integer.
    \item The \emph{complementary series}, $\mathcal{C}_{\Delta}$, for which $\Delta \in \open{0}{1}$. Here $n$ must be an integer, so the complementary series contains no fermionic representations. 
    \item The \emph{discrete series}, $\mathcal{D}_\Delta^{\pm}$, which has $\Delta \in \frac{1}{2}\Z_{>0}$. Half-integer values of $\Delta$ give half-integer $n$, and hence fermionic representations, while integer $\Delta$ give bosonic representations. The label $\pm$ records whether the representation contains a highest- or a lowest-weight state annihilated by $L_{\pm}$, respectively. 
\end{itemize}
Note that the Casimir is invariant under the shadow transformation $\Delta \leftrightarrow (1-\Delta)$. To avoid overcounting, it is customary to restrict the principal series to $\Im\Delta>0$ and the complementary series to $\Delta>\frac{1}{2}$.

Starting from global $\ds_2$, as in \cref{eq:global-dS}, one can obtain the metric of the two-dimensional sphere by Wick rotation. In particular, the map $t \rightarrow \ii\qty(\vartheta - \pi/2)$ gives the round metric of $S^2$:
\begin{equation}
    \dd{s}^2 = \lds^2\qty(\dd{\vartheta}^2+\sin^2\vartheta \dd{\varphi}^2)~, \qquad 0\leq \vartheta\leq\pi~, \quad \varphi\sim\varphi+2\pi~. 
\end{equation}
Euclidean de~Sitter is therefore compact, which makes the analysis of Euclidean de~Sitter QFT more tractable. In particular, one can often compute observables in Euclidean signature and then analytically continue the result back to Lorentzian dS, in order to extract real-time properties. This is the strategy adopted in this paper.

The basic observables of interest are correlation functions of local operators evaluated in the Euclidean, or Bunch--Davies, vacuum of de~Sitter spacetime \cite{Chernikov:1968zm,Spindel:1976,Bunch:1978yq}. Such correlators can be defined by analytic continuation from correlation functions computed on the Euclidean sphere; see \cite{Higuchi:2010xt} for a detailed discussion. 
As with states, de~Sitter covariance strongly constrains correlation functions. For instance, scalar two-point functions can only depend on the de~Sitter-invariant distance%
\footnote{Correlators of non-zero spin depend also on the allowed tensor and spinor structures; see e.g. \cite{Allen:1985wd,Camporesi:1995fb,Letsios:2020twa,Schaub:2023scu}.}
\begin{align}\label{eq:thermal-uE-def}
	\u_{xy} \coloneqq 1-\cos\Theta_{xy} = 2\sin^2\frac{\Theta_{xy}}{2}~ \in [0 ,2]~,
\end{align}
where \(\Theta_{xy}\) is the geodesic angle between the two points. The limit $\u \rightarrow 0$ corresponds to coincident points, while $\u \rightarrow 2$ describes antipodal points on the sphere. Analytic continuation to Lorentzian de~Sitter sends the Euclidean invariant distance to
\begin{equation}
    \u_{xy} \rightarrow \uL_{xy} = \frac{\cos(T_x-T_y)-\cos(\varphi_x-\varphi_y)}{\sin(T_x)\sin(T_y)}~.
\end{equation}
The range $\uL_{xy}>0$ corresponds to spacelike separation, $\uL_{xy}=0$ is null, while $\uL_{xy}<0$ corresponds to timelike separation. When there is no risk of confusion, we will drop the subscripts \(x,y\)
and denote the invariant distance simply by \(u\). It will sometimes also be convenient to use the rescaled variable
\begin{equation}
    X \coloneqq \frac{u}{2}~,
\end{equation}
instead.

One of the most interesting theoretical challenges of de~Sitter physics is to understand correlators with insertions taken on the late-time surface, also called future boundary. In planar coordinates, it is the codimension-one submanifold of $\ds_2$ reached in the limit $\eta \to 0^-$. These are the so-called \emph{cosmological correlators}. Because of the \ds\ isometries, correlation functions on the future boundary transform under the Euclidean conformal group. This implies that cosmological correlators have a precise dependence on the Lorentzian de~Sitter-invariant distance $\uL_{xy}$. For example, two-point cosmological correlators have the general late-time behaviour:
\begin{equation}
    \ev{\cO(x) \cO(y)} \xrightarrow{\uL \rightarrow \infty} \sum_{\tilde\Delta}c_{\tilde\Delta} \qty(\uL_{xy})^{-\tilde\Delta} + \cdots~, 
\end{equation}
where the dots represent subleading terms in $\uL$. Generically, $\tilde{\Delta}$ may range over both discrete and continuous values, and it can include the value $\tilde{\Delta}=0$, which accounts for disconnected contributions. See \cite{SalehiVaziri:2024joi} for details. In fact, under reasonable assumptions, one can show that the operators $\cO(x)$ appearing in the two-point functions can be expanded around the future boundary in terms of boundary operators transforming as primaries and descendants under $\SO(2,1)$ \cite{Loparco:2023rug}.

Importantly, the scaling dimensions $\tilde{\Delta}$ need not obey the usual unitarity bounds of Euclidean
CFT. Instead, unitarity is imposed in the dS UIRs \cite{Sengor:2019mbz,Penedones:2023uqc,Cohen:2024anu,VasilisGizem}, and may result in complex late-time dimensions \cite{Guijosa:2003ze,Anous:2020dbn,}. One of the goals of this work is to determine these dimensions non-perturbatively for various cosmological correlators.

\section{The Ising model and its free field realisation}\label{sec:Isingmodel}

The model we analyse in this paper is the two-dimensional Ising model perturbed by the thermal operator. As is well known --- and reviewed below --- this theory has a dual description in terms of a massive Majorana fermion coupled to a discrete gauge field. In fact, in the case of interest the gauge field decouples and we are left with a massive Majorana. This duality gives us non-perturbative control over various observables, which we can then compare with the perturbative expansion around the UV conformal fixed point. Before turning to the perturbed theory, let us briefly review some properties of the Ising CFT.

\subsection{Ising CFT}

The two-dimensional Ising CFT has central charge $c=1/2$ and it is the first unitary minimal model. Its primary operators are
\begin{equation}
    \unit ~, \qquad \sigma(x)_{\frac{1}{16},\frac{1}{16}}~, \qquad \varepsilon(x)_{\frac{1}{2},\frac{1}{2}}~,
\end{equation}
where the subscripts denote the holomorphic and anti-holomorphic conformal dimensions $h$ and $\bar{h}$. The operator $\sigma(x)$ is the \emph{spin operator}. It is the continuum limit of the spin degrees of freedom in the statistical-mechanics lattice definition of the Ising model. The operator $\varepsilon(x)$ is the  \emph{thermal} or \emph{energy operator}; we will use both names interchangeably. The names are motivated by the fact that, on the one hand, $\epsilon$ induces the thermal deformation, driving the theory away from the critical temperature, while on the other hand it corresponds to the trace of the stress tensor outside criticality. 

This theory possesses an (invertible) $\Z_2$ global symmetry generated by a topological line, $\eta$. The spin operator $\sigma(x)$ is the only local primary operator that is charged under this symmetry. As is customary in 2d CFTs, given a global symmetry, one can consider the Hilbert space twisted by the symmetry action. Under the state-operator map, states in the twisted Hilbert space correspond to twist operators, i.e. local operators living at the endpoint of a topological symmetry line. For the $\Z_2$ symmetry of the Ising CFT, the $\Z_2$-twist operators are
\begin{equation}
    \mu_{\frac{1}{16},\frac{1}{16}}~, \qquad \qty(\psi_\t{L})_{\frac{1}{2},0} ~, \qquad \qty(\psi_\t{R})_{0,\frac{1}{2}}~. 
\end{equation}
Here $\mu$ is the \emph{disorder} operator, and is closely connected to $\sigma$ as will be made clear shortly. The operators $\psi_\t{L,R}$ are left- and right-moving fermions, whose names originate in the fermionisation duality which we will review below. 

Besides the invertible $\Z_2$ symmetry, the Ising CFT has a non-invertible symmetry: Kramers--Wannier (KW) duality \cite{Frohlich:2004ef}. KW duality acts on the localised operators as
\begin{equation}\label{eq:KW-def}
    \quad\sigma \leftrightarrow \mu~, \qquad \varepsilon \mapsto -\varepsilon~, \qquad \psi_\t{L}\mapsto-\psi_\t{L}~, \qquad \psi_\t{R}\mapsto \psi_\t{R}~. 
\end{equation}
Note that this symmetry mixes local and twist operators. This is a smoking-gun signal of its non-invertibility. In particular, it implies that $\sigma$ is annihilated when surrounded by a loop of the non-invertible topological line \cite{Chang:2018iay}.\footnote{Since KW is a symmetry of the Ising CFT, one can also consider the corresponding twisted Hilbert space as well as twist operators. We will not discuss them in this paper.}

Remarkably, the Ising CFT can be mapped to the theory of a single free massless Majorana fermion. This map is obtained by gauging the $\Z_2$ symmetry, with fermionic discrete torsion given by the Arf invariant; see \cite{Karch:2019lnn} for more details. Discrete gaugings in two dimensions can be understood in two equivalent ways. The first, more traditional, approach is to couple the theory to a $\Z_2$ gauge field and then path-integrate over it. The second approach reinterprets gauging as the insertion of a composite topological line $\mathcal{A} := 1 + \eta$, known as a Frobenius algebra \cite{Bhardwaj:2017xup}, that wraps all non-contractible cycles of the spacetime manifold.%
\footnote{In this second picture, the discrete torsion is a non-zero weight assigned to junctions making up the network of topological lines.}
As we will see, since $S^2$ has no non-contractible cycles, the gauging procedure will simplify considerably in the case of interest.

In the gauged theory, the original $\Z_2$ symmetry gives way to a new $\Z_2$ global symmetry: the $(-1)^F$ fermion-number symmetry of the fermionic theory. Moreover, the gauging reshuffles twisted and untwisted sectors. The fermionic operators $\psi_\t{L}$ and $\psi_\t{R}$ now become genuine local operators describing the left- and right-moving components of the Majorana fermion $\Psi$. The spin and disorder operators, $\sigma$ and $\mu$, become twist operators that live at the end of a $(-1)^F$ topological line. In string theory language, they can be understood as operators creating states in the Ramond sector \cite{Runkel:2020zgg}. Finally, the thermal operator $\varepsilon$, being neutral under the $\mathbb{Z}_2$ symmetry, remains a local operator and becomes the fermion bilinear $\varepsilon\propto \Psi^\sfT C\, \Psi$. 

Here, our interest lies in placing this model (and its deformation) on de~Sitter space. Since Euclidean $\ds_2$ is isomorphic to the sphere, the gauging operation simplifies drastically. On the two-dimensional sphere without operator insertions there are no non-contractible one-cycles. Therefore, as far as the partition function is concerned, the gauging procedure is trivial and the partition functions of the Ising and the Majorana CFTs agree. Operator insertions create non-contractible cycles around each insertion point. Now, the effect of the gauging is to reshuffle twisted and untwisted operators, in the way described before. Thus one also gets a one-to-one map between correlation functions.

The main advantage of the fermionic frame is that, being free, it is exactly solvable. This implies solvability of the Ising CFT as well. In particular, all conformal data and therefore all correlation functions are known explicitly. The relevant two- and three-point functions of local Ising operators on the plane are
\begin{equation}
    \begin{split}
       \ev{\sigma(z_1)\sigma(z_2)}_\C &= \abs{z_{12}}^{-1/4}~, \\
\ev{\epsilon(z_1)\epsilon(z_2)}_\C &=\abs{z_{12}}^{-2}~, \\
\ev{\sigma(z_1)\sigma(z_2)\epsilon(z_3)}_\C &=
\frac{1}{2}
\abs{z_{12}}^{3/4}\,
\abs{z_{13}}^{-1}\, \abs{z_{23}}^{-1}~, \\ 
\ev{\sigma\epsilon\epsilon}_\C =
\ev{\sigma\sigma\sigma}_\C =
\ev{\epsilon\epsilon\epsilon}_\C &= 0~,
    \end{split}\label{eq: 2ptf ising}
\end{equation}
where $z_{ij} \coloneqq z_i-z_j$. We will also need the following four-point
functions:
\begin{equation}
	\begin{aligned}
		\ev{
			\varepsilon(z_1)
			\varepsilon(z_2)
			\varepsilon(z_3)
			\varepsilon(z_4)
		}_{\C}^{\t{c}}
		&=
		\frac{
			G_{\varepsilon\varepsilon\varepsilon\varepsilon}^{\t{c}}
			(\eta,\bar{\eta})
		}{
			\abs{z_{13}}^2
			\abs{z_{24}}^2
		}~,
        \qquad 
        &&G_{\varepsilon\varepsilon\varepsilon\varepsilon}^{\t{c}}
		(\eta,\bar{\eta})
		~\coloneqq
		\frac{
			(\eta-\bar{\eta})^2
		}{
			\abs{\eta}^2
			\abs{1-\eta}^2
		}~,
        \\
		\ev{
			\varepsilon(z_1)
			\varepsilon(z_2)
			\sigma(z_3)
			\sigma(z_4)
		}_{\C}^\t{c}
		&=
		\frac{
			G_{\varepsilon\varepsilon\sigma\sigma}^{\t{c}}
			(\eta,\bar{\eta})
		}{
			\abs{z_{12}}^2
			\abs{z_{34}}^{1/4}
		}~, \qquad
   		&&G_{\varepsilon\varepsilon\sigma\sigma}^{\t{c}}
		(\eta,\bar{\eta})
		\coloneqq
		\frac{1}{4}
		\qty(
			\frac{\abs{2-\eta}^2}{\abs{1-\eta}}
			-4
		)~,
	\end{aligned}
	\label{eq: 4pt Ising}
\end{equation}
where
\begin{equation}
	\eta
	\coloneqq
	\frac{z_{12}z_{34}}{z_{13}z_{24}}
\end{equation}
is the cross-ratio and the superscript $\t{c}$ denotes connected correlators.

\subsection{The thermal deformation}

The main focus of this work is the Ising model away from criticality. From the operator content of the CFT, it is clear that there are two relevant deformations one can turn on: the spin operator $\sigma$, whose dimension is $\Delta_{\sigma} = \frac{1}{8}$, and the thermal operator $\varepsilon$, with dimension $\Delta_{\varepsilon}=1$. When only one of the two deformations is turned on, the theory on flat space is integrable. In both cases, the theory flows to a trivially gapped vacuum and the entire flat-space spectrum, as well as the exact scattering matrix, is known exactly \cite{Zamolodchikov:1989cf,Delfino:2004yva}. The flow triggered by the thermal operator $\varepsilon$ is simpler, since this operator has a local description in the dual fermionic presentation of the theory. This is the deformation we will study here.

The action of the theory can be written in Euclidean signature as
\begin{equation}\label{eq: ising action}
    S_\t{Ising} = S_\t{CFT} + \tau \int \dd[2]{x}\, \sqrt{g} \;\varepsilon(x)~. 
\end{equation}
Under the fermionisation map, this becomes the Majorana action
\begin{equation}\label{eq:majorana-action}
	S_\t{Maj}[\Psi] = \frac{1}{2}\int \dd^2 x\,\sqrt{g}\;
	\Psi^\sfT C\, \qty(\dsl + m)\,\Psi~, 
\end{equation}
where $\dsl = \gamma^\mu\nabla_\mu$ is the curved-space Dirac operator, $C$ is the charge-conjugation matrix 
\begin{equation}
    C = \mat{0 & 1 \\ -1 & 0}~,
\end{equation}
and $m$ is proportional to $\tau$. The precise relation between them will be fixed in \cref{sec:sphere partition function} by
matching the two sphere partition functions. The Majorana field $\Psi$ is a two-component real fermion $\Psi^\sfT=\qty(\psi_\t{L}\,,\;\psi_\t{R})$. Let us emphasise that, although the theory has a free realisation, this does not make all observables equally simple. Only correlators of operators that are local with respect to the fermion
field \(\Psi\) can be computed directly by Wick contractions. Correlators involving non-local fields, such as \(\sigma\) or \(\mu\), are much more subtle.

Let us also comment on Kramers--Wannier duality. At the critical point, it is a genuine non-invertible symmetry of the theory, as explained above. The thermal operator, however, is charged under it and changes sign under the duality action. Away from criticality, this implies that the massive theory at coupling $\tau$ is mapped to the theory at coupling $-\tau$, together with the rest of the transformations in \cref{eq:KW-def}. This duality will be useful for some of the arguments below. Note also that although the massive theory has no non-invertible topological defects, \cite{Ambrosino:2025pjj} proposed a replacement, in the form of ``translation-invariant'' defects. It would be interesting to recast some of our arguments on a symmetry basis in terms of these defects.

On flat space, the theory \cref{eq: ising action} does not have any dimensionless parameter and the relevant parameter $\tau$ (or $m$) simply sets the mass scale of the spectrum. In contrast, on $\ds_2$ the \ds\ radius $\lds$ allows one to form the dimensionless  combination $\nu = m\lds \propto \tau\, \lds$, which can be tuned arbitrarily. In particular, in the regime $\nu \ll 1$ one can study the perturbative expansion around the UV conformal fixed point. 

Since many observables can be computed exactly in the free field realisation of the theory, this is a perfect toy model to address the well-known issues of perturbation theory in de~Sitter spacetime. This will be one of the main themes of the rest of the paper. It is worth noting, in passing, that recent solvable fermionic models in de~Sitter have mostly involved Dirac fermions, as in the cases of \cite{Anninos:2024fty,Aguilera-Damia:2026dbk}. The Ising model instead provides the first solvable example in which Majorana fermions play a direct role.

\section{The sphere partition function}\label{sec:sphere partition function}

We begin with the simplest observable, the sphere partition function. Through this computation, we will set up the conformal perturbative expansion that we will use in the rest of the paper. Moreover, it will allow us to pin down the precise map between the Ising coupling $\tau$ and the Majorana mass $m$.

Let us start from the fixed point $\tau = 0$, which corresponds to $m = 0$ in the fermionic frame. By conformal invariance, the $\lds$ dependence of the partition function is completely fixed by the Weyl anomaly as
\begin{equation}
    \dv{\log\lds}\, \log\parti_\t{CFT}[\lds] = \frac{c}{24\pi}\int_{S^2_{\lds}} \dd[2]{x}\, \sqrt{g}\; \cR = \frac{c}{3}~, 
\end{equation}
where $\cR$ is the Ricci scalar of the sphere. Integrating this equation gives
\begin{equation}
    \log\parti_\t{CFT}[\lds] = \frac{c}{3}\log(\frac{\lds}{\lds_0})~, 
\end{equation}
where $\lds_0$ is a scheme-dependent integration constant. As usual, the CFT partition function depends only on the central charge $c$ of theory. Setting $c=1/2$, we get the Ising/Majorana partition function.

Let us now move to the massive theory. We will denote the sphere partition function of the deformed theory at coupling $\tau$ by $\parti_\t{Ising}[\lds,\tau]$, with $\parti_\t{Ising}[\lds,0] \equiv \parti_\t{CFT}[\lds]$. By the fermionisation duality, we also have 
\begin{align}
    \parti_\t{Ising}[\lds,\tau] = \parti_\t{Maj}[\lds,m]~,
\end{align}
in terms of the Majorana partition function. The full Ising sphere partition function can be written as
\begin{equation}
    \log\parti_\t{Ising}[\lds,\tau] = \log\parti_\t{CFT}[\lds] + \log\frac{\parti_\t{Ising}[\lds,\tau]}{\parti_\t{Ising}[\lds,0]} = \frac{1}{6}\log(\frac{\lds}{\lds_0}) + \log\frac{\parti_\t{Ising}[\lds,\tau]}{\parti_\t{Ising}[\lds,0]}~.
\end{equation}
The new contribution can be computed exactly in the Majorana frame. Performing the Gaussian integral gives%
\footnote{For fermion path integrals one also needs to keep track of the phase of the partition function \cite{Witten:2015aba}. The Majorana Pfaffian is real, but its sign is determined by the Arf invariant \cite{Tong:2019bbk}. On the sphere there is a unique spin structure, so there is no sign ambiguity.}
\begin{equation}
    \parti_\t{Maj}[\lds,m] = \pf\!\qty\big(C(\dsl + m))~. 
\end{equation}
The spectrum of the Dirac operator on $\S^2_\lds$ is \cite{Camporesi:1995fb}
\begin{align}
	\dsl\,\psi_{\pm,n,k}^{(s)} = \pm \frac{\ii\, n}{\lds}\,\psi_{\pm,n,k}^{(s)}~, 
	\qquad  
	n=1,2,\dots~, \ 
	k=0,\dots,n-1~, \ s=\pm~. 
\end{align}
with degeneracy $d_n=2n$ for each sign. See \cref{app:dirac-op} for more details. Hence we have
\begin{align}
	\log\frac{\parti_\t{Maj}[\lds,m]}{\parti_\t{Maj}[\lds,0]} = \sum_{n=1}^\infty n \log(1+\frac{\nu^2}{n^2})~, \qquad \nu = m\lds~. 
\end{align}
This sum is divergent. We can isolate the divergent contribution by rewriting it as
\begin{equation}
    \log\frac{\parti_\t{Maj}[\lds,m]}{\parti_\t{Maj}[\lds,0]} =\sum_{n=1}^{\infty}\frac{\nu^2}{n}+ \sum_{n=1}^\infty\left( n \log(1+\frac{\nu^2}{n^2})-\frac{\nu^2}{n}\right)~. 
\end{equation}
The second term is now finite, while the first gives the divergent contribution 
\begin{equation}
    \sum_{n=1}^{\infty}\frac{\nu^2}{n} = \nu^2\, \zeta(1)~.
\end{equation}
Nonetheless, this divergence can be removed by a local counterterm, namely a cosmological constant counterterm. We write
\begin{equation}
\begin{split}
     S_\t{Maj}[\Psi] &= \int_{\S^1_\lds} \dd^2 x\,\sqrt{g}\qty(\frac{1}{2} \Psi^\sfT C\, \qty(\dsl + m)\,\Psi+\delta \Lambda)  \\ &= \frac{1}{2}\int \dd^2 x\,\sqrt{g}\;
	\Psi^\sfT C\, \qty(\dsl + m)\,\Psi + 4\pi \lds^2\, \delta \Lambda~. 
\end{split}
\end{equation}
This gives 
\begin{equation}
    \log\parti_\t{Maj}[\lds,m] = \log\parti_\t{Maj}[\lds,0] +\log\frac{\parti_\t{Maj}[\lds,m]}{\parti_\t{Maj}[\lds,0]} -4\pi \lds^2\, \delta \Lambda~, 
\end{equation}
and results in a finite partition function by setting 
\begin{equation}
    \delta \Lambda = \frac{\zeta(1)-\alpha}{4\pi}m^2~, \qquad \alpha \in \R~, 
\end{equation}
where $\alpha$ is a finite scheme-dependent constant. Therefore, we finally get the renormalised sphere partition function
\begin{equation}\label{eq: exact Z}
    	\log \frac{\parti_\t{Maj}^\t{ren}[\lds,m]}{\parti_\t{Maj}[\lds,0]} = \alpha\,\nu^2 + \sum_{n=1}^\infty \qty[
		n\log\!\left(1+\frac{\nu^2}{n^2}\right)-\frac{\nu^2}{n}
	]~.
\end{equation}

\subsection{Conformal perturbation theory on the sphere}

We now match the exact result for the partition function with the perturbative expansion around the strongly coupled Ising model. To this end, let us first expand the non-perturbative result \cref{eq: exact Z} in powers of $\nu$, for $\abs{\nu}<1$. Using
\begin{equation}  
	\log(1+\frac{\nu^2}{n^2}) = \sum_{k=1}^\infty \frac{(-1)^{k+1}}{k}\frac{\nu^{2k}}{n^{2k}}~,
\end{equation}
we get
\begin{equation}
    \log\frac{\parti_\t{Maj}^\t{ren}[\lds,m]}{\parti_\t{Maj}[\lds,0]}
	=
	\alpha\,\nu^2
	+
	\sum_{k=2}^\infty
	\frac{(-1)^{k+1}}{k}\,\zeta(2k-1)\,\nu^{2k} = \alpha\,\nu^2
	-\frac{\zeta(3)}{2}\nu^4 + \order{\nu^6}~. \label{eq:Zmaj-nu4}
\end{equation}
This simple result is actually a highly non-trivial prediction of Ising/Majorana duality.

Let us now turn to the Ising frame, described by the action \cref{eq: ising action}. The sphere partition function can be written as
\begin{equation}
    \frac{\parti_\t{Ising}[\lds,\tau]}{\parti_\t{Ising}[\lds,0]}
	=
	\ev{\exp(- \tau \int_{\S^2_\lds} \dd[2]{x}\,\sqrt{g}\,\varepsilon(x))}_{\kern-3pt0},
\end{equation}
where $\ev{\cdots}_{0}$ denotes expectation values in the critical theory. For $\tau\lds \ll 1$, we can expand this correlator in conformal perturbation theory; see \cite{Zamolodchikov:2001dz}. This gives
\begin{equation}
    \log \frac{\parti_\t{Ising}[\lds,\tau]}{\parti_\t{Ising}[\lds,0]}
	=
	\sum_{n=1}^\infty
	\frac{(-\tau)^n}{n!}
	\int_{\qty(\S^2_\lds)^n}
	\prod_{i=1}^n \dd[2]{x_i}\,\sqrt{g(x_i)}\;
	\ev{
		\vphantom{A^A}
		\varepsilon(x_1)\cdots \varepsilon(x_n)
	}^\t{c}_\t{0}~. 
\end{equation}
Here, $\ev{\cdots}^\t{c}_0$ denotes the connected correlator in the Ising CFT. Since $\varepsilon(x)$ is odd under Kramers--Wannier duality, only correlation functions with an even number of insertions of $\varepsilon$ can contribute to the above expansion.

The $\order{\tau^2\lds^2}$ contribution comes from the integrated two-point function. Explicitly,
\begin{equation}
\begin{split}
    \log \frac{\parti_\t{Ising}[\lds,\tau]}{\parti_\t{Ising}[\lds,0]} &= \frac{\tau^2}{2}\int_{\qty(S^2_\lds)^2} \dd[2]{x_1} \sqrt{g(x_1)} \;\dd[2]{x_2} \sqrt{g(x_2)}\; \langle \varepsilon(x_1) \varepsilon(x_2)\rangle + \order{(\tau \lds)^4} 
    \\ 
    &\eqqcolon \frac{(\tau\lds)^2}{2}\cI_2+\order{(\tau \lds)^4}~.
\end{split}\label{eq:Z-CPT-nu2}
\end{equation}
It is convenient to introduce stereographic coordinates on the sphere, in terms of which the metric reads:
\begin{align}\label{eq:stereo-1}
	\dd{s}^2
	= \Omega(z)^2 \dd{z} \dd{\bar{z}}~, \qq{with}
	\Omega(z) = \frac{2\lds}{1+\abs{z}^2}~. 
\end{align}
The conformal factor can be removed by a Weyl transformation, but we must keep track of how the energy operator transforms. Since $\epsilon$ has dimension $\Delta=1$, its transformation is given by
\begin{equation}
	\varepsilon_{\S^2_\lds}(z,\bar{z})=\Omega(z)^{-1}\,\varepsilon_{\C}(z,\bar z)~.
	\label{eq:app-epsilon-weyl}
\end{equation}
Therefore, the integral $\cI_2$ from \cref{eq:Z-CPT-nu2}, controlling the $\order{\nu^2}$ contribution, can be written as
\begin{equation}\label{eq:I2-reduced-main}
\begin{split}
      \cI_2 &= \frac{1}{\lds^{2}}\int_{\C^2}\dd[2]{z_1}\, \dd[2]{z_2} \;\Omega(z_1)\,\Omega(z_2)\;\langle \varepsilon(z_1)\varepsilon(z_2)\rangle_{\C} \\
      &= 4 \int_{\C^2} \dd[2]{z_1}\, \dd[2]{z_2}\;\frac{1}{(1+\abs{z_1}^2)(1+\abs{z_2}^2)\abs{z_1-z_2}^2}~.
\end{split}
\end{equation}
As in the Majorana frame, this contribution is divergent and renormalises the cosmological constant counterterm. In particular, as shown in \cref{app: nu2},
\begin{equation}\label{eq:I2-final}
    \cI_2 = 8\pi^2 \zeta(1)~. 
\end{equation}
Therefore
\begin{equation}
    \log \frac{\parti_\t{Ising}[\lds,\tau]}{\parti_\t{Ising}[\lds,0]} = (2\pi\tau\lds)^2 \zeta(1)+\order{\tau^4\lds^4}~.
\end{equation}
Matching with the Majorana computation, we conclude that the correct duality map is
\begin{equation}\label{eq:tau-m}
    \tau = \frac{m}{2\pi}~,
\end{equation}
and so we also identify $\nu \equiv m\lds = 2\pi\,\tau\lds$.

Let us now explicitly compute the $\order{\nu^4}$ correction to check the validity of this prediction. With the identification \cref{eq:tau-m}, the perturbative expansion takes the form
\begin{equation}
    \begin{split}
    \log \frac{\parti_\t{Ising}[\lds,\tau]}{\parti_\t{Ising}[\lds,0]} 
    &= \frac{\nu^2}{8\pi^2}\, \cI_2 +\frac{\nu^4}{4!(2\pi)^4}\cI_4 + \order{\nu^6}~,
\end{split}
\end{equation}
where 
\begin{equation}
    \cI_4 \coloneqq \frac{1}{\lds^4}
	\int_{\qty(\S^2_\lds)^4}
	\prod_{i=1}^4 \dd[2]{x_i}\,\sqrt{g(x_i)}\;
	\ev{
		\vphantom{A^A}
    \varepsilon(x_1)\varepsilon(x_2)\varepsilon(x_3)\varepsilon(x_4)
	}^\t{c}_{\S^2_\lds}~.
\end{equation}
Using the known explicit form of the connected energy four-point function in \cref{eq: 4pt Ising} and again performing a Weyl transformation from the sphere to the plane, we get
\begin{equation}\label{eq:I4-main-pre}
    \cI_4 = \frac{1}{\lds^{4}} \int_{\C^4}\prod_{i=1}^4 \dd[2]{x_i}\,
	\frac{\Omega(x_1)\Omega(x_2)\Omega(x_3)\Omega(x_4)}{\abs{x_{13}}^2\,\abs{x_{24}}^2}\frac{(\eta-\bar\eta)^2}{\abs{\eta}^2\,\abs{1-\eta}^2}~. 
\end{equation}
This integral is manifestly finite. We devote \cref{app:s2nu4} to its evaluation. When the dust settles, we find
\begin{equation}\label{eq:I4-final}
    \cI_4 = -12\, (2\pi)^4 \,\zeta(3)~,
\end{equation}
in complete agreement with the exact result in \cref{eq:Zmaj-nu4}. Note that this matching is highly non-trivial. In particular, the value $\zeta(3)$ appears only as a result of an integral involving the Bloch--Wigner dilogarithm \cite{zagier2007dilogarithm}, which also has a
surprising connection to the geometry of hyperbolic three-manifolds \cite{thurston2022geometry}.

\section{Energy two-point function}\label{sec: energy 2ptf}

We now turn to more interesting observables, namely de~Sitter correlation functions. The simplest example is the two-point function of the energy operator, $\ev{\varepsilon(x)\varepsilon(y)}$. This correlator is particularly tractable, as $\varepsilon(x)$ is a local bilinear of the free fermionic field $\Psi$. Explicitly,
\begin{equation}
    \varepsilon(x) = \pi\,\Psi(x)^\sfT C\,\Psi(x)~. 
\end{equation}
The full non-perturbative correlator therefore follows by Wick contractions in the fermion picture. The connected correlator on the sphere is:
\begin{equation}
    G_\varepsilon\qty(x,y) \coloneqq \ev{\varepsilon(x)\varepsilon(y)}^\t{c}_{\S^2_\lds}
	=
	-2\pi^2\,\tr[S_m(x,y)S_m(y,x)]~, 
	\qquad
	x \neq y~. 
\end{equation}
Here, $S_m(x,y)$ is the free fermion propagator, the trace is taken over spinor indices and the factor of $2$ comes from the antisymmetric Majorana bilinear. The propagator satisfies the differential equation
\begin{equation}   \label{eq: green eq}
	\qty(\dsl_x + m)\,S_m(x,y) = \frac{\delta^{(2)}(x,y)}{\sqrt{g_x}}\,\unit~. 
\end{equation}
Using the $\SO(3)$ isometries, we are free to place $y$ at the north pole of the sphere, $y=\t{N}$. In \cref{app:dirac-op} we show that the propagator takes the form:
\begin{align}\label{eq:thermal prop ansatz}
	S_m(x) \coloneqq S_m(x,\t{N})
	=
	\begin{pmatrix}
		\ex{\ii\varphi/2}\,a(\u) & \ex{-\ii\varphi/2}\,b(\u) \\
		\ex{\ii\varphi/2}\,b(\u) & \ex{-\ii\varphi/2}\,a(\u)
	\end{pmatrix}~, 
\end{align}
where we have made explicit the two chiral components of the Majorana field. See also \cite{Mueck:1999efk,Letsios:2020twa,Pethybridge:2021eoh,Schaub:2023scu,Letsios:2025pqo} for details. Solving \cref{eq: green eq} determines the functions $a(\u)$ and $b(\u)$ as
\begin{align}\label{eq:thermal-a-closed}
	a(\u)
	&=
	\frac{\nu\,\Gamma(1+\ii\nu)\Gamma(1-\ii\nu)}{4\pi\lds}
	\sqrt{1-\frac{\u}{2}}\;
	{}_2F_1\!\qty(1+\ii\nu,1-\ii\nu;2;1-\frac{\u}{2})~, 
	\\
	\label{eq:thermal-b-closed}
	b(\u)
	&=
	\frac{\Gamma(1+\ii\nu)\Gamma(1-\ii\nu)}{4\pi\lds}
	\sqrt{\frac{\u}{2}}\;
	{}_2F_1\!\qty(1+\ii\nu,1-\ii\nu;1;1-\frac{\u}{2})~. 
\end{align}
Combining everything together gives the two-point function
\begin{align}\label{eq:thermal 2pt exact}
	\begin{split}
	G_\varepsilon\qty(\u_{xy}) 
	&=
	\frac{\Gamma(1+\ii\nu)^2\Gamma(1-\ii\nu)^2}{4\lds^2}
	\left\{
	\frac{\u_{xy}}{2}\,
	\qty[{}_2F_1\!\qty(1+\ii\nu,1-\ii\nu;1;1-\frac{\u_{xy}}{2})]^2 \right.
	\\
	&\hspace{6em}
	\left.-\nu^2\qty(1-\frac{\u_{xy}}{2})\,
	\qty[{}_2F_1\!\qty(1+\ii\nu,1-\ii\nu;2;1-\frac{\u_{xy}}{2})]^2\,
	\right\}~~,
	\end{split}
\end{align}
in agreement with results reported in the literature \cite{Loparco:2024ibp}. Note that two-point function contains the square of the hypergeometric functions, since $\epsilon$ is itself quadratic in the Majorana fermions, so the computation effectively involves a fermion four-point function.

\paragraph{UIRs inside $\ev{\varepsilon\varepsilon}$.}

It is useful to recast \cref{eq:thermal 2pt exact} in the language of \(\ds_2\) representation theory. In two dimensions, a massive Majorana fermion of real mass $m$ belongs to a principal series representation \cite{Letsios:2020twa,Letsios:2023qzq} with weight $\Delta$ and spin $s$
	\begin{align}\label{eq:Delta-s-majorana}
		\Delta = \half+\ii\nu~, 
		\qquad
		s=\half~, 
	\end{align}
	where, as above, \(\nu=m\lds\). The exact two-point function \cref{eq:thermal 2pt exact} can then be written as
	\begin{align}\label{eq:thermal-2pt-Delta-s}
		\begin{split}
		\ev{\varepsilon(x)\varepsilon(y)}^\t{c}_{\S^2_\lds}
	&=
	\frac{\Gamma(\Delta+s)^2\Gamma(1-\Delta+s)^2}{4\lds^2}
	\Bigg[
	\frac{\u_{xy}}{2}\,
	{}_2F_1\!\qty(\Delta+s,1-\Delta+s;2s;1-\frac{\u_{xy}}{2})^2
	\\
	&
	-(\Delta-s)(1-\Delta-s)\qty(1-\frac{\u_{xy}}{2})
	{}_2F_1\!\qty(\Delta+s,1-\Delta+s;1+2s;1-\frac{\u_{xy}}{2})^2
	\Bigg]~. 
	\end{split}
\end{align}
    The hypergeometric parameters are now fixed directly by the principal-series data \((\Delta,s)\). The two distinct terms reflect the two invariant structures of the spin-\(\half\) propagator. Hence, while \(\varepsilon\) is a scalar operator, its two-point function reveals its composition as a bilinear of fields with $(\Delta,s)$ given in \cref{eq:Delta-s-majorana}. Moreover, the relative minus sign between the two terms encodes the fermionic nature of the underlying fields.

    Let us briefly comment on other UIRs. As discussed in \cref{sec:review-dS-QFT}, apart from the principal series, there can be also discrete-series fermionic representations. These require an \emph{imaginary} fermion mass, $m = \ii k/\lds$, with $k\in\Z$ \cite{Anninos:2023lin,Letsios:2025pqo} and have $\Delta = \frac{1}{2}-k$. Although Euclidean signature does not allow for imaginary-mass fermionic modes obeying the Majorana condition, Lorentzian signature allows them \cite{Letsios:2025pqo}. Let us therefore entertain this possibility. 
    
    A few observations are in order. First, if we naively analytically continue \cref{eq:thermal-2pt-Delta-s}, to $\Delta=\frac{1}{2}-k$, the two-point function develops a double pole due to the product of Gamma functions. This can be traced back to the fact that the discrete-series fermions have fermionic shift symmetries \cite{Bonifacio:2023prb} and, correspondingly, zero-modes. Relatedly, the partition function \cref{eq: exact Z} vanishes for the same reason. To compute meaningful observables these zero modes must be saturated in one way or another. One possibility would be to gauge the fermionic shift symmetries. 
    
    On the Ising side, the discrete series representations would appear as a deformation away from the critical temperature, in the imaginary direction: $\tau = \ii k/(2\pi\lds)$. Note that these are finite shifts, hence access to the perturbative expansion is lost. Nonetheless, the Ising model with a complex thermal deformation has been studied extensively on the lattice, starting with \cite{Fisher:1965rna}. The zeroes of the partition function in the complex-temperature plane, known as Fisher zeroes, are of particular importance \cite{Itzykson:1983gb}. Strikingly, on spherical lattices, the Fisher zeroes were found to be compatible with purely imaginary temperature \cite{Hoelbling:1996gv}. It would be very interesting to understand the interrelation between these observations.

\paragraph{All-order perturbative expansion.}

The exact correlator \cref{eq:thermal 2pt exact} is non-perturbative in the coupling $\tau$, or equivalently in $\nu$. For comparison with other computations of this observable, in particular CPT, it is useful to expand the exact answer at  small mass $\nu \ll 1$, keeping the invariant distance $\u_{xy}$ fixed. One can obtain this expansion to any desired order using a computer algebra system such as \texttt{Mathematica}. In this case, we can actually obtain the all-order expansion analytically. 

The Gamma-function prefactor can be expanded as
\begin{equation}
    \Gamma(1+\ii\nu)^2\Gamma(1-\ii\nu)^2
	= \qty(\frac{\pi\nu}{\sinh(\pi\nu)})^2 = 
    \sum_{k=0}^\infty p_k \nu^{2k}~, \qq{with} p_k = -\frac{(2k-1) (2\pi)^{2k} B_{2k}}{(2k)!}~,
\end{equation}
where $B_{2k}$ are the Bernoulli numbers.
Moreover, it follows from \cite{Zagier-polylogs} that the two hypergeometric functions can be expanded in powers of $\nu^2$, in terms of the multiple polylogarithms: 
\begin{alignat}{2}
	L_r(z) &\coloneqq \Li_{\underbrace{\scriptstyle 2,\ldots,2}_{r}}(z) &&= \sum_{n_1>n_2>\cdots>n_r >0} \frac{z^{n_1}}{n_1^2\cdots n_r^2}~, \\
	\qquad
	\tilde{L}_r(z) &\coloneqq \Li_{1,\underbrace{\scriptstyle 2,\ldots,2}_{r}}(z) &&= \sum_{n_0>n_1>\cdots>n_r >0} \frac{z^{n_0}}{n_0\,n_1^2\cdots n_r^2}~, 
\end{alignat}
with \(L_0(z)=1\) and \(\tilde{L}_0(z)=\Li_1(z)=-\log(1-z)\). Explicitly, setting 
\begin{equation}
    X\coloneqq \frac{\u}{2}~,
\end{equation}
and defining 
\begin{equation}
    F_1(\nu; X)\coloneqq {}_2F_1\!\qty(1+\ii\nu,1-\ii\nu;1;1-X) \qq{,}
F_2(\nu; X) \coloneqq {}_2F_1\!\qty(1+\ii\nu,1-\ii\nu;2;1-X)~, 
\end{equation}
we get
\begin{align}
	F_1(\nu;X) = 
	\frac{1}{X}\sum_{r=0}^\infty \nu^{2r}\,L_r(1-X)~
	\qq{and} 
	F_2(\nu;X) = \frac{1}{1-X}\sum_{r=0}^\infty \nu^{2r}\, \tilde{L}_r(1-X)~. 
\end{align}
Combining these ingredients, the all-order $\nu$-expansion of the two-point function follows:
\begin{align}\label{eq:G-ferm-all-order}
    G_\varepsilon(X) &= \sum_{n=0}^\infty g_{2n}(X)\, \nu^{2n}~, 
\end{align}
where
\begin{align}
    g_{2n}(X) 
    = 
    \frac{1}{4\lds^2} \sum_{k=0}^n p_k\, f_{n-k}(X)~, 
    \qquad
    f_n(X) 
    = 
    \frac{1}{X}\sum_{k=0}^n L_k\, L_{n-k} - \frac{1}{1-X} \sum_{k=0}^{n-1} \tilde{L}_k\, \tilde{L}_{n-1-k}~, 
\end{align}
with the polylogarithms evaluated at $1-X$, i.e. $L_k\equiv L_k(1-X)$ and $\tilde{L}_k\equiv\tilde{L}_k(1-X)$. For example, up to $\nu^4$, we have:
\begin{align}
    g_0(X) &= \frac{1}{4\lds^2}\frac{1}{X} \label{eq:g0} \\
    g_2(X) &= \frac{1}{4\lds^2}\qty(\frac{2\operatorname{Li}_2(1-X)}{X} - \frac{\log[2](X)}{1-X}-\frac{\pi^2}{3\,X}) \label{eq:g2} \\
    g_4(X) &= \frac{1}{4\lds^2}\qty(\frac{2 \operatorname{Li}_{2,2} + \operatorname{Li}_2^2 + - \frac{2\pi^2}{3}\operatorname{Li}_2 + \frac{\pi^4}{15}}{X}+ \frac{2\log(X)\;\operatorname{Li}_{1,2} + \frac{\pi^3}{3}\log[2](X)}{1-X})~, 
\end{align}
where all multiple polylogarithms are evaluated at $1-X\equiv1-\u/2$.

\subsection{Late-time behaviour}

We return to the exact form \cref{eq:thermal 2pt exact} of the two-point function of the energy operator. Analytic continuation of this observable grants us access to late-time physics. In particular, we need to analytically continue the Euclidean $\SO(3)$-invariant distance $\u_{xy}$ as
\begin{equation}\label{eq:uE-to-uL}
    \u_{xy} \mapsto \uL_{xy} = \frac{\cos(T_x-T_y)-\cos(\varphi_x-\varphi_y)}{\sin(T_x)\sin(T_y)}~. 
\end{equation}
We will focus on equal-time insertions, $T_x = T_y = T$,\footnote{An advantage of this choice is that it avoids the need for a time-ordering prescription.} and take the late-time limit as $T \rightarrow 0 $. In this limit, the invariant distance can be expanded as
\begin{equation}
    \uL_{xy} = \frac{2\sin^2((\varphi_x-\varphi_y)/2)}{\sin^2(T)} = \frac{2\sin^2(\Delta \varphi/2)}{T^2}+\order{T^0}~. 
\end{equation}
At late times, equivalently $X\gg 1$, the asymptotic behaviour of the relevant functions appearing in the energy two-point function is given as
\begin{equation}\label{eq:hypergeom-short}
F_1(\nu; X)\sim A_-\, X^{-1-\ii\nu}+A_+\, X^{-1+\ii\nu},
\qquad
F_2(\nu; X)\sim B_-\, X^{-1-i\nu}+B_+\, X^{-1+i\nu},
\end{equation}
with
\begin{equation}
    B_\pm = \frac{\Gamma(\pm 2\ii \nu)}{\Gamma(1\pm \ii \nu)^2}~ \qq{and} A_\pm = \pm\ii\nu\, B_\pm~. 
\end{equation}

Substituting these asymptotics into \cref{eq:thermal 2pt exact} gives the leading late-time behaviour of the thermal correlator:
\begin{equation}\label{eq:eps-eps-exact-leading}
    G_\varepsilon\qty(\uL_{xy})
\sim \frac{1}{4\lds^2}\frac{2\pi\nu}{\sinh(2\pi\nu)} \frac{1}{X} = \frac{1}{4\lds^2}\frac{2\pi\nu}{\sinh(2\pi\nu)} \qty(\frac{\uL_{xy}}{2})^{-1}~. 
\end{equation}
Interestingly, we find that the late-time cosmological two-point function of the energy operator $\varepsilon$ decays with a scaling dimension that is independent of the dimensionless parameter $\nu$. In particular, this dimension coincides with the CFT value obtained at $\nu=0$, and the full $\nu$-dependence appears only in the overall normalisation of the two-point function.

\paragraph{Descendants and subleading behaviour.}

Apart from the leading behaviour, it is useful to record the next term in the large-$X$ expansion of the correlator. Expanding $F_1(X)$ and $F_2(X)$ one order further gives:
\begin{align}
    G_\varepsilon\qty(X)
    &\sim \frac{1}{4\lds^2}\qty[\frac{2\pi\nu}{\sinh(2\pi\nu)}\qty(\frac{1}{X} + \frac{1}{2 X^2}) 
    - \frac{1}{2} \Re\qty(\frac{X^{-2-2\ii\nu}}{1+2\ii\nu}\,\frac{\Gamma(1+\ii\nu)^2\Gamma(1-2\ii \nu)^2}{\Gamma(1-\ii\nu)^2})]
    \\
    &= \frac{1}{4\lds^2}\frac{2\pi\nu}{\sinh(2\pi\nu)}\qty\Bigg[\frac{1}{X} + \frac{1-\cos(2\nu\log(4 X)+\arctan(2\nu))}{2 X^2}]~. 
\end{align}
We observe that at subleading order at late times, the correlator reveals late-time oscillations, compatible with the exchange of principal series fields. 

In fact, one can isolate such contributions by acting on the two-point function with $\cD_X = 1+X\pd_X$. This removes the leading $X^{-1}$ tail, leaving the order $X^{-2}$ as the leading contribution. Equivalently, there is a local observable that displays such oscillatory behaviour at leading order at late times. This is a correlator of the thermal operator with the ``descendant'' operator,
\begin{align}\label{eq:descendant}
    \varepsilon_T(x) = \varepsilon_T(T_x, \varphi_x) \coloneqq \qty(1-T_x \pd_{T_x})\varepsilon(x)~. 
\end{align}
By \cref{eq:uE-to-uL}, at late times this has the same effect as $\cD_X$. Hence the mixed correlator behaves as:
\begin{align}\label{eq:mixed-correlator-exact}
    \ev{\varepsilon_T(x)\,\varepsilon(y)}^\t{c}_{\ds_2} 
    &\sim \frac{1}{4\lds^2}\frac{2\pi\nu}{\sinh(2\pi\nu)}\,\frac{\sqrt{1+4\nu^2}\,\cos(2\nu\log(4 X))-1}{2 X^2}~, 
\end{align}
at $X\gg 1$.

\subsection{Energy two-point function from CPT}

As we have just seen, in the free Majorana frame, the two-point function of the thermal operator $\varepsilon(x)$ is computed by Wick contractions of free fermion propagators. At weak coupling, the same result should arise from a perturbative expansion around the Ising CFT. Here we check this statement, and further assess the reliability of the perturbative computation for different regimes of interest. 

We compute $G_\varepsilon(\u_{xy}) = \ev{\varepsilon(x) \varepsilon(y)}^\t{c}$ from the Ising field theory as
\begin{align}
    G_\varepsilon^{(\t{Ising})}\qty(\u_{xy})  
    &= 
     \frac
     {\ev{\epsilon(x) \epsilon(y) \exp(-\tau \int_{\S^2_\lds} \dd{\mu_z}\, \epsilon(z))}_0}
     {\ev{\exp(-\tau \int_{\S^2_\lds} \dd{\mu_z}\, \epsilon(z))}_0}
     \nn 
     &\phantom{=~} -\frac
     {\ev{\epsilon(x) \exp(-\tau \int_{\S^2_\lds} \dd{\mu_z}\, \epsilon(z))}_0 \ev{\epsilon(y) \exp(-\tau \int_{\S^2_\lds} \dd{\mu_z}\, \epsilon(z))}_0}
     {\qty\Big(\ev{\exp(-\tau \int_{\S^2_\lds} \dd{\mu_z}\, \epsilon(z))}_0)^2}~. 
\end{align}
Here $\ev{\cdots}_0$ denotes expectation values in the critical Ising CFT, and $\dd{\mu_z}$ stands for  $\dd[2]{z} \sqrt{g(z)}$. In conformal perturbation theory we compute the correlator for small $\tau$. Equivalently, in view of \cref{eq:tau-m} this corresponds to $\nu = 2\pi \tau \lds \ll 1$. Expanding to second order in $\tau$ leaves a single integrated four-point function:
\begin{align}
    G_\epsilon^{(\t{Ising})}\qty(\u_{xy}) =  \ev{\epsilon(x)\epsilon(y)}_0 + \frac{\tau^2}{2} \int_{\S^2_\lds} \dd{\mu_z} \dd{\mu_w} \ev{\epsilon(x)\epsilon(y)\epsilon(z)\epsilon(w)}_0^\t{c} + \order{\tau^4}~. 
\end{align}
Once computed order-by-order, the two-point function can be readily compared with \cref{eq:G-ferm-all-order}.

It is convenient to pass again to stereographic coordinates, as in \cref{eq:stereo-1,eq:app-epsilon-weyl}. As before, by $\SO(3)$-invariance the correlator depends only on the invariant distance $\u_{xy}$. As such, we can place the two external points at $x=0$ and $y=r\in\R_+$, where $x$ and $y$ are now in stereographic coordinates. The location of the second insertion can be expressed in terms of the chordal distance $X=\u_{xy}/2$ as 
\begin{align}
    r = \sqrt{\frac{X}{1-X}} \quad\eqv\quad X =\frac{r^2}{1+r^2}~. 
\end{align}

At the critical point, $\tau=0$, the sphere two-point function is 
\begin{align}
    G_\epsilon^{(0)}(X) \coloneqq \ev{\epsilon(0)\epsilon(r)}_0 = \frac{1}{4\lds^2}\frac{1}{X}~, 
\end{align}
which matches exactly \cref{eq:g0}. By the Weyl transformation \cref{eq:app-epsilon-weyl}, the second-order correction,
\begin{align}
    G_\epsilon^{(2)}(X) \coloneqq \frac{\tau^2}{2} \int_{\S^2_\lds} \dd{\mu_z}\dd{\mu_w} \ev{\epsilon(0)\epsilon(r)\epsilon(z)\epsilon(w)}_0^\t{c}~, 
\end{align}
can be written  as
\begin{align}
    G_\epsilon^{(2)}(X) &= \frac{\tau^2}{2} \Omega(0)^{-1} \Omega(r)^{-1}\, \cI_\t{c}(X)~, \qq{with} \\
    \cI_\t{c}(X) &\coloneqq \int_{\C} \dd[2]{z}\dd[2]{w} \Omega(z) \Omega(w)\; \ev{\epsilon(0)\epsilon(r)\epsilon(z)\epsilon(w)}_\C^\t{c}~, 
\end{align}
reducing the computation to an integrated correlator of a CFT on the plane.

On general grounds, a four-point function $\ev{\cO_1(z_1)\cdots\kern1pt\cO_4(z_4)}$ in a two-dimensional CFT can only depend on the cross-ratio 
\begin{align}
    \eta = \frac{z_{12}z_{34}}{z_{13}z_{24}}~, \qquad z_{ij}\coloneqq z_i -z_j~, 
\end{align}
apart from kinematically fixed prefactors, whose $z_i$-dependence is fixed by conformal invariance. In this case, with insertions at $0$, $r$, $z$ and $w$, the cross-ratio is:
\begin{align}
    \eta = \frac{r(z-w)}{z(r-w)}~. 
\end{align}
The plane four-point function of $\epsilon$ is determined by the Belavin--Polyakov--Zamolodchikov (BPZ) equation \cite{Belavin:1984vu}. See for instance \cite{DiFrancesco:1987ez} for explicit expressions. Subtracting the disconnected piece, the connected four-point function reads:
\begin{align}
    \ev{\epsilon(0)\epsilon(r)\epsilon(z)\epsilon(w)}_\C^\t{c} = \frac{\qty(\eta-\bar{\eta})^2}{\abs{z}^2\abs{r-w}^2\abs{\eta}^2\abs{1-\eta}^2}~. 
\end{align}
Hence the calculation of $G_\epsilon^{(2)}$ boils down to performing the integral 
\begin{align}\label{eq:intc-main}
    \cI_\t{c}(X) = 4\lds^2 \int_{\C} \dd[2]{z}\dd[2]{w} \frac{1}{1+\abs{z}^2}\,\frac{1}{1+\abs{w}^2}\,\frac{\qty(\eta-\bar{\eta})^2}{\abs{z}^2\abs{r-w}^2\abs{\eta}^2\abs{1-\eta}^2}~. 
\end{align}
After an elaborate sequence of algebraic manipulations, detailed in \cref{app:eps-eps-app}, this integral can be brought to the form: 
\begin{align}\label{eq:intc-main-goal}
    \cI_\t{c}(X) &= -8\pi^2\lds^2\, \xi\, \cJ(\xi)~, 
\end{align}
where we defined $\xi = 1/r^2 = (1-X)/X$. The function $\cJ(\xi)$ is given as
\begin{align}\label{eq:J(xi)-intermediate}
    \cJ(\xi) &= \int_0^1 \dd{s} \int_0^1\dd{t} \frac{1}{t(1-s\,t)}\qty[\log(1-\frac{\xi\, s\, t}{1+\xi\,s})-\log(1-t)]~. 
\end{align}
Inspired by differential-equation methods for Feynman integrals \cite{Kotikov:1990kg,Remiddi:1997ny,Henn:2013pwa,Henn:2014qga}, we are led to take a derivative of $\cJ(\xi)$ with respect to its argument. This gives the much simpler integral:
\begin{align}
    \cJ'(\xi) = -\int_0^1\dd{s}\int_0^1\dd{t} \frac{s}{(1+\xi\,s)(1-s\,t)(1+\xi\,s(1-t))}~, 
\end{align}
which can be easily computed to give:
\begin{align}
    \cJ'(\xi) = -\frac{\log[2](1+\xi)}{\xi^2}~. 
\end{align}
After integrating over $\xi$ the final result follows:
\begin{align}
    \cJ(\xi) = -2\operatorname{Li}_2\!\qty(\frac{\xi}{1+\xi}) + \frac{\log[2](1+\xi)}{\xi}+\frac{\pi^2}{3}~, 
\end{align}
where the integration constant $\pi^2/3$ is fixed by evaluating $\cJ(0)$ in \cref{eq:J(xi)-intermediate}. Altogether, $\cI_\t{c}(X)$ reads:
\begin{align}
    \cI_\t{c}(X) = 8\pi^2\lds^2(1-X)\qty[\frac{2\operatorname{Li}_2(1-X)}{X} - \frac{\log[2](X)}{1-X}-\frac{\pi^2}{3\,X}]~, 
\end{align}
giving the final expression
\begin{align}
    G_\epsilon^{(2)}(X) = \pi^2\tau^2 \qty[\frac{2\operatorname{Li}_2(1-X)}{X} - \frac{\log[2](X)}{1-X}-\frac{\pi^2}{3\,X}]~, 
\end{align}
in complete agreement with the contribution to this order of the non-perturbative computation. Explicitly, from \cref{eq:g2} and recalling that $\tau=m/(2\pi)$, we find that
\begin{align}
    G^{(2)}_\epsilon(X) = g_2(X)\, \nu^2~. 
\end{align}

\subsection{The late-time fate of perturbation theory}

As explained above, sphere correlators can be analytically continued to Lorentzian observables. Having computed the sphere two-point function of the energy operator, we can take the late time limit $\uL_{xy}\to\infty$, with $\uL_{xy}$ as in \cref{eq:uE-to-uL}. Equivalently, this corresponds to $X\to\infty$.

Up to subleading order in $X$, the desired correlator behaves as
\begin{align}
    G_\epsilon^{(\t{Ising})}(X) \sim \frac{1}{4\lds^2}\frac{1}{X} +\frac{\nu^2}{\lds^2}\qty(-\frac{\pi^2}{6\,X}+\frac{2+2\log(X)+\log[2](X)}{4\,X^2}) + \order{\nu^4,X^{-3}}~, 
\end{align}
at $X\gg 1$. While at the $X^{-1}$ order we find full agreement with the non-perturbative prediction \cref{eq:eps-eps-exact-leading}, we note the appearance of suspicious logarithms at subleading order. 

To isolate the potentially dangerous behaviour, we place our focus on the mixed correlator, $\ev{\epsilon_T\, \epsilon}$, with $\epsilon_T$ the descendant operator defined in \cref{eq:descendant}. The late-time behaviour of this observable, as computed in conformal perturbation theory, is
\begin{align}
    \ev{\epsilon_T(x)\,\epsilon(y)}_\t{Ising}^\t{c} \sim \qty(1+X\pd_X)G_\epsilon^{(\t{Ising})}(X) \sim -\frac{\nu^2}{\lds^2}\frac{\log[2](X)}{4\, X^2}~,
\end{align}
at $X \gg 1$. 

\begin{figure}[t]
    \centering
     \includegraphics[width=1\linewidth]{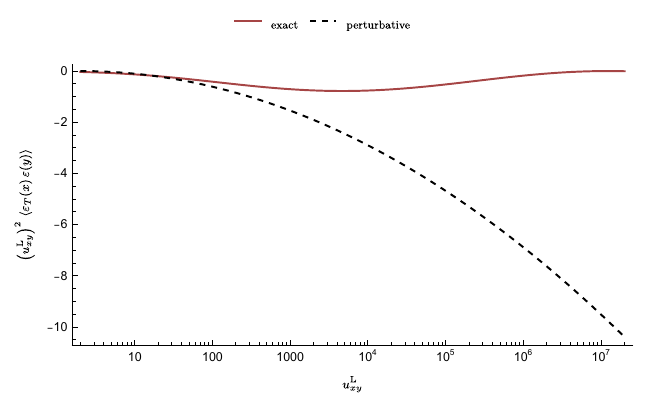}
    \caption{Exact (solid red line) and $\order{\nu^2}$ perturbative (dashed black line) plots of $(\uL_{xy})^2\ev{\epsilon_T(x)\epsilon(y)}$ at $\nu=1/5$. The perturbative correlator displays secular growth. The exact correlator resums all the secular terms into a finite, oscillating expression.}
    \label{fig:epsilon 2pt oscill}
\end{figure}

Stripping off the denominator, we see that conformal perturbation theory predicts secular ``late-time logarithms'' for certain observables. However, these are entirely an artefact of pushing the perturbative expansion beyond its regime of validity. Indeed, comparing with the exact late-time behaviour from \cref{eq:mixed-correlator-exact}, it is evident that the dangerous terms are happily resummed into an oscillatory behaviour, characteristic of principal-series fields. In \cref{fig:epsilon 2pt oscill} we plot the perturbatively predicted and the exact correlator, showing a good match at short times, but a wild difference once we go to late times.

\section{Spin two-point function}
\label{sec:spin 2pt}

So far, we have studied observables in the de Sitter Ising model that are local in the free fermion field. This property made it possible to derive exact expressions for the correlators, and to compare these results directly with perturbation theory. The Ising model also contains operators that are subtler in the fermion picture. This is the case, for instance, of the spin field $\sigma(x)$ and the disorder operator $\mu(x)$, both of which are non-local in terms of the fermions. These two operators, and their correlation functions, are closely tied by Kramers--Wannier (KW) duality, which acts as
\begin{equation}
   \text{KW}:\qquad \sigma(x) \leftrightarrow\mu(x)~, \qquad \nu\leftrightarrow-\nu~. 
\end{equation} 
In this section we analyse the two-point functions \begin{equation}
    \Gs\qty(\u_{xy}) \coloneqq \ev{\sigma(x)\, \sigma(y)}_{\S^2}~, \qquad \Gm\qty(\u_{xy})\coloneqq \ev{\mu(x)\,\mu(y)}_{\S^2}~, 
\end{equation}
and their analytic continuation to $\ds_2$, with particular emphasis on their non-perturbative behaviour at late Lorentzian times.

Since both $\sigma(x)$ and $\mu(x)$ are non-local functions of the fermion field $\Psi(x)$, their two-point functions cannot be computed directly by standard free-fermion methods. Indeed, even in flat space, no closed-form analytic expressions for these correlators are known, although they are exactly characterised by the celebrated Painlevé III equations \cite{Wu:1975mw}. Remarkably, using a set of special Ward identities \cite{Fonseca:2003ee}, Doyon and Fonseca \cite{Doyon:2004fv} generalised this approach to the Euclidean sphere and derived a system of differential equations for $\Gs$ and $\Gm$. While these equations were also obtained on the Euclidean two-sphere in \cite{Doyon:2004fv}, their solutions were not analysed.%
\footnote{Rather, in \cite{Doyon:2004fv}, the spin and disorder two-point functions in the Euclidean two-dimensional AdS were analysed in more detail.} 
In terms of the invariant distance $\u=\u_{xy}$, the differential equations read:
\begin{alignat}{4}
u\qty(\dGs^{2} - \Gs\,\ddGs + \dGm^{2}-\Gm\,\ddGm)
&-\qty(\Gs\,\dGs+\Gm\,\dGm)
=0~, 
\label{eq:E1u}\\[4pt]
(2-u)\qty(\dGs^{2}-\Gs\,\ddGs-\dGm^{2}+\Gm\,\ddGm)
&+\qty(\Gs\,\dGs-\Gm\,\dGm)
=0~, 
\label{eq:E2u}\\[4pt]
u(2-u)\qty(\Gm\,\ddGs+\Gs\,\ddGm-2\dGs\,\dGm)
&+2(1-u)\qty(\Gm\,\dGs+\Gs\,\dGm)
-\frac{4\nu^2+1}{4}\,\Gm\,\Gs 
=0~, 
\label{eq:E3u}
\end{alignat}
where the dot denotes differentiation with respect to $\u$. As with the thermal two-point function, the physical correlators are singled out by boundary conditions. They are the solutions of this system that obey the following requirements:
\begin{itemize}
\item At short distances, $\u\to 0$, their singularities are fixed by the CFT correlators \cref{eq: 2ptf ising}:
\begin{equation}\label{eq:Gs-Gm-short-distance}
    \Gs(u),\Gm(u) \xrightarrow[]{u\rightarrow0}\Gs^{(\nu=0)}(u)= \Gm^{(\nu=0)}(u) = \qty[\frac{1}{4\lds^2}\,\qty(\frac{2}{\u})]^{1/8}~. 
\end{equation}
\item They are smooth at antipodal points, $u=2$.
\end{itemize}

Although no closed analytic solution in terms of elementary or standard special functions is known, we can extract the non-perturbative late-time behaviour at $u\rightarrow \infty$.%
\footnote{Following closely the analysis of \cite{Doyon:2004fv}, these equations can be recast into a system of Painlevé equations. However, this reformulation will not be needed for our purposes.}
To this end, we use the ansätze: 
\begin{equation}
    \Gs(\u)  \sim A\, u^{-\sds}~, \qq{and} \Gm(\u) \sim B\, u^{-\sdm}~, 
\end{equation}
for some constants $A$ and $B$ and late-time exponents $\sds$ and $\sdm$. This behaviour is expected from the conformal structure of cosmological correlators. Substituting our ansätze into the ODEs, we see that \cref{eq:E1u,eq:E2u} are automatically satisfied for any $A$, $B$ and $\sds$, $\sdm$. However, the third equation gives
\begin{equation}\label{eq: pq const.}
    \sds+\sdm-\qty(\sds-\sdm)^2 = \frac{4\nu^2+1}{4}~, 
\end{equation}
which is a non-perturbative constraint on the decay rates of the two correlators. To determine $\sds$ and $\sdm$ we can either solve the system of differential equations numerically, or compute them in perturbation theory in $\nu$.

\begin{figure}[thb]
    \centering
     \includegraphics[width=1\linewidth]{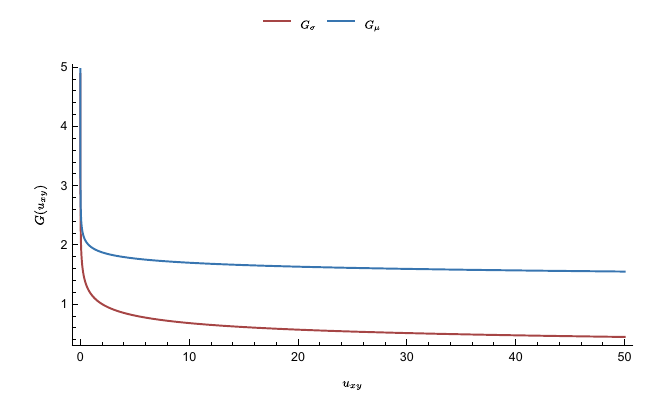}
    \caption{Numerical spin and disorder two-point functions, $\Gs(\u_{xy})$ (red) and $\Gm(\u_{xy})$ (blue) as a function of the \ds\ invariant distance $\u_{xy}$, at $\nu=1/5$. Their large-$\u$ power-law tails determine the late-time exponents $\sds$ and $\sdm$.}
    \label{fig:G and Gtilde correlators}
\end{figure}

\begin{figure}[thb]
    \centering
    \begin{subfigure}[t]{0.75\textwidth}
        \centering
        \includegraphics[width=\textwidth]{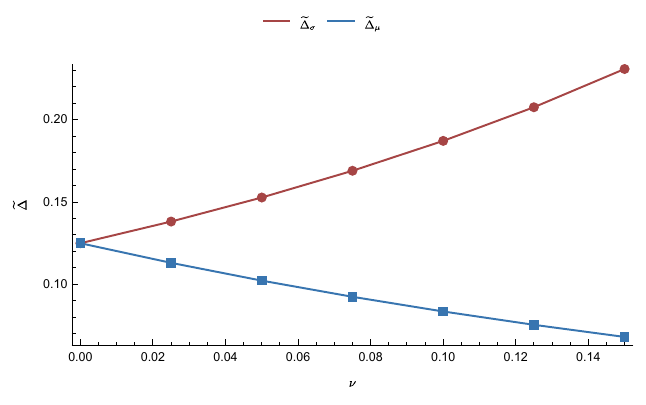}

    \end{subfigure}
    \hfill
    \begin{subfigure}[t]{0.75\textwidth}
        \centering
        \includegraphics[width=\textwidth]{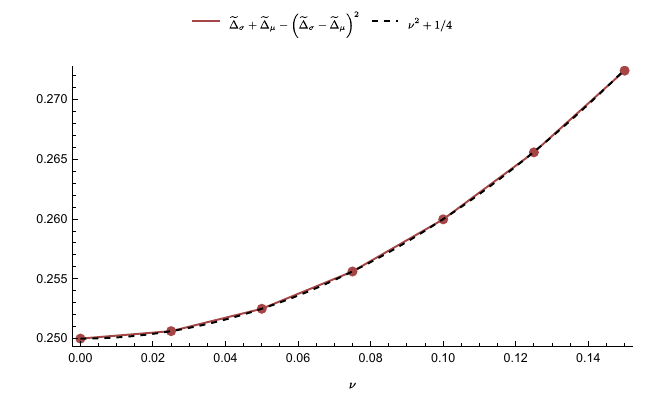}
    \end{subfigure}
    \caption{
    Late-time exponents extracted from the numerical solutions. Top: $\sds$ (red circles) and $\sdm$ (blue squares) as functions of $\nu$. Bottom: the numerical combination $\sds+\sdm-(\sds-\sdm)^2$ (red) agrees with the exact non-perturbative prediction $\nu^2+1/4$ (black dashed line), as a check of \cref{eq: pq const.}.}
    \label{fig:pq}
\end{figure}

Let us start with the numerical solution. A numerical evaluation of $\Gs$ and $\Gm$ is not difficult. The only complication is imposing the boundary conditions, since imposing the divergent behaviour at $u=0$ is inefficient. Luckily, the free fermion picture allows us to determine the exact ratio $\Gs/\Gm$ when evaluated at antipodal points on the Euclidean sphere, i.e. at $u=2$. See \cref{app:antipodal} for a derivation. We get
\begin{equation}\label{eq:GoverGtilde}
    \frac{\Gm(2)}{\Gs(2)}=\ex{\pi \nu}~. 
\end{equation}
Using this boundary condition, we obtain numerically the solution for $\Gs$ and $\Gm$, shown in \cref{fig:G and Gtilde correlators}. As expected, for any value of $\nu$, the decay rates $\sds$ and $\sdm$ satisfy the non-perturbative constraint \cref{eq: pq const.}; see \cref{fig:pq}. We note here that the spin two-point function shows no oscillatory behaviour at late times.

\begin{figure}[thb]
    \centering
     \includegraphics[width=0.8\linewidth]{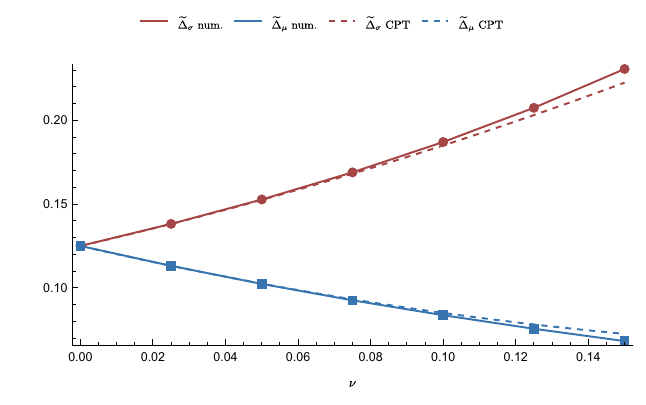}
    \caption{Numerical late-time exponents (solid) compared with the $\order{\nu^2}$ expansions in \cref{eq:sds-sdm-pert} (dotted). The perturbative curves agree at small $\nu$, with deviations increasing as $\nu$ grows.}
    \label{fig:pqpert}
\end{figure}

We can also derive a perturbative expression for $\sds$ and $\sdm$ when $\nu \ll 1$.  By Kramers--Wannier duality 
\begin{equation}\label{eq:KW-Gs-Gm}
    \Gs^{(\nu)}(\u) = \Gm^{(-\nu)}(\u)~, 
\end{equation}
we can parametrise the two correlators as
\begin{equation}
    \begin{split}
        \Gs(\u)&=\Gs^{(\nu=0)}(\u)\qty[1+\nu\, f(u) + \order{\nu^2}] \\
        \Gm(\u)&=\Gs^{(\nu=0)}(\u)\qty[1-\nu\, f(u) + \order{\nu^2}]~, 
    \end{split}
\end{equation}
where $\Gs^{(\nu=0)}(\u)$ is given by \cref{eq:Gs-Gm-short-distance}. Plugging this ansatz into \cref{eq:E1u,eq:E2u,eq:E3u}, we find that to first order in $\nu$, the function $f(u)$ has to satisfy
\begin{equation}
    (2-u)\ddot{f} - \dot{f} + \frac{f}{2u^2}=0~. 
\end{equation}
The solution to this equation that is smooth at $u=2$ and gives $\Gm(2)/\Gs(2) \kern-1pt =\kern-1pt 1 + \pi \nu + \order{\nu^2}$ is
\begin{equation}
    f(u) =-\sqrt{\frac{u}{2}}\; K\qty(1-\frac{u}{2})~, 
\end{equation}
where $K(x)$ is the complete elliptic integral of the first kind, defined as
\begin{equation}
    K(x) := \frac{\pi}{2}\; {}_2F_1\qty(\frac{1}{2},\frac{1}{2};1;x)~. 
\end{equation}
Thus
\begin{equation}
    \begin{split}
        \Gs(u) &= \Gs^{(\nu=0)}(\u)\qty[1-\nu \sqrt{\frac{u}{2}}\,K\qty(1-\frac{u}{2})+ \order{\nu^2}] \\
        \Gm(u) &= \Gs^{(\nu=0)}(\u)\qty[1+\nu \sqrt{\frac{u}{2}}\, K\qty(1-\frac{u}{2})+ \order{\nu^2}]~. 
    \end{split}
\end{equation}
From the known large-$u$ expansion of $K(u)$ it follows that
\begin{equation}\label{eq:K(u)-large-u}
    \sqrt{\frac{u}{2}}\,K\qty(1-\frac{u}{2})\xrightarrow[]{u\rightarrow\infty}\frac{1}{2}\log u~. 
\end{equation}
Therefore, we deduce that
\begin{equation}
    \sds = \frac{1}{8} + \frac{\nu}{2}+\order{\nu^2}~, \qquad  \sdm = \frac{1}{8} - \frac{\nu}{2}+\order{\nu^2}~. 
\end{equation}
Finally, from Kramers--Wannier duality \cref{eq:KW-Gs-Gm} and the non-perturbative relation \cref{eq: pq const.}, we get for free the order-$\nu^2$ term:
\begin{equation}\label{eq:sds-sdm-pert}
     \sds = \frac{1}{8} + \frac{\nu}{2}+\nu^2+\order{\nu^3}~, \qquad \sdm = \frac{1}{8} - \frac{\nu}{2}+\nu^2+\order{\nu^3}~. 
\end{equation}
We can now compare the numerical and perturbative results, as shown in \cref{fig:pqpert}.

\subsection{Spin two-point function from CPT}

The perturbative result obtained by combining KW duality with the non-perturbative ODEs \cref{eq:E1u,eq:E2u,eq:E3u} can also be derived directly by conformal perturbation theory around the Ising CFT.

As in the $\varepsilon$ two-point function, expanding for $\nu \ll 1$ gives
\begin{align}
\Gs\qty(\u_{xy})
&=
\frac
{
    \ev{
        \sigma(x)\sigma(y)\,
        \exp(-\tau\int_{S^2_\lds} \dd{\mu_z}\,\epsilon(z))
    }_0
}
{
    \ev{
        \exp(-\tau\int_{S^2_\lds} \dd{\mu_z}\,\epsilon(z))
    }_0
}
\nonumber\\[4pt]
&=
\ev{\sigma(x)\sigma(y)}_0
-
\tau\int_{S^2_\lds} \dd{\mu_z}\,
\ev{\sigma(x)\sigma(y)\varepsilon(z)}_0
+\order{\tau^2}~, 
\label{eq:first-order-CPT-general}
\end{align}
where, again, $\dd{\mu_z}$ stands for $\dd[2]{z} \sqrt{g(z)}$. We need to evaluate the integrated three-point function. We start by setting $x=0$ and $y=r$, using the $\SO(3)$-invariance of the sphere. As before, $r$ is related to the \ds-invariant distance as 
\begin{equation}
    \frac{\u_{xy}}{2}= \frac{r^2}{1+r^2}~. 
\end{equation}
The three-point function on the plane is 
\begin{equation}
    \ev{\sigma(0)\sigma(r)\varepsilon(z)}_{\C} = \frac{1}{2}\frac{r^{3/4}}{\abs{z}\abs{z-r}}~,
\end{equation}
therefore
\begin{equation}
    \frac{\ev{\sigma(0)\sigma(r)\varepsilon(z)}_{\C}}{\ev{\sigma(0)\sigma(r)}_{\C}} = \frac{1}{2}\frac{r}{\abs{z}\abs{z-r}}~. 
\end{equation}
A Weyl rescaling, as in \cref{eq:stereo-1}, gives the three-point function on the sphere:
\begin{equation}
         \frac{\ev{\sigma(0)\sigma(r)\varepsilon(z)}_{\S^2}}{\ev{\sigma(0)\sigma(r)}_{\S^2}} = \Omega(z)^{-1} \frac{\ev{\sigma(0)\sigma(r)\varepsilon(z)}_{\C}}{\ev{\sigma(0)\sigma(r)}_{\C}} = \frac{1+\abs{z}^2}{2\lds}\frac{r}{2\abs{z}\abs{z-r}}
\end{equation}
The first-order term in the CPT expansion is therefore
\begin{equation}
    \frac{\delta\Gs}{\Gs} =- \tau \int_{\C}\dd[2]{z}\;\Omega(z)^2\,  \frac{\ev{\sigma(0)\sigma(r)\varepsilon(z)}_{\S^2}}{\ev{\sigma(0)\sigma(r)}_{\S^2}} = -\tau\lds r \int_{\C}\frac{\dd[2]{z}}{\qty(1+\abs{z}^2)\abs{z}\abs{z-r}}~. 
\end{equation}
Performing the integral and using $\tau = m/(2\pi)$, we find
\begin{equation}
    \frac{\delta \Gs}{\Gs} = -\nu \sqrt{\frac{u}{2}}\,K\left(1-\frac{u}{2}\right)~, 
\end{equation}
in complete agreement with the ODE result above.

Let us end by noting that, also in this case, perturbation theory is trustworthy only over a finite range of times. In the small-$\nu$ expansion, logarithmic terms as in \cref{eq:K(u)-large-u} cause $\delta \Gs/\Gs^{(\nu = 0)}$ to grow without a bound at late times, $u \gg 1$, and the perturbative expansion breaks down. By contrast, the exact result remains well-behaved and decays at late times, as expected. However, in contrast to the $\varepsilon$ two-point function, the presence of these secular terms in the spin two-point function is not caused by truncating late-time oscillations. Rather, it manifests the explicit $\nu$-dependence of the late-time scaling dimensions.

\section{Outlook}\label{sec:outlook}

In this work, we studied the two-dimensional Ising model in de~Sitter space, making use of its well-known duality with a free massive Majorana fermion. This dual description made it possible to obtain exact and non-perturbative results for several observables, including the sphere partition function and the two-point functions of the energy, spin, and disorder operators. We used these exact answers to study the small-mass perturbative expansion, as well as the late-time limit of de~Sitter dynamics. A recurring lesson, made sharp through this model, is that these two limits do not generically commute.

Having an explicitly solvable and easily tractable model in de Sitter space opens several directions for future work. We list some of them below. 

\paragraph{Spectral decomposition of two-point functions.}

As shown in \cite{Bros:2010rku} and explained in more detail in \cite{Loparco:2023rug}, de Sitter two-point functions admit a K{\"a}ll{\'e}n--Lehmann-type decomposition in terms of UIRs of the de Sitter isometry group. Applying this decomposition to the exact correlators studied in this work would allow one to extract the corresponding spectral densities and check their conjectural analyticity \cite{DiPietro:2021sjt}. This would be especially interesting for the spin and disorder operators, whose two-point functions are not obtained via Wick contractions of free field propagators.

\paragraph{Adding the magnetic field.} 

As reviewed in \cref{sec:Isingmodel}, the Ising CFT has another relevant deformation parametrised by the spin operator $\sigma(x)$. In flat space this deformation is integrable,%
\footnote{Turning on both $\sigma(x)$ and $\varepsilon(x)$ results in a strongly coupled, non-solvable model, usually referred to as Ising field theory.} 
but, unlike the thermal deformation, the resulting massive theory has a richer spectrum and an interacting S-matrix \cite{Zamolodchikov:1989cf}. It would be nice to explore this deformation in de Sitter space and 	see if some remnant of integrability can be seen for finite values of $\lds$.%
\footnote{The results of \cite{Antunes:2025iaw} suggest that the infinite number of local conserved charges present in flat space, does not survive in de~Sitter space. However, infinitely many conserved charges constructed non-locally may still exist. We thank Tarek Anous for discussions on this point.}

\paragraph{Spontaneous symmetry breaking.} 

The two-dimensional Ising model is the paradigmatic example of spontaneous symmetry breaking. In flat space for $\tau<0$ the $\Z_2$ global symmetry is spontaneously broken and the theory has two degenerate ground states. In contrast, for $\tau>0$ the $\mathbb{Z}_2$ symmetry is preserved and the vacuum is unique. In de~Sitter space, it is usually argued that spontaneous symmetry breaking (SSB) cannot occur because spatial slices are compact, although it is not clear if this conclusion persists all the way to future infinity, where the spatial volume becomes infinite. As argued in \cite{PhysRevD.33.2833}, and recently verified in several computable models \cite{DiPietro:2023inn,Letsios:2026ypo}, strong infrared effects may be responsible for the absence of SSB in de Sitter space. It would be interesting to understand whether the same mechanism precludes discrete symmetry breaking too.\footnote{Recently, it was shown in \cite{Aguilera-Damia:2026dbk} that QFT in de Sitter space can exhibit SSB of discrete symmetries in the presence of mixed discrete 't Hooft anomalies. By contrast, the SSB observed in the Ising model is not tied to any 't Hooft anomaly, but rather arises dynamically, making it a less robust phenomenon.} The Ising model seems to be the natural setup to test this assertion. 

\paragraph{Line defects.} 

The Ising model contains a zoo of line defects \cite{Oshikawa:1996ww,Oshikawa:1996dj,Quella:2006de}, making it a very rich model to study the effects of defects --- topological or not --- in de~Sitter space. For instance, at its critical point it contains the non-invertible Kramers--Wannier defect, which becomes non-topological in the deformed model \cite{Ambrosino:2025pjj,Antinucci:2026uuh}. Remarkably, it also contains families of line defects that do not spoil the exact solvability of the model \cite{Delfino:1994nx}. In de~Sitter space, such line operators can be interpreted as the worldlines of infinitely heavy probe particles. They therefore provide a natural framework for modeling aspects of observer physics, within an exactly solvable setting. Relatedly, such defects may also shed light on the somewhat elusive edge modes that appear in sphere path integrals \cite{Anninos:2020hfj}. It would be interesting to explore this direction in future work.

\paragraph{Quantum gravity.} 

Throughout this paper the effects of quantum gravity were switched off. Arguably the most interesting direction is to couple this model to a dynamical metric. In conformal gauge, this amounts to coupling the thermally deformed Ising model to Liouville gravity and gravitationally dressing the thermal deformation, a question going back to \cite{Brezin:1989db}. Aspects of this question were analysed from the Majorana point of view in \cite{deLacroix:2023uem,Namuduri:2023smx}. However, Ising matter alone is not expected to lead to a gravitational saddle with positive cosmological constant, at least at small thermal coupling.%
\footnote{At large thermal coupling a \ds\ saddle is still not expected, but it is harder to exclude convincingly.}
To obtain such a saddle, one may attempt to couple to spectator matter, with total central charge $c_\t{tot}>25$ (at the critical point and thermally deforming after), which would then couple to \emph{timelike} Liouville theory.  It would be interesting to follow this lead, similarly to \cite{Anninos:2021ene,Anninos:2024iwf}, and attempt to compute exact observables in the combined system. Remarkably, the Ising model coupled to 2D gravity admits a two-matrix model description \cite{Brezin:1989db} (see also \cite{Anninos:2020ccj}), which invites for a non-perturbative definition of de~Sitter gravity.

\paragraph{Acknowledgements.}

It is a great pleasure to thank Dionysios Anninos, Tarek Anous, Alan Rios Fukelman, for helpful discussions and comments on an earlier draft, and Vasileios Letsios and Kamran Salehi Vaziri for helpful discussions. S.V. would like to thank the organisers and the participants of the workshop \href{https://indico.sns.it/event/129/}{\emph{Expanding thoughts on de Sitter}} for stimulating discussions.

The research of G.G. is funded through an ARC advanced project and further supported by IISN-Belgium (convention 4.4503.15). The research of S.V. is supported by a Marina Solvay fellowship and by the Fonds de la Recherche Scientifique (FNRS) under grant no. 4.4503.15.

\appendix
\crefalias{section}{appendix}
\crefalias{subsection}{appendix}
\addtocontents{toc}{\protect\setcounter{tocdepth}{1}}

\section{Spectral properties of the Dirac operator on the sphere}
\label{app:dirac-op}

We collect here some spectral properties of the Dirac operator, \(\dsl\), on \(\S^2_\lds\) that enter the main text. The formulas follow from the construction of Camporesi and Higuchi \cite{Camporesi:1995fb}, specialised to the two-dimensional case. The Dirac operator is defined as \(\dsl=\gamma^\mu\nabla_\mu\), where \(\gamma^\mu\) are the Dirac matrices, obeying the Clifford algebra:
\begin{align}
	\acomm{\gamma^\mu}{\gamma^\nu}=2\delta^{\mu \nu}\unit~. 
\end{align} 
In two dimensions, an explicit representation of the Dirac matrices is in terms of \(2\times 2\) matrices, 
\begin{align}
	\gamma^1 = \mqty(\admat[0]{\ii ,-\ii})~, \qquad \gamma^2 = \mqty(\admat[0]{1,1})~. 
\end{align}
In Euclidean signature, \(\dsl\) is anti-Hermitian. Its eigenvalues are therefore imaginary.

It is enough to work on the unit sphere, with metric:
\[
    \dd s^2 = \dd\vartheta^2+\sin^2\vartheta\,\dd\varphi^2~, 
\]
where the angles \(\vartheta\in\closed{0}{\pi}\) and \(\varphi\in\ropen{0}{2\pi}\) are the polar and azimuthal angles, respectively.
The round sphere of radius \(\lds\) is obtained by multiplying the metric by \(\lds^2\). The Dirac operator scales as
\begin{align}
    \dsl_{\S^2_\lds}=\frac{1}{\lds}\,\dsl_{\S^2}~, 
\end{align}
while normalised eigenspinors scale as \(\psi_{\S^2_\lds}=\lds^{-1}\psi_{\S^2}\).

In the above coordinate system, the Dirac operator on \(\S^2\) takes the form 
\begin{align}\label{eq:CH-S2-dirac-operator}
    \dsl_{\S^2}
    =
    \qty(\partial_\vartheta+\frac{1}{2}\cot\vartheta)\gamma^2
    +
    \frac{1}{\sin\vartheta}\,\pd_\varphi\,\gamma^1~, 
\end{align}
as follows from restricting the recursive formula of \cite{Camporesi:1995fb} to two dimensions, with the
\(\S^1\) Dirac operator equal to \(\pd_\varphi\).

\paragraph{Dirac equation on the sphere.} 
We are aiming to solve the eigenvalue equation 
\begin{align}\label{eq:Dirac-eigeneq}
	\dsl_{\S^2} \psi = \ii\, \lambda\, \psi~, 
\end{align}
where the spinor \(\psi\) has components: 
\begin{align}
    \psi=
    \mqty(
        \phi_+ \\
        \phi_-
	)~. 
\end{align}
In the above coordinate system, \cref{eq:Dirac-eigeneq} gives the coupled first-order equations
\begin{align}
    \qty(\partial_\vartheta+\half\cot\vartheta
    +\frac{\ii}{\sin\vartheta}\partial_\varphi)\phi_-
    &=
    \ii\lambda\,\phi_+~, \\
    \qty(\partial_\vartheta+\half\cot\vartheta
    -\frac{\ii}{\sin\vartheta}\partial_\varphi)\phi_+
    &=
    \ii\lambda\,\phi_-~. 
\end{align}
In order to solve the above equation, consider the separation of variables, 
\begin{align}
	\phi_\pm(\vartheta,\varphi) = f_k(\vartheta)\chi_k^{(\pm)}(\varphi)~, 
\end{align}
with \(\chi_k^{(\pm)}(\varphi)\) normalised spinor modes on \(\S^1\), i.e. eigenfunctions of the operator \(\pd_\varphi\): 
\begin{align}
    \chi_k^{(\pm)}(\varphi)
    =
    \frac{1}{\sqrt{2\pi}}\,
    \ex{\pm\ii(k+1/2)\varphi}~, 
    \qquad
    k=0,1,2,\ldots~. 
\end{align}
The half-integer momentum enforces the spinorial sign change under
\(\varphi\mapsto\varphi+2\pi\). Eliminating \(\phi_-\) gives a second-order equation for \(f_k\): 
\begin{align}\label{eq:CH-S2-theta-equation}
    \cD_k f_k=-\lambda^2 f_k~, 
\end{align}
where
\begin{align}
    \cD_k
    =
    \qty(\partial_\vartheta+\half\cot\vartheta)^2
    -\frac{(k+\half)^2}{\sin^2\vartheta}
    +(k+\half)\frac{\cos\vartheta}{\sin^2\vartheta}~. 
\end{align}
Equation \cref{eq:CH-S2-theta-equation} can be solved with the help of Jacobi polynomials, 
\begin{align}
	P^{(a,b)}_n(x) = \frac{\Gamma(n+a+1)}{\Gamma(n+1)\Gamma(a+1)}\;{}_2F_1\qty(-n,n+a+b+1,a+1;\frac{1-x}{2})~. 
\end{align}
Its solutions that are regular at \(\vartheta=0,\pi\) are
\begin{align}\label{eq:fk-solns}
	\begin{split}
    f^{(1)}_{nk}(\vartheta)
    &=
    \qty(\cos\frac{\vartheta}{2})^{k+1}
    \qty(\sin\frac{\vartheta}{2})^k
    P_{n-k-1}^{(k,k+1)}(\cos\vartheta)~, \\
    f^{(2)}_{nk}(\vartheta)
    &=
    \qty(\cos\frac{\vartheta}{2})^k
    \qty(\sin\frac{\vartheta}{2})^{k+1}
    P_{n-k-1}^{(k+1,k)}(\cos\vartheta)~, 
	\end{split}
\end{align}
both of which have eigenvalues:
\begin{align}
	\lambda_n^2 = n^2~, \qquad n = 1,2,\ldots~. 
\end{align}
Importantly, regularity of \(f_{nk}^{(i)}\) restricts the range of the index \(k\) to \(k \leq n-1\). 

Hence, in terms of the solutions to the \(\vartheta\)-problem, the eigenspinors of \(\dsl_{\S^2}\) can be found as
\begin{align}
    \psi_{\pm,n,k}^{(-)}
    &=
    \frac{c_{nk}}{\sqrt{2}}
    \mqty(
        f^{(1)}_{nk}(\vartheta)\,\chi_k^{(-)}(\varphi) \\[2ex]
        \pm \ii\,f^{(2)}_{nk}(\vartheta)\,\chi_k^{(-)}(\varphi)
	)~, \\[2ex]
    \psi_{\pm,n,k}^{(+)}
    &=
    \frac{c_{nk}}{\sqrt{2}}
    \mqty(
        \ii\,f^{(2)}_{nk}(\vartheta)\,\chi_k^{(+)}(\varphi) \\[2ex]
        \pm f^{(1)}_{nk}(\vartheta)\,\chi_k^{(+)}(\varphi)
    )~. 
\end{align}
In the above, \(c_{nk}\) is a normalisation constant, ensuring that \(\psi^{(s)}_{\pm,n,k}\) are orthonormalised with respect to the standard inner product on \(\S^2\). On the unit sphere it is fixed as
\begin{align}
    \abs{c_{nk}}^{-2}
    =
    \frac{\Gamma(n)^2}{(n-k-1)!(n+k)!}~. 
\end{align}

\paragraph{Spectrum and degeneracy.}
Restoring the radius, the spectrum is thus
\begin{align}\label{eq:dirac-spectrum-S2}
    \dsl\,\psi_{\pm,n,k}^{(s)}
    =
    \pm \frac{\ii n}{\lds}\,
    \psi_{\pm,n,k}^{(s)}~, 
    \qquad
    n=1,2,\ldots~, 
    \qquad
    k=0,\ldots,n-1~, 
    \qquad
    s=\pm~. 
\end{align}
Note here that \(n\) starts from \(1\), so the spectrum has no zero mode.

For each sign of the eigenvalue there are \(2n\) complex eigenspinors: \(n\)
values of \(k\) and two values of \(s\). This agrees with the general
degeneracy formula on \(\S^{N}\) (for even $N$) \cite{Camporesi:1995fb}: 
\begin{align}
    D_N(q)=\frac{2^{N/2}(N+q-1)!}{q!\,(N-1)!}~, 
\end{align}
upon setting \(N=2\) and relabelling \(n=q+1\). The same degeneracy is the
dimension of the \(\SU(2)\) representation with \(j_n=n-1/2\), namely 
\begin{align}
    2j_n+1=2n~. 
\end{align}

\paragraph{The north pole.} 
In the main text many formulas simplify by fixing \(y\) to be the north pole, \(y=\t{N}\). At the north pole only the \(k=0\) sector survives, as follows from \cref{eq:fk-solns}. For \(k=0\),
\begin{align}
    f^{(1)}_{n0}(\vartheta)
    &=
    \cos\frac{\vartheta}{2}\,
    P_{n-1}^{(0,1)}(\cos\vartheta)~, \qquad 
    f^{(2)}_{n0}(\vartheta)
    =
    \sin\frac{\vartheta}{2}\,
    P_{n-1}^{(1,0)}(\cos\vartheta)~, 
\end{align}
while the normalisation is \(c_{n0}=\sqrt{n}\). We recognise these functions as Wigner \(d\)-functions in the spin-\(j_n\) representation \cite{Varshalovich:1988ye}: 
\begin{align}
    f_{n0}^{(1)}(\vartheta)=d^{j_n}_{\frac{1}{2},\frac{1}{2}}(\vartheta)~, 
    \qquad
    f_{n0}^{(2)}(\vartheta)=-d^{j_n}_{\frac{1}{2},-\frac{1}{2}}(\vartheta)~. 
\end{align}

\paragraph{The massive propagator.}
The massive Green's function on \(\S^2_\lds\) has the spectral representation
\begin{align}
    S_m(x,y)
    =
    \frac{1}{\lds^2}
    \sum_{\sigma=\pm}
    \sum_{n=1}^{\infty}
    \sum_{k=0}^{n-1}
    \sum_{s=\pm}
    \frac{
    \psi_{\sigma,n,k}^{(s)}(x)
    \psi_{\sigma,n,k}^{(s)}(y)^\dagger
    }{
    m+\sigma\,\ii n/\lds
    }~. 
\end{align}
As in the main text, we fix \(y\) to be the north
pole, \(y=\t{N}\). Then only \(k=0\)
contributes, with \(f^{(1)}_{n0}(0)=1\) and \(f^{(2)}_{n0}(0)=0\).
The spherical coordinate system \((\vartheta,\varphi)\) is singular at the pole, as the angle \(\varphi\) is not defined there. To circumvent this problem, one can temporarily offset \(y\) from the north pole, but keep only the \(k=0\) modes --- since we know only these will contribute --- and return to \(y=\t{N}\) at the end. This is done in \cite{Letsios:2025pqo} using the spinor parallel propagator of \cite{Mueck:1999efk}. As it turns out, it is equivalent to take
conventionally \(\chi_0^{(\pm)}(\varphi)\chi_0^{(\pm)}(\t{N})^* = \ex{\mp \ii \varphi/2}/(2\pi)\). With that convention we get
\begin{align}
    \psi_{\sigma,n,0}^{(-)}(x)
    \psi_{\sigma,n,0}^{(-)}(\t{N})^\dagger
    &=
    \frac{n}{4\pi}\,
    \ex{\ii\varphi/2}
    \mqty(
        f^{(1)}_{n0}(\vartheta) & 0 \\
        \sigma\,\ii f^{(2)}_{n0}(\vartheta) & 0
    )~, \\
    \psi_{\sigma,n,0}^{(+)}(x)
    \psi_{\sigma,n,0}^{(+)}(\t{N})^\dagger
    &=
    \frac{n}{4\pi}\,
    \ex{-\ii\varphi/2}
    \mqty(
        0 & \sigma\,\ii f^{(2)}_{n0}(\vartheta) \\
        0 & f^{(1)}_{n0}(\vartheta)
    )~. 
\end{align}
Hence, the propagator finally takes the form used in \cref{eq:thermal prop ansatz}:
\begin{align}
    S_m(x,\t{N})
    =
    \begin{pmatrix}
        \ex{\ii\varphi/2}\,a(\vartheta) &
        \ex{-\ii\varphi/2}\,b(\vartheta) \\
        \ex{\ii\varphi/2}\,b(\vartheta) &
        \ex{-\ii\varphi/2}\,a(\vartheta)
    \end{pmatrix}~, 
\end{align}
with
\begin{align}\label{eq:a-mode-sum}
    a(\vartheta)
    &=
    \frac{\nu}{2\pi\lds}\cos\frac{\vartheta}{2}
    \sum_{n=1}^{\infty}
    \frac{n}{n^2+\nu^2}
    P_{n-1}^{(0,1)}(\cos\vartheta)~, \\
    \label{eq:b-mode-sum}
    b(\vartheta)
    &=
    \frac{1}{2\pi\lds}\sin\frac{\vartheta}{2}
    \sum_{n=1}^{\infty}
    \frac{n^2}{n^2+\nu^2}
    P_{n-1}^{(1,0)}(\cos\vartheta)~, 
\end{align}
where \(\nu=m\lds\). Equivalently, the propagator can be rewritten as 
\begin{align}
    S_m(x,\t{N}) = \qty(a(\vartheta)\, \unit + b(\vartheta)\, \sigma_1) \Lambda_\varphi~, \qq{with} \Lambda_\varphi = \ex{\frac{\ii \varphi}{2}\sigma_3}~, 
\end{align}
where $\sigma_{i}$ denote the Pauli matrices. The matrix $\Lambda_\varphi$ can be identified with the spinor parallel propagator of \cite{Mueck:1999efk}. It follows that
\begin{align}
    S_m(\t{N},x) = \Lambda_\varphi^{-1}\,\qty(a(\vartheta)\,\unit - b(\vartheta)\,\sigma_1)~, 
\end{align}
and so 
\begin{align}\label{eq:trSS}
    \tr[S_m(x,\t{N})S_m(\t{N},x)] = 2\qty(a(\vartheta)^2-b(\vartheta)^2)~. 
\end{align}
See also \cite{Letsios:2020twa} for details.

\section{Details of the partition function from CPT}\label{app:sphere-parti}

In this appendix we provide details on the computation of the sphere partition function, needed for \cref{sec:sphere partition function}.

\subsection{Order $\nu^2$ contribution}\label{app: nu2}

In the main text we reduced the order-$\nu^2$ contribution to the sphere partition function to the integral \cref{eq:I2-reduced-main}
\begin{equation}
    \cI_2 \coloneqq 4 \int_{\C^2} \dd[2]{z_1}\, \dd[2]{z_2}\;\frac{1}{\qty(1+\abs{z_1}^2)\qty(1+\abs{z_2}^2)\abs{z_1-z_2}^2}~. 
\end{equation}
Let us show how this integral evaluates to the value claimed in \cref{eq:I2-final}.

First we pass to polar coordinates $z_1 = r\,\ex{\ii\vartheta}$ and $z_2=s\,\ex{\ii\varphi}$, so that
\begin{equation}
\begin{split}
     \cI_2 =& 4 \int_0^\infty \dd{r}\,\dd{s}\frac{rs}{\qty(1+r^2)\qty(1+s^2)}\;\int_0^{2\pi} \dd{\vartheta} \,\dd{\varphi} \; \frac{1}{r^2+s^2-2\,rs\cos(\vartheta-\varphi)}\\
     &= 4\pi^2 \int_0^\infty \dd{r}\,\dd{s}\,\frac{4rs}{\qty(1+r^2)\qty(1+s^2)\abs{r^2-s^2}}~. 
\end{split}
\end{equation}
We now perform the change of variables $t = r^2/\qty(1+r^2)$ and $u = s^2/\qty(1+s^2)$. We get
\begin{equation}
    \cI_2 = 4\pi^2 \int_0^1\dd{t} \int_0^1 \dd{u} \frac{1}{\abs{t-u}}= 8\pi^2 \int_0^1 \dd{t}\int_0^t \dd{u}
\frac{1}{t-u} = 8\pi^2 \int_0^1 \dd{t}\int_0^1 \dd{v}\frac{1}{1-v}~,
\end{equation}
where $u = t\,v$. Now using the integral representation of the Riemann zeta function:
\begin{align}
    \zeta(s) = \frac{1}{\Gamma(s)}\int_0^\infty \dd{x}\, \frac{x^{s-1}}{\ex{x} - 1}~,    
\end{align}
and evaluating at $s=1$, after the change of variables $\ex{-x}=v$, gives
\begin{equation}\label{eq:zeta(1)}
\zeta(1) = \int_0^1 \frac{\dd{v}}{1-v}~.
\end{equation}
Since both sides of the above expression are divergent, the equality should be understood formally. One can perform the above manipulations in the presence of a regulator and remove the cutoff at the end. Substituting \cref{eq:zeta(1)} into $\cI_2$, we get the desired result:
\begin{equation}
    \cI_2 = 8\pi^2 \zeta(1)~. 
\end{equation}

\subsection{Order $\nu^4$ contribution}\label{app:s2nu4}
The order-$\nu^4$ contribution to the sphere partition function amounts to evaluating the integral \cref{eq:I4-main-pre}, repeated here for convenience:
\begin{equation}\label{eq: I4app}
    \cI_4 = \frac{1}{\lds^{4}}\int_{\C^4}\prod_{i=1}^4 \dd[2]{x_i}\,
	\frac{\Omega(x_1)\Omega(x_2)\Omega(x_3)\Omega(x_4)}{\abs{x_{13}}^2\,\abs{x_{24}}^2}\frac{(\eta-\bar\eta)^2}{\abs{\eta}^2\,\abs{1-\eta}^2}~. 
\end{equation}
We now show how to derive the result claimed in \cref{eq:I4-final}. The cross-ratio $\eta$ is defined as
\begin{equation}
	\eta=\frac{x_{12}\,x_{34}}{x_{13}\,x_{24}}
	=
	\frac{x_{12}(x_3-x_4)}{(x_1-x_3)x_{24}}~,
	\label{eq:app-cross-ratio-equation}
\end{equation}
where $x_{ij} \coloneqq x_i - x_j$. Equivalently, we have
\begin{equation}
    x_3(\eta)=\frac{x_1\, x_{24}\, \eta+ x_4\, x_{12}}{x_{12}+\eta\, x_{24}}.
	\label{eq:app-x3-solved}
\end{equation}
We can therefore trade the integration over $x_3$ for an integration over the cross-ratio $\eta$. The Jacobian of this change of variable is
\begin{equation}
	\frac{\dd[2]{x_3}}{\abs{x_{13}}^2\abs{x_{24}}^2}
	=
	\frac{\dd[2]{\eta}}{\abs{x_{12}+\eta\,x_{24}}^2}.
	\label{eq: jacobian4}
\end{equation}
Substituting \cref{eq: jacobian4} into \cref{eq: I4app}, we arrive at
\begin{equation}
	\cI_4
	=
	\frac{1}{\lds^{4}} \int_{\C} \dd[2]{\eta}\; K(\eta,\bar \eta)\, G^\t{c}_{\epsilon\epsilon\epsilon\epsilon}(\eta,\bar{\eta})~, 
	\label{eq: I4reduces app}
\end{equation}
with
\begin{equation}
	K(\eta,\bar \eta)
	=
	\int_{\C^3}
	\dd[2]{x_1}\, \dd[2]{x_2}\, \dd[2]{x_4}\;
	\frac{\Omega(x_1)\,\Omega(x_2)\,\Omega\qty\big(x_3(\eta))\,\Omega(x_4)}
	{\abs{x_{12}+\eta\,x_{24}}^2}~,
	\label{eq:app-K-three-point-integral}
\end{equation}
and $G^\t{c}_{\epsilon\epsilon\epsilon\epsilon}(\eta,\bar{\eta})$ as in \cref{eq: 4pt Ising}. 
Let us now evaluate $K(\eta,\bar \eta)$. The strategy is to use the $\SO(3)$ isometries of the sphere to map the generic points $(x_1,x_2,x_3,x_4)$ to $(\infty,1,\eta,0)$. The usual Möbius transformation that achieves this is
\begin{equation}
    g(w)=\frac{a\,w+b}{c\,w+d}~,
	\label{eq: mobius}
\end{equation}
with
\begin{equation}
	a = x_1 \qty(\frac{x_{24}}{x_{12}\,x_{14}})^{1/2}, 
	\quad
	b= x_4 \qty(\frac{x_{12}}{x_{14}\,x_{24}})^{1/2}, 
	\quad
	c= \qty(\frac{x_{24}}{x_{12}\,x_{14}})^{1/2}, 
	\quad
	d= \qty(\frac{x_{12}}{x_{14}\,x_{24}})^{1/2}.
	\label{eq: abcd}
\end{equation}
As a check these satisfy the determinant condition $a\,d-b\,c=1$. Moreover, one can check that
\begin{equation}
	g(\infty)=x_1~, \qquad g(1)=x_2~, \qquad g(0)=x_4~, \qq{and} g(\eta)=x_3~. 
	\label{eq:app-xi-as-gw}
\end{equation}
We change variables from $\qty(x_1,x_2,x_4)$ to $\qty(x_4,c,d)$. The Jacobian is
\begin{equation}
    \dd[2]{x_1}\,\dd[2]{x_2}\,\dd[2]{x_4} = 2\,\abs{x_{12}x_{24}x_{14}}^2 \dd[2]{x_4}\,\dd[2]{c}\,\dd[2]{d}~,
\end{equation}
and combined with the denominator of \cref{eq:app-K-three-point-integral} gives
\begin{equation}
    \frac{\abs{x_{12}\,x_{24}\,x_{14}}^2}{\abs{x_{12}+\eta\, x_{24}}^2}=\abs{g'(\infty)}\,\abs{g'(1)}\,\abs{g'(\eta)}\,\abs{g'(0)}~.
\end{equation}
Here 
\begin{equation}
    g'(w)=\frac{1}{(cw+d)^2}~,
\end{equation}
so the kernel can be rewritten as
\begin{equation}
    K(\eta,\bar\eta)=2\, \int_{\C^3} \dd[2]{x_4}\,\dd[2]{c}\,\dd[2]{d}\; \prod_{w = 0,1,\eta,\infty}\Omega(g(w))\,\abs{g'(w)}~. 
\end{equation}
The combination $\Omega(g(w))\abs{g'(w)}$ takes a simple form in terms of
\begin{equation}
   \rho_0  \coloneqq  \frac{1}{\abs{a}^2+\abs{c}^2} \in \R_+~, 
	\qq{and}
	\varrho  \coloneqq  -\frac{b\,\bar{a}+d\,\bar{c}}{\abs{a}^2+\abs{c}^2} \in \C~. 
\end{equation}
Specifically we have:
\begin{equation}
\begin{split}
    \Omega(g(w))\abs{g'(w)} 
    &= 
	\frac{2\lds\,\rho_0}{\rho_0^2+\abs{\varrho-w}^2}~, \qquad w=0,1,\eta~, 
    \\
    \Omega(g(\infty))\abs{g'(\infty)}
    &=
    2\lds \rho_0~. 
\end{split}
\end{equation}
It is useful to change variables again, from $(x_4,d)$ to $(\varrho,a)$. This transformation has a unit Jacobian. We parametrise the coordinates $(a,c)$ as $a=r\,u$ and $c=r\,v$ such that $r = \rho_0^{-1/2}$, with $\abs{u}^2+\abs{v}^2=1$. This implies
\begin{equation}
    \dd[2]{x_4}\,\dd[2]{c}\, \dd[2]{d} = \dd[2]{\varrho}\, \dd[2]{a}\, \dd[2]{c} =  \frac{1}{2}\dd[2]{\varrho}\;\frac{\dd{\rho_0}}{\rho_0^3}\; \dd{\Omega_{S^3}}~,
\end{equation}
where $\dd{\Omega_{\S^3}}$ is the volume form of the unit $\S^3$. The kernel reduces nicely to
\begin{equation}
    K\qty(\eta,\bar{\eta}) =  2\pi^2\,(2\lds)^4
    \int_0^\infty \dd{\rho_0}\; \rho_0
    \int_{\C}\dd[2]{\varrho}\,
    \frac{1}{
    \qty(\rho_0^2+\abs{\varrho}^2)
    \qty(\rho_0^2+\abs{\varrho-\eta}^2)
    \qty(\rho_0^2+\abs{\varrho-1}^2)
    }~.
\end{equation}
This integral can be simplified with a Schwinger parametrisation as
\begin{equation}
\begin{split}
& \frac{1}{
    \qty(\rho_0^2+\abs{\varrho}^2)
    \qty(\rho_0^2+\abs{\varrho-\eta}^2)
    \qty(\rho_0^2+\abs{\varrho-1}^2)
    }
    \\
    &= 
	\int_0^\infty \dd{t_1}\,\dd{t_2}\,\dd{t_3}\; 
	\exp[-t_1\qty(\rho_0^2+\abs{\varrho}^2)-t_2 \qty(\rho_0^2+\abs{\varrho-\eta}^2)-t_3 \qty(\rho_0^2+\abs{\varrho-1}^2)]~. 
\end{split}
\end{equation}
The integral over $\varrho$ and $\rho_0$ is now Gaussian and yields
\begin{equation}
    K\qty(\eta,\bar{\eta})
	=
	\pi^3(2\lds)^4
	\int_0^\infty
	\frac{\dd{t_1}\,\dd{t_2}\,\dd{t_3}}{\qty(t_1+t_2+t_3)^2}
	\exp(
	-\frac{t_1\,t_2\,\abs{\eta}^2+t_1\,t_3+t_2\,t_3\,\abs{1-\eta}^2}{t_1+t_2+t_3}).
\end{equation}
We recognise this as one of the $D$-functions appearing in the computation of Witten diagrams \cite{DHoker:1999kzh,Penedones:2010ue}. More precisely, it is
\begin{equation}
	K\qty(\eta,\bar{\eta})
	=
	16\pi^3\,\lds^4\, D_{1111}(\eta,\bar \eta)~, 
	\label{eq:app-K-D1111}
\end{equation}
where
\begin{equation}
\begin{split}
	D_{1111}(\eta,\bar \eta)
	&=
	\int_0^\infty
	\dd{\alpha_1} \int_0^\infty\dd{\alpha_2}\int_0^\infty\dd{\alpha_3}\;
	\frac{\delta(1-\alpha_1-\alpha_2-\alpha_3)}{
		\alpha_1\alpha_2\abs{\eta}^2+\alpha_1\alpha_3+\alpha_2\alpha_3\abs{1-\eta}^2
	} \\
    &=
	\frac{2\,D_\t{BW}(\eta)}{\Im \eta}~. 
\end{split}
\end{equation}
In the last line we rewrote the $D_{1111}$ function in terms of the Bloch--Wigner dilogarithm 
\begin{equation}
    D_\t{BW}(\eta)
	\coloneqq
	\Im(\Li_2(\eta))+\arg(1-\eta)\log\abs{\eta}~. 
\end{equation}
With this rewriting the computation of the integral is finally straightforward. Combining the final result for the kernel $K(\eta,\bar \eta)$ with the connected four-point function $G^\t{c}_{\epsilon\epsilon\epsilon\epsilon}$ from \cref{eq: 4pt Ising}, we get
\begin{equation}\label{eq: I4 intermediate}
    \cI_4
	=
	-128\pi^3
	\int_\C \dd[2]{\eta}\,
	\frac{\Im \eta\, D_\t{BW}(\eta)}{\abs{\eta}^2\abs{1-\eta}^2}.
\end{equation}
Since the integrand is symmetric under $\eta \mapsto \bar{\eta}$, we can restrict the integration to the upper half-plane, and multiply by 2. 

The Bloch--Wigner dilogarithm is central in the computation of volumes of hyperbolic 3-manifolds \cite{zagier2007dilogarithm,thurston2022geometry}. One of the standard proofs \cite{Neumann:1985yjj,thurston2022geometry}, leads us to consider an angular parametrisation
\begin{equation}
	\eta=\frac{\sin\beta}{\sin(\alpha+\beta)}\,\ex{\ii\alpha},
	\qquad
	\alpha>0,\quad \beta>0,\quad \alpha+\beta<\pi.
\end{equation}
With this parametrisation the integration measure is flat
\begin{equation}
	\frac{\dd[2]{\eta}\;\Im\eta}{\abs{\eta}^2\abs{1-\eta}^2}
	=
	\dd{\alpha}\,\dd{\beta},
\end{equation}
and $D_\t{BW}$ takes the form 
\begin{equation}
	D_\t{BW}(\eta)=\Lambda(\alpha)+\Lambda(\beta)+\Lambda(\pi-\alpha-\beta)~,
\end{equation}
where $\Lambda(\vartheta)$ is the Lobachevsky function defined as
\begin{align}
	\Lambda(\vartheta)\coloneqq-\int_0^\vartheta \dd{t}\,\log(2\sin t)~. 
\end{align}
Therefore, 
\cref{eq: I4 intermediate}  becomes
\begin{align}
	\cI_4
	 & = -256\pi^3
	\int_0^\pi \dd{\alpha}
	\int_0^{\pi-\alpha}\dd{\beta}\;
	\qty[\Lambda(\alpha)+\Lambda(\beta)+\Lambda(\pi-\alpha-\beta)] \nn
	 & = -768\pi^3
	\int_0^\pi \dd{\alpha}\,(\pi-\alpha)\,\Lambda(\alpha)~.
	\label{eq:app-I4-Lobachevsky}
\end{align}
In the second line we used the symmetry of the triangular region
$\alpha>0$, $\beta>0$, $\alpha+\beta<\pi$ under permutations of the three
angles $\alpha$, $\beta$, and $\pi-\alpha-\beta$. The Lobachevsky function has a simple form in Fourier space:
\begin{equation}
	\Lambda(\vartheta)=\frac{1}{2}\sum_{n=1}^\infty \frac{\sin(2n\vartheta)}{n^2}~.
\end{equation}
Through its integral, the Riemann zeta function comes alive:
\begin{align}
	\int_0^\pi \dd{\alpha}\,(\pi-\alpha)\,\Lambda(\alpha)
	 & = \frac{1}{2}\sum_{n=1}^\infty \frac{1}{n^2}
	\int_0^\pi \dd{\alpha}\,(\pi-\alpha)\sin(2n\alpha) = \frac{1}{2}\sum_{n=1}^\infty \frac{1}{n^2}\,\frac{\pi}{2n}
	 = \frac{\pi}{4}\,\zeta(3).
\end{align}
Substituting this into \cref{eq:app-I4-Lobachevsky}, we then obtain
\begin{equation}
	\cI_4=-12\,(2\pi)^4\,\zeta(3)~.
\end{equation}
This is the value claimed in \cref{eq:I4-final}. We stress again that the value $\zeta(3)$ ultimately appears by the remarkable properties of the Bloch--Wigner dilogarithm \cite{zagier2007dilogarithm} and is related to the geometry of hyperbolic manifolds \cite{thurston2022geometry}.

\section{Details of the energy two-point function from CPT}\label{app:eps-eps-app}

In this appendix we provide details on the integral \cref{eq:intc-main} controlling the $\ev{\epsilon\epsilon}$ correlator in conformal perturbation theory. In particular, we fill in the steps between \cref{eq:intc-main} and \cref{eq:intc-main-goal}, after which the integral can be evaluated exactly.

We repeat the desired integral here for convenience:
\begin{align}\label{eq:int-app-def}
    \cI_\t{c}(X) = 4\lds^2 \int_{\C^2} \dd[2]{z}\dd[2]{w} \frac{(\eta-\bar{\eta})^2}{(1+\abs{z}^2)(1+\abs{w}^2)\,\abs{z}^2\abs{r-w}^2\abs{\eta}^2\abs{1-\eta}^2}~. 
\end{align} 
The first useful step is to replace integration over \(z\) by integration over the conformal cross-ratio \(\eta\). The relevant change of variables is
\begin{align}
    z=\frac{r w}{r(1-\eta)+\eta\, w}~, 
\end{align}
under which
\begin{align}
    \frac{\dd[2]{z}}{\abs{z}^2 \abs{r-w}^2}=\frac{\dd[2]{\eta}}{\abs{r(1-\eta)+\eta\, w}^2}~. 
\end{align}
It is useful to package the remaining \(w\)-integral into the ``kernel'' \(K_r(\eta,\bar{\eta})\), defined by
\begin{align}\label{eq:kernel-def}
    K_r(\eta,\bar{\eta})=4\lds^2 \int_\C \frac{\dd[2]{w}}{\qty(1+\abs{w}^2)\qty(\abs{r(1-\eta)+\eta\, w}^2+r^2\abs{w}^2)}~. 
\end{align}
With this definition, the integral becomes
\begin{align}
    \cI_\t{c}(X) = \int_\C \dd[2]{\eta} \frac{(\eta-\bar{\eta})^2}{\abs{\eta}^2\abs{1-\eta}^2}\, K_r(\eta,\bar{\eta})~. 
\end{align}

\paragraph{The kernel.}

We now evaluate the kernel \(K_r(\eta,\bar{\eta})\). It is useful to introduce a Feynman parameter \(\lambda\) by writing
\begin{align}
    \frac{1}{A\,B} = \int_0^\infty \frac{\dd{\lambda}}{\qty(A + \lambda\,B)^2}~, 
\end{align}
with
\begin{align}
    A=1+\abs{w}^2~, \qquad B=\abs{r(1-\eta)+\eta\, w}^2+r^2\abs{w}^2~, 
\end{align}
so that the kernel becomes
\begin{align}
    K_r(\eta,\bar{\eta})=4\lds^2 \int_0^\infty \dd{\lambda} \int_\C \frac{\dd[2]{w}}{\qty(A+\lambda\, B)^2}~. 
\end{align}
The \(w\)-integral can now be computed directly, since both \(A\) and \(B\) are quadratic in \(w\). We make use of the integral formula,
\begin{align}
    \int_\C \frac{\dd[2]{w}}{\qty(a\abs{w}^2+b\, w+\bar{b}\,\bar{w}+c)^2} = \frac{\pi}{a\, c - \abs{b}^2}~, 
\end{align} 
which holds for $a,c>0$, $b\in\C$ with $a\,c>\abs{b}^2$. The formula can be easily proved by completing the square and passing to polar coordinates.  
Overall, \(K_r(\eta,\bar{\eta})\) takes the form
\begin{align}
    K_r(\eta,\bar{\eta}) &= 4\pi\,\lds^2 \int_{0}^\infty \frac{\dd{\lambda}}{1 + \qty[\abs{\eta}^2+r^2\qty(1+\abs{1-\eta}^2)]\,\lambda + r^4\abs{1-\eta}^2\,\lambda^2} \\
    &= 
    \frac{4\pi\lds^2}{r^2} \int_0^1 \frac{\dd{\varpi}}{(1-\varpi)+ \varpi\abs{1-\eta}^2 +\frac{\varpi(1-\varpi)}{r^2} \abs{\eta}^2}~, 
\end{align}
where the second line follows from the further change of variables \(\varpi=r^2\, \lambda/(1+r^2\,\lambda)\). The total integral is therefore
\begin{align}
    \cI_\t{c}(X) = \frac{4\pi \lds^2}{r^2} \int_0^1 \dd{\varpi} \int_\C \dd[2]{\eta} \frac{(\eta-\bar{\eta})^2}{\abs{\eta}^2\abs{1-\eta}^2\qty((1-\varpi)+ \varpi\abs{1-\eta}^2 +\frac{\varpi(1-\varpi)}{r^2} \abs{\eta}^2)}~. 
\end{align}

\paragraph{The cross-ratio integral.}

The integral over the conformal cross-ratio \(\eta\) can be reduced with another Feynman parameterisation:
\begin{align}
    \frac{1}{A\,B\,C} = 2\int_{[0,1]^3} \dd{\alpha}\dd{\beta}\dd{\gamma}\frac{\delta\qty(1-\alpha-\beta-\gamma)}{\qty(\alpha A + \beta B + \gamma C)^3}~, 
\end{align}
where
\begin{align}
    A = \abs{\eta}^2~, \quad B = \abs{1-\eta}^2~, \quad C = (1-\varpi)+ \varpi\abs{1-\eta}^2 +\frac{\varpi(1-\varpi)}{r^2} \abs{\eta}^2~. 
\end{align}
It follows that
\begin{align}\label{eq:intc-app-intermediate-copy}
    \cI_\t{c}(X) = -\frac{32\pi \lds^2}{r^2} \int_0^1 \dd{\varpi} 
    \int_{[0,1]^3} \dd{\alpha}\dd{\beta}\dd{\gamma}\;\delta\qty(1-\alpha-\beta-\gamma) 
    \int_\C \dd[2]{\eta} \frac{(\Im\eta)^2}{\cD_r^3}~, 
\end{align}
with
\begin{align}
    \cD_r = \abs{\eta}^2\,\alpha+ \abs{1-\eta}^2 \beta + \qty((1-\varpi)+ \varpi\abs{1-\eta}^2 +\frac{\varpi(1-\varpi)}{r^2} \abs{\eta}^2)\,\gamma~. 
\end{align}
From here, we can perform the integral over \(\eta\) explicitly. It is convenient to view it as an integral over \(\R^2\), with \(\eta = \eta_x +\ii \eta_y\), and \(\dd[2]{\eta}=\dd{\eta_x}\dd{\eta_y}\). A direct calculation gives
\begin{align}
    \int_\C \dd[2]{\eta} \frac{(\Im\eta)^2}{\cD_r^3} = \frac{\pi}{4\qty(\tilde{A}+\tilde{B})\qty[\tilde{A}\,\tilde{B} + \tilde{C}\qty(\tilde{A}+\tilde{B})]}~, 
\end{align}
with
\begin{align}
    \tilde{A} = \alpha + \gamma\, \frac{\varpi(1-\varpi)}{r^2}~, \quad \tilde{B} = \beta + \gamma\, \varpi~, \qq{and} \tilde{C} = \gamma(1-\varpi)~, 
\end{align}
For the above we made use of the standard integral
\begin{align}
    \int_{\R^2} \dd{\eta_x}\dd{\eta_y}\;\frac{\eta_y^2}{\qty(\eta_x^2+\eta_y^2+\Delta)^3} = \frac{\pi}{4\Delta}~, 
\end{align}
familiar from one-loop momentum integrals in QFT.

\paragraph{Reduction to a simplex integral.}

The delta function in \cref{eq:intc-app-intermediate-copy} enforces that \(\alpha\), \(\beta\) and \(\gamma\) live on a simplex. Resolving the delta function by setting \(\beta = 1-\alpha-\gamma\), we get
\begin{align}
    \cI_\t{c}(X) = -\frac{8 \pi^2\lds^2}{r^2} \int_0^1 \dd{\varpi} \int_0^1 \dd{\alpha} \int_0^{1-\alpha} \dd{\gamma} \frac{1}{\Upsilon_+\qty(\Upsilon_0+\gamma\,\Upsilon_-- \gamma^2)}~, 
\end{align}
where
\begin{align}
    \Upsilon_\pm(\alpha,\varpi) &= 1 - \alpha\varpi \pm \frac{\alpha\, \varpi(1-\varpi)}{r^2} \\ 
    \Upsilon_0(\alpha,\varpi) &= \qty(1 - \alpha\varpi + \frac{1-\varpi}{r^2})\alpha\,\varpi~, 
\end{align}
after also relabelling $\varpi\to 1-\varpi$. The integral over $\gamma$ can be performed directly, after further trading $\alpha$ for $\delta = \alpha\varpi/(1-\gamma)$, with $0\leq \delta\leq \varpi$. This gives
\begin{align}
    \cI_\t{c}(X) = -\frac{8 \pi^2\lds^2}{r^2} \int_0^1 \frac{\dd{\varpi}}{\varpi} \int_0^\varpi \frac{\dd{\delta}}{(1-\delta)^2} \log(\frac{1-\delta+\delta(1-\varpi)/r^2}{\delta\qty(1-\delta+(1-\varpi)/r^2)})~. 
\end{align}
After one final change of variables to $t=1-\delta$ and $s=(1-\varpi)/(1-\delta)$, so that $0\leq s,t\leq 1$, the integral is brought to its final form: 
\begin{align}
    \cI_\t{c}(X) = -\frac{8 \pi^2\lds^2}{r^2} \int_0^1 \dd{s} \int_0^1 \dd{t} \frac{1}{t(1-s\,t)}\qty[\log(1-\frac{s\,t}{r^2+s})-\log(1-t)]~, 
\end{align}
which is \cref{eq:intc-main-goal,eq:J(xi)-intermediate}.

\section{Spin two-point function at antipodal points}\label{app:antipodal}

In this appendix we prove the result claimed in \cref{eq:GoverGtilde}, namely that
\begin{equation}
    \frac{\Gm(2)}{\Gs(2)}=\ex{\pi \nu}~,
\end{equation}
where $\Gs(u),\Gm(u)$ are the sphere two-point functions of the spin and disorder operators, respectively. For this purpose, it is useful to use the free Majorana presentation of the Ising model. In this frame both $\sigma(x)$ and $\mu(x)$ create states in the Ramond sectors of the theory. Equivalently, the composite operators
\begin{equation}
    \sigma(x_0)\Psi(x) \qq{and} \mu(x_0)\Psi(x)
\end{equation}
acquire a minus sign under a $2\pi$-rotation of $x$ around $x_0$.

To compute this ratio, we first observe that the sphere with two antipodal points removed is conformally equivalent to the cylinder. We write the metric on the sphere as
\begin{equation}
\dd{s}^2_{S^2_\lds}
=
\lds^2\qty(
\dd{\vartheta}^2+\sin^2\vartheta\,\dd{\varphi}^2
)~,
\qquad
\vartheta\in\closed{0}{\pi},
\quad
\varphi\in\ropen{0}{2\pi}~.
\end{equation}
We then introduce the cylinder coordinate
\begin{equation}
t=\log\tan\frac{\vartheta}{2}~,
\label{eq:sphere-cylinder-coordinate}
\end{equation}
which gives
\begin{equation}
\sin\vartheta=\frac{1}{\cosh t}~,
\qquad
\dd{\vartheta}=\frac{\dd{t}}{\cosh t}~. 
\end{equation}
Hence
\begin{equation}
\dd{s}^2_{S^2_\lds}
=
\frac{\lds^2}{\cosh^2 t}
\qty(
\dd{t}^2+\dd{\varphi}^2
).
\label{eq:sphere-cylinder-metric}
\end{equation}
The north and south poles correspond respectively to
\begin{equation}
\vartheta=0
\quad\Longleftrightarrow\quad
t=-\infty~, \qq{and}
\vartheta=\pi
\quad\Longleftrightarrow\quad
t=+\infty.
\end{equation}

We can remove the conformal factor from the metric by a Weyl rescaling as
\begin{equation}
g_{\alpha\beta}
=
\ex{2\omega(t)}\widehat g_{\alpha\beta},
\qq{with}
\ex{\omega(t)}
=
\frac{\lds}{\cosh t}~,
\end{equation}
where
\begin{equation}
\dd{\widehat{s}}^{\,2}=\dd{t}^2+\dd{\varphi}^2
\end{equation}
is the flat-cylinder metric. However, since the massive theory is not conformally invariant, this Weyl rescaling produces a different Lagrangian. The kinetic term of the Majorana fermion is obviously Weyl invariant, while the mass term transforms as
\begin{equation}
\int \dd[2]{x}\,\sqrt{g}\; 
m\, \overline{\Psi}_g\,\Psi_g
=
\int \dd[2]{x}\,\sqrt{\widehat{g}}\; 
M(t)\, \overline{\Psi}_\t{cyl}\,\Psi_\t{cyl}~,
\end{equation}
where
\begin{equation}
M(t)
=
m\, \ex{\omega(t)}
=
\frac{m\lds}{\cosh t}
=
\frac{\nu}{\cosh t}~.
\label{eq:cylinder-mass-profile}
\end{equation}
Thus the original constant mass on the sphere becomes a \emph{time-dependent} but \emph{spatially homogeneous} mass on the cylinder.%
\footnote{Note that for a two-point function with non-antipodal insertions, the corresponding mass profile would be both time- and space-dependent.}

The insertions of the spin operator $\sigma(x)$ or disorder operator $\mu(x)$ at $t=-\infty$ and $t=\infty$ change the monodromy of the Majorana field, and $\Psi$ is now in the Ramond sector. To quantise the theory in this sector, we follow \cite{Fonseca:2003ee} and expand the fermionic field as
\begin{equation}
\begin{split}
    \psi(x) &= \sum_{n=-\infty}^{\infty} \qty\Big(a_n u_{-n}(x)+\bar{a}_n v_{-n}(x))~,\\
    \overline{\psi}(x) &= \sum_{n=-\infty}^{\infty} \qty\Big(a_n \bar u_{-n}(x)+\bar{a}_n \bar v_{-n}(x))~, 
\end{split}
\end{equation}
where $\psi$ and $\overline{\psi}$ denote the holomorphic and anti-holomorphic components of the Majorana spinor. The operators $a_{n}$ and $\bar{a}_n$ obey the algebra
\begin{equation}
\acomm{a_n}{a_m}=\delta_{n+m,0}~,
\qquad
\acomm{\bar{a}_n}{\bar{a}_m}=\delta_{n+m,0}~,
\qquad
\acomm{a_n}{\bar{a}_m}=0~.
\label{eq:ramond-mode-algebra}
\end{equation}
In particular, the zero modes satisfy
\begin{equation}\label{eq:zero-mode-modes}
a_0^2=\bar{a}_0^2=\frac{1}{2}~.
\end{equation}

From the above properties it follows that there are two distinguished states in the Ramond sector. These are the states defined by the action of $\sigma$ and $\mu$ on the \ds-invariant Bunch--Davies vacuum, $\ket{0}$. From a Euclidean path-integral perspective, they can be viewed as states prepared by performing a hemisphere path integral with an insertion of $\sigma$ or $\mu$ at the south pole. We denote these order and disorder states by $\ket{\sigma}$ and $\ket{\mu}$, respectively. They are special because they are both annihilated by all positive Ramond modes:
\begin{equation}
a_n\ket{\sigma}
	=
\bar{a}_n\ket{\sigma}
	=
a_n\ket{\mu}
	=
\bar{a}_n\ket{\mu}
	=
0~,
\qquad n>0~.
\label{eq:positive-mode-vacua}
\end{equation}
Their difference lies only in the representation of the zero-mode algebra they furnish.

The zero modes act on the two Ramond ground states as
\begin{equation}
\begin{aligned}
a_0\ket{\sigma}
&=
\frac{\omega}{\sqrt{2}}\ket{\mu}~,
&
\bar{a}_0\ket{\sigma}
&=
\frac{\bar{\omega}}{\sqrt{2}}\ket{\mu}~,
		\\
a_0\ket{\mu}
&=
\frac{\bar{\omega}}{\sqrt{2}}\ket{\sigma}~,
&
\bar{a}_0\ket{\mu}
&=
\frac{\omega}{\sqrt{2}}\ket{\sigma}~,
	\end{aligned}
\label{eq:zero-mode-action}
\end{equation}
where $\omega = \ex{\ii \pi/4}$, with $\bar{\omega}=\ex{-\ii\pi/4}$~.
It is useful to introduce the Ramond zero-mode operator
\begin{equation}
Q_\t{R} \coloneqq 2\ii\, a_0\,\bar{a}_0~.
\label{eq:zero-mode-Q}
\end{equation}
Note that by \cref{eq:ramond-mode-algebra,eq:zero-mode-modes}, it follows that $Q_\t{R}$ squares to the identity. 
Successive application of \cref{eq:zero-mode-action} reveals that
\begin{equation}
Q_\t{R}\ket{\sigma}=+\ket{\sigma}~,
\qquad
Q_\t{R}\ket{\mu}=-\ket{\mu}~.
\label{eq:Q-eigenvalues}
\end{equation}
Thus $\sigma$ and $\mu$ correspond to the two eigenstates of the Ramond zero-mode operator $Q_\t{R}$.

Now, the crucial observation is that, since the Majorana mass on the flat cylinder is still $\varphi$-independent, angular momentum is conserved and different Fourier modes do not mix. The full Hamiltonian can be written as
\begin{equation}
    H = H_0 + H_\t{osc}~,
\end{equation}
where
\begin{equation}
    H_0 = \frac{M(t)}{2}Q_\t{R}~, 
\end{equation}
and $H_\t{osc}$ is the non-zero-mode contribution
\begin{equation}
    H_\t{osc} = \sum_{n>0}\qty\Big[n\qty\big(a_{-n} a_n + \bar{a}_{-n}\bar{a}_n) + \ii\, M(t) \qty\big(a_{-n}\bar{a}_n + a_n\bar{a}_{-n})]~.
\end{equation}
In the cylinder geometry, the antipodal two-point functions of the spin and disorder operators are given by the expectation values of the Euclidean time-evolution operator 
\begin{align}
    U(-\infty,+\infty) \coloneqq \operatorname{T}\!\exp(-\int_{-\infty}^{+\infty} \dd{t} H(t))~,
\end{align}
in the states $\ket{\sigma}$ and $\ket{\mu}$, respectively. In the above, $\operatorname{T}$ stands for time ordering, since the Hamiltonian is time-dependent. Now, since the oscillator contribution acts in exactly the same way on $\ket{\sigma}$ and $\ket{\mu}$, its contribution to the two-point function cancels in the ratio. We thus obtain
\begin{equation}
    \frac{\Gm(2)}{\Gs(2)}= \frac{\exp(\frac{1}{2}\int_{-\infty}^{\infty}\dd{t}\; M(t))}{\exp(-\frac{1}{2}\int_{-\infty}^{\infty}\dd{t}\; M(t))}=\exp(\int_{-\infty}^{\infty}\dd{t}\; M(t)) = \ex{\pi \nu}~.
\end{equation}

\printbibliography

@article{Anninos:2020hfj,
    author = "Anninos, Dionysios and Denef, Frederik and Law, Y. T. Albert and Sun, Zimo",
    title = "{Quantum de Sitter horizon entropy from quasicanonical bulk, edge, sphere and topological string partition functions}",
    eprint = "2009.12464",
    archivePrefix = "arXiv",
    primaryClass = "hep-th",
    doi = "10.1007/JHEP01(2022)088",
    journal = "JHEP",
    volume = "01",
    pages = "088",
    year = "2022"
}

@article{deLacroix:2023uem,
    author = "de Lacroix, Corinne and Erbin, Harold and Lahoche, Vincent",
    title = "{Gravitational action for a massive Majorana fermion in 2d quantum gravity}",
    eprint = "2308.08342",
    archivePrefix = "arXiv",
    primaryClass = "hep-th",
    doi = "10.1007/JHEP01(2024)068",
    journal = "JHEP",
    volume = "01",
    pages = "068",
    year = "2024"
}

@article{Anninos:2024fty,
    author = "Anninos, Dionysios and Anous, Tarek and Rios Fukelman, Alan",
    title = "{De Sitter at all loops: the story of the Schwinger model}",
    eprint = "2403.16166",
    archivePrefix = "arXiv",
    primaryClass = "hep-th",
    doi = "10.1007/JHEP08(2024)155",
    journal = "JHEP",
    volume = "08",
    pages = "155",
    year = "2024"
}

@article{Karch:2019lnn,
    author = "Karch, Andreas and Tong, David and Turner, Carl",
    title = "{A Web of 2d Dualities: ${\bf Z}_2$ Gauge Fields and Arf Invariants}",
    eprint = "1902.05550",
    archivePrefix = "arXiv",
    primaryClass = "hep-th",
    doi = "10.21468/SciPostPhys.7.1.007",
    journal = "SciPost Phys.",
    volume = "7",
    pages = "007",
    year = "2019"
}

@article{DHoker:1999kzh,
    author = "D'Hoker, Eric and Freedman, Daniel Z. and Mathur, Samir D. and Matusis, Alec and Rastelli, Leonardo",
    title = "{Graviton exchange and complete four point functions in the AdS / CFT correspondence}",
    eprint = "hep-th/9903196",
    archivePrefix = "arXiv",
    reportNumber = "MIT-CTP-2843, UCLA-99-TEP-2",
    doi = "10.1016/S0550-3213(99)00525-8",
    journal = "Nucl. Phys. B",
    volume = "562",
    pages = "353--394",
    year = "1999"
}

@article{Zamolodchikov:2001dz,
    author = "Zamolodchikov, A.",
    title = "{Scaling Lee-Yang model on a sphere. 1. Partition function}",
    eprint = "hep-th/0109078",
    archivePrefix = "arXiv",
    reportNumber = "LPM-01-30",
    doi = "10.1088/1126-6708/2002/07/029",
    journal = "JHEP",
    volume = "07",
    pages = "029",
    year = "2002"
}

@article{Pethybridge:2021eoh,
    author = "Pethybridge, Ben and Schaub, Vladimir",
    title = "{Tensors and spinors in de Sitter space}",
    eprint = "2111.14899",
    archivePrefix = "arXiv",
    primaryClass = "hep-th",
    doi = "10.1007/JHEP06(2022)123",
    journal = "JHEP",
    volume = "06",
    pages = "123",
    year = "2022"
}

@article{Anous:2020dbn,
    author = "Anous, Tarek and Skulte, Jim",
    title = "{An invitation to the principal series}",
    doi = "10.21468/SciPostPhys.9.3.028",
    journal = "SciPost Phys.",
    volume = "9",
    number = "3",
    pages = "028",
    year = "2020"
}

@article{Wu:1975mw,
    author = "Wu, Tai Tsun and McCoy, Barry M. and Tracy, Craig A. and Barouch, Eytan",
    title = "{Spin spin correlation functions for the two-dimensional Ising model: Exact theory in the scaling region}",
    doi = "10.1103/PhysRevB.13.316",
    journal = "Phys. Rev. B",
    volume = "13",
    pages = "316--374",
    year = "1976"
}

@article{Fonseca:2003ee,
    author = "Fonseca, P. and Zamolodchikov, A.",
    title = "{Ward identities and integrable differential equations in the Ising field theory}",
    eprint = "hep-th/0309228",
    archivePrefix = "arXiv",
    reportNumber = "RUNHETC-2003-28",
    month = "9",
    year = "2003"
}

@article{Anninos:2023lin,
    author = {Anninos, Dionysios and Anous, Tarek and Pethybridge, Ben and {\c{S}}eng{\"o}r, Gizem},
    title = "{The discreet charm of the discrete series in dS$_{2}$}",
    eprint = "2307.15832",
    archivePrefix = "arXiv",
    primaryClass = "hep-th",
    doi = "10.1088/1751-8121/ad14ad",
    journal = "J. Phys. A",
    volume = "57",
    number = "2",
    pages = "025401",
    year = "2024"
}

@article{Chernikov:1968zm,
    author = "Chernikov, N. A. and Tagirov, E. A.",
    title = "{Quantum theory of scalar field in de Sitter space-time}",
    journal = "Ann. Inst. H. Poincare Phys. Theor. A",
    volume = "9",
    number = "2",
    pages = "109--141",
    year = "1968"
}

@article{Bunch:1978yq,
    author = "Bunch, T. S. and Davies, P. C. W.",
    title = "{Quantum Field Theory in de Sitter Space: Renormalization by Point Splitting}",
    doi = "10.1098/rspa.1978.0060",
    journal = "Proc. Roy. Soc. Lond. A",
    volume = "360",
    pages = "117--134",
    year = "1978"
}

@article{Anninos:2012qw,
    author = "Anninos, Dionysios",
    title = "{De Sitter Musings}",
    eprint = "1205.3855",
    archivePrefix = "arXiv",
    primaryClass = "hep-th",
    doi = "10.1142/S0217751X1230013X",
    journal = "Int. J. Mod. Phys. A",
    volume = "27",
    pages = "1230013",
    year = "2012"
}

@article{Bhardwaj:2017xup,
    author = "Bhardwaj, Lakshya and Tachikawa, Yuji",
    title = "{On finite symmetries and their gauging in two dimensions}",
    eprint = "1704.02330",
    archivePrefix = "arXiv",
    primaryClass = "hep-th",
    reportNumber = "IPMU-17-0049",
    doi = "10.1007/JHEP03(2018)189",
    journal = "JHEP",
    volume = "03",
    pages = "189",
    year = "2018"
}

@article{DiPietro:2021sjt,
    author = "Di Pietro, Lorenzo and Gorbenko, Victor and Komatsu, Shota",
    title = "{Analyticity and unitarity for cosmological correlators}",
    eprint = "2108.01695",
    archivePrefix = "arXiv",
    primaryClass = "hep-th",
    reportNumber = "CERN-TH-2021-118",
    doi = "10.1007/JHEP03(2022)023",
    journal = "JHEP",
    volume = "03",
    pages = "023",
    year = "2022"
}

@article{Doyon:2004fv,
    author = "Doyon, Benjamin and Fonseca, Pedro",
    title = "{Ising field theory on a Pseudosphere}",
    eprint = "hep-th/0404136",
    archivePrefix = "arXiv",
    reportNumber = "RUNHETC-2003-37, SPHT-T04-031",
    doi = "10.1088/1742-5468/2004/07/P07002",
    journal = "J. Stat. Mech.",
    volume = "0407",
    pages = "P07002",
    year = "2004"
}

@article{Penedones:2010ue,
    author = "Penedones, Joao",
    title = "{Writing CFT correlation functions as AdS scattering amplitudes}",
    eprint = "1011.1485",
    archivePrefix = "arXiv",
    primaryClass = "hep-th",
    doi = "10.1007/JHEP03(2011)025",
    journal = "JHEP",
    volume = "03",
    pages = "025",
    year = "2011"
}

@article{Zamolodchikov:1989cf,
  author = {Zamolodchikov, A. B.},
  title = {Integrals of motion and {$S$}-matrix of the (scaled) {$T=T_c$} Ising model with magnetic field},
  journal = {Int. J. Mod. Phys. A},
  volume = {4},
  pages = {4235--4248},
  year = {1989},
  doi = {10.1142/S0217751X8900176X}
}

@article{Delfino:2004yva,
  author = {Delfino, Gesualdo},
  title = {Integrable field theory and critical phenomena: The Ising model in a magnetic field},
  eprint = {hep-th/0312119},
  archivePrefix = {arXiv},
  primaryClass = {hep-th},
  journal = {J. Phys. A},
  volume = {37},
  pages = {R45--R78},
  year = {2004},
  doi = {10.1088/0305-4470/37/14/R01}
}

@article{Camporesi:1995fb,
    author = "Camporesi, Roberto and Higuchi, Atsushi",
    title = "{On the eigenfunctions of the Dirac operator on spheres and real hyperbolic spaces}",
    eprint = "gr-qc/9505009",
    archivePrefix = "arXiv",
    doi = "10.1016/0393-0440(95)00042-9",
    journal = "J. Geom. Phys.",
    volume = "20",
    pages = "1--18",
    year = "1996"
}

@book{Varshalovich:1988ye,
  author    = {Varshalovich, D. A. and Moskalev, A. N. and Khersonskii, V. K.},
  title     = {Quantum Theory of Angular Momentum},
  publisher = {World Scientific},
  address   = {Singapore},
  year      = {1988}
}

@article{Letsios:2025pqo,
    author = "Letsios, Vasileios A. and Pethybridge, Ben and Rios Fukelman, Alan",
    title = "{Quite discrete for a fermion}",
    eprint = "2501.03724",
    archivePrefix = "arXiv",
    primaryClass = "hep-th",
    doi = "10.1007/JHEP07(2025)016",
    journal = "JHEP",
    volume = "07",
    pages = "016",
    year = "2025"
}

@article{Mueck:1999efk,
    author = "Mueck, Wolfgang",
    title = "{Spinor parallel propagator and Green's function in maximally symmetric spaces}",
    eprint = "hep-th/9912059",
    archivePrefix = "arXiv",
    doi = "10.1088/0305-4470/33/15/308",
    journal = "J. Phys. A",
    volume = "33",
    pages = "3021--3026",
    year = "2000"
}

@article{DiPietro:2023inn,
    author = "Di Pietro, Lorenzo and Gorbenko, Victor and Komatsu, Shota",
    title = "{Cosmological Correlators at Finite Coupling}",
    eprint = "2312.17195",
    archivePrefix = "arXiv",
    primaryClass = "hep-th",
    month = "12",
    year = "2023"
}

@article{Allen:1985ux,
    author = "Allen, Bruce",
    title = "{Vacuum States in de Sitter Space}",
    reportNumber = "UCSB-TH-3-1985",
    doi = "10.1103/PhysRevD.32.3136",
    journal = "Phys. Rev. D",
    volume = "32",
    pages = "3136",
    year = "1985"
}

@article{HartleHawking,
  title = {Wave function of the Universe},
  author = {Hartle, J. B. and Hawking, S. W.},
  journal = {Phys. Rev. D},
  volume = {28},
  issue = {12},
  pages = {2960--2975},
  numpages = {0},
  year = {1983},
  month = {12},
  publisher = {American Physical Society},
  doi = {10.1103/PhysRevD.28.2960},
  url = {https://link.aps.org/doi/10.1103/PhysRevD.28.2960}
}

@article{Baumann:2022jpr,
    author = "Baumann, Daniel and Green, Daniel and Joyce, Austin and Pajer, Enrico and Pimentel, Guilherme L. and Sleight, Charlotte and Taronna, Massimo",
    title = "{Snowmass White Paper: The Cosmological Bootstrap}",
    eprint = "2203.08121",
    archivePrefix = "arXiv",
    primaryClass = "hep-th",
    doi = "10.21468/SciPostPhysCommRep.1",
    journal = "SciPost Phys. Comm. Rep.",
    volume = "2024",
    pages = "1",
    year = "2024"
}

@article{Miller:2025jbz,
    author = "Miller, Noah",
    title = "{Path integral games with de Sitter {\ensuremath{\alpha}}-vacua}",
    eprint = "2503.13701",
    archivePrefix = "arXiv",
    primaryClass = "hep-th",
    doi = "10.1007/JHEP10(2025)097",
    journal = "JHEP",
    volume = "10",
    pages = "097",
    year = "2025"
}

@article{Mottola:1984ar,
    author = "Mottola, E.",
    title = "{Particle Creation in de Sitter Space}",
    reportNumber = "NSF-ITP-84-123",
    doi = "10.1103/PhysRevD.31.754",
    journal = "Phys. Rev. D",
    volume = "31",
    pages = "754",
    year = "1985"
}

@article{Letsios:2026ypo,
    author = "Letsios, Vasileios A. and Vitouladitis, Stathis",
    title = "{Axions on de Sitter space}",
    eprint = "2606.28858",
    archivePrefix = "arXiv",
    primaryClass = "hep-th",
    month = "6",
    year = "2026"
}

@article{Aguilera-Damia:2026dbk,
    author = "Aguilera-Damia, Jeremias and Anninos, Dionysios and Anous, Tarek and Gleeson, Johnny and Rios Fukelman, Alan",
    title = "{de Sitter Vacua {\&} pUniverses}",
    eprint = "2605.02883",
    archivePrefix = "arXiv",
    primaryClass = "hep-th",
    month = "5",
    year = "2026"
}

@article{Hogervorst:2021uvp,
    author = "Hogervorst, Matthijs and Penedones, Jo{\~a}o and Vaziri, Kamran Salehi",
    title = "{Towards the non-perturbative cosmological bootstrap}",
    eprint = "2107.13871",
    archivePrefix = "arXiv",
    primaryClass = "hep-th",
    doi = "10.1007/JHEP02(2023)162",
    journal = "JHEP",
    volume = "02",
    pages = "162",
    year = "2023"
}

@article{Sleight:2020obc,
    author = "Sleight, Charlotte and Taronna, Massimo",
    title = "{From AdS to dS exchanges: Spectral representation, Mellin amplitudes, and crossing}",
    eprint = "2007.09993",
    archivePrefix = "arXiv",
    primaryClass = "hep-th",
    doi = "10.1103/PhysRevD.104.L081902",
    journal = "Phys. Rev. D",
    volume = "104",
    number = "8",
    pages = "L081902",
    year = "2021"
}

@article{Loparco:2024ibp,
    author = "Loparco, Manuel",
    title = "{RG flows in de Sitter: C-functions and sum rules}",
    eprint = "2404.03739",
    archivePrefix = "arXiv",
    primaryClass = "hep-th",
    doi = "10.21468/SciPostPhys.17.3.079",
    journal = "SciPost Phys.",
    volume = "17",
    number = "3",
    pages = "079",
    year = "2024"
}

@article{Loparco:2023rug,
    author = "Loparco, Manuel and Penedones, Joao and Salehi Vaziri, Kamran and Sun, Zimo",
    title = {{The K{\"a}ll{\'e}n-Lehmann representation in de Sitter spacetime}},
    eprint = "2306.00090",
    archivePrefix = "arXiv",
    primaryClass = "hep-th",
    doi = "10.1007/JHEP12(2023)159",
    journal = "JHEP",
    volume = "12",
    pages = "159",
    year = "2023"
}

@article{Antunes:2025iaw,
    author = "Antunes, Ant{\'o}nio and Levine, Nat and Meineri, Marco",
    title = "{Demystifying integrable QFTs in AdS: No-go theorems for higher-spin charges}",
    eprint = "2502.06937",
    archivePrefix = "arXiv",
    primaryClass = "hep-th",
    doi = "10.21468/SciPostPhys.20.3.088",
    journal = "SciPost Phys.",
    volume = "20",
    pages = "088",
    year = "2026"
}

@article{Senatore:2009cf,
    author = "Senatore, Leonardo and Zaldarriaga, Matias",
    title = "{On Loops in Inflation}",
    eprint = "0912.2734",
    archivePrefix = "arXiv",
    primaryClass = "hep-th",
    doi = "10.1007/JHEP12(2010)008",
    journal = "JHEP",
    volume = "12",
    pages = "008",
    year = "2010"
}

@article{Antinucci:2026uuh,
    author = "Antinucci, Andrea and Copetti, Christian and Galati, Giovanni and Rizi, Giovanni",
    title = "{A Twist on Scattering from Defect Anomalies}",
    eprint = "2605.13961",
    archivePrefix = "arXiv",
    primaryClass = "hep-th",
    month = "5",
    year = "2026"
}

@article{Polyakov:2012uc,
    author = "Polyakov, A. M.",
    title = "{Infrared instability of the de Sitter space}",
    eprint = "1209.4135",
    archivePrefix = "arXiv",
    primaryClass = "hep-th",
    month = "9",
    year = "2012"
}

@article{TSAMIS19951,
title = {Strong Infrared Effects in Quantum Gravity},
journal = {Annals of Physics},
volume = {238},
number = {1},
pages = {1-82},
year = {1995},
issn = {0003-4916},
doi = {https://doi.org/10.1006/aphy.1995.1015},
url = {https://www.sciencedirect.com/science/article/pii/S0003491685710159},
author = {N.C. Tsamis and R.P. Woodard},
}

@article{Antoniadis:1986,
  title = {Quantum Instability of de Sitter Space},
  author = {Antoniadis, I. and Iliopoulos, J. and Tomaras, T. N.},
  journal = {Phys. Rev. Lett.},
  volume = {56},
  issue = {13},
  pages = {1319--1322},
  numpages = {0},
  year = {1986},
  month = {3},
  publisher = {American Physical Society},
  doi = {10.1103/PhysRevLett.56.1319},
  url = {https://link.aps.org/doi/10.1103/PhysRevLett.56.1319}
}

@article{Ford:1985,
  title = {Quantum instability of de Sitter spacetime},
  author = {Ford, L. H.},
  journal = {Phys. Rev. D},
  volume = {31},
  issue = {4},
  pages = {710--717},
  numpages = {0},
  year = {1985},
  month = {2},
  publisher = {American Physical Society},
  doi = {10.1103/PhysRevD.31.710},
  url = {https://link.aps.org/doi/10.1103/PhysRevD.31.710}
}

@article{Anninos:2014lwa,
    author = "Anninos, Dionysios and Anous, Tarek and Freedman, Daniel Z. and Konstantinidis, George",
    title = "{Late-time Structure of the Bunch-Davies De Sitter Wavefunction}",
    eprint = "1406.5490",
    archivePrefix = "arXiv",
    primaryClass = "hep-th",
    reportNumber = "MIT-CTP-4561",
    doi = "10.1088/1475-7516/2015/11/048",
    journal = "JCAP",
    volume = "11",
    pages = "048",
    year = "2015"
}

@article{Penedones:2023uqc,
    author = "Penedones, Joao and Salehi Vaziri, Kamran and Sun, Zimo",
    title = "{Hilbert space of quantum field theory in de Sitter spacetime}",
    eprint = "2301.04146",
    archivePrefix = "arXiv",
    primaryClass = "hep-th",
    doi = "10.1103/PhysRevD.111.045001",
    journal = "Phys. Rev. D",
    volume = "111",
    number = "4",
    pages = "045001",
    year = "2025"
}

@article{PhysRevD.33.2833,
  title = {Global symmetry breaking in two-dimensional flat spacetime and in de Sitter spacetime},
  author = {Ford, L. H. and Vilenkin, Alexander},
  journal = {Phys. Rev. D},
  volume = {33},
  issue = {10},
  pages = {2833--2839},
  numpages = {0},
  year = {1986},
  month = {5},
  publisher = {American Physical Society},
  doi = {10.1103/PhysRevD.33.2833},
  url = {https://link.aps.org/doi/10.1103/PhysRevD.33.2833}
}

@article{Delfino:1994nx,
    author = "Delfino, G. and Mussardo, G. and Simonetti, P.",
    title = "{Statistical models with a line of defect}",
    eprint = "hep-th/9403049",
    archivePrefix = "arXiv",
    reportNumber = "SISSA-30-94-EP",
    doi = "10.1016/0370-2693(94)90439-1",
    journal = "Phys. Lett. B",
    volume = "328",
    pages = "123--129",
    year = "1994"
}

@article{Letsios:2020twa,
    author = "Letsios, Vasileios A.",
    title = "{The eigenmodes for spinor quantum field theory in global de Sitter space{\textendash}time}",
    eprint = "2011.07875",
    archivePrefix = "arXiv",
    primaryClass = "gr-qc",
    doi = "10.1063/5.0038651",
    journal = "J. Math. Phys.",
    volume = "62",
    number = "3",
    pages = "032303",
    year = "2021"
}

@article{Zagier-polylogs,
  title = {Multiple Zeta Values of Fixed Weight, Depth, and Height},
  author = {Ohno, Yasuo and Zagier, Don},
  date = {2001-12-17},
  journaltitle = {Indagationes Mathematicae},
  volume = {12},
  number = {4},
  pages = {483--487},
  issn = {0019-3577},
  doi = {10.1016/S0019-3577(01)80037-9},
}

@online{Schaub:2023scu,
  title = {Spinors in ({{Anti-}})de {{Sitter Space}}},
  author = {Schaub, Vladimir},
  date = {2023-06-22},
  eprint = {2302.08535},
  eprinttype = {arXiv},
  eprintclass = {hep-th},
  doi = {10.48550/arXiv.2302.08535},
  }

@article{Belavin:1984vu,
    author = "Belavin, A. A. and Polyakov, Alexander M. and Zamolodchikov, A. B.",
    editor = "Khalatnikov, I. M. and Mineev, V. P.",
    title = "{Infinite Conformal Symmetry in Two-Dimensional Quantum Field Theory}",
    reportNumber = "CERN-TH-3827",
    doi = "10.1016/0550-3213(84)90052-X",
    journal = "Nucl. Phys. B",
    volume = "241",
    pages = "333--380",
    year = "1984"
}

@article{DiFrancesco:1987ez,
    author = "Di Francesco, P. and Saleur, H. and Zuber, J. B.",
    title = "{Critical Ising Correlation Functions in the Plane and on the Torus}",
    reportNumber = "SACLAY-SPH-T-87-097",
    doi = "10.1016/0550-3213(87)90202-1",
    journal = "Nucl. Phys. B",
    volume = "290",
    pages = "527",
    year = "1987"
}

@article{Chang:2018iay,
    author = "Chang, Chi-Ming and Lin, Ying-Hsuan and Shao, Shu-Heng and Wang, Yifan and Yin, Xi",
    title = "{Topological Defect Lines and Renormalization Group Flows in Two Dimensions}",
    eprint = "1802.04445",
    archivePrefix = "arXiv",
    primaryClass = "hep-th",
    reportNumber = "CALT-TH-2017-067, CALT-TH 2017-067, PUPT-2546",
    doi = "10.1007/JHEP01(2019)026",
    journal = "JHEP",
    volume = "01",
    pages = "026",
    year = "2019"
}

@article{Henn:2013pwa,
  title = {Multiloop Integrals in Dimensional Regularization Made Simple},
  author = {Henn, Johannes M.},
  date = {2013},
  journaltitle = {Phys. Rev. Lett.},
  volume = {110},
  pages = {251601},
  doi = {10.1103/PhysRevLett.110.251601},
}

@article{Henn:2014qga,
  title = {Lectures on Differential Equations for {{Feynman}} Integrals},
  author = {Henn, Johannes M.},
  date = {2015},
  journaltitle = {J. Phys. A},
  volume = {48},
  pages = {153001},
  doi = {10.1088/1751-8113/48/15/153001},
}

@article{Kotikov:1990kg,
  title = {Differential Equations Method. {{New}} Technique for Massive {{Feynman}} Diagram Calculation},
  author = {Kotikov, A. V.},
  date = {1991-01-17},
  journaltitle = {Physics Letters B},
  shortjournal = {Physics Letters B},
  volume = {254},
  number = {1},
  pages = {158--164},
  issn = {0370-2693},
  doi = {10.1016/0370-2693(91)90413-K},
}

@article{Remiddi:1997ny,
  title = {Differential {{Equations}} for {{Feynman Graph Amplitudes}}},
  author = {Remiddi, Ettore},
  date = {1997-12},
  journaltitle = {Il Nuovo Cimento A},
  shortjournal = {Il Nuovo Cimento A (1971-1996)},
  volume = {110},
  number = {12},
  eprint = {hep-th/9711188},
  eprinttype = {arXiv},
  pages = {1435--1452},
  issn = {1826-9869},
  doi = {10.1007/BF03185566},
}

@article{Starobinsky:1980te,
    author = "Starobinsky, Alexei A.",
    title = "{A New Type of Isotropic Cosmological Models Without Singularity}",
    doi = "10.1016/0370-2693(80)90670-X",
    journal = "Phys. Lett. B",
    volume = "91",
    pages = "99--102",
    year = "1980"
}

@article{Guth:1980zm,
    author = "Guth, Alan H.",
    title = "{The Inflationary Universe: A Possible Solution to the Horizon and Flatness Problems}",
    doi = "10.1103/PhysRevD.23.347",
    journal = "Phys. Rev. D",
    volume = "23",
    pages = "347--356",
    year = "1981"
}

@article{Linde:1981mu,
    author = "Linde, Andrei D.",
    title = "{A New Inflationary Universe Scenario: A Possible Solution of the Horizon, Flatness, Homogeneity, Isotropy and Primordial Monopole Problems}",
    doi = "10.1016/0370-2693(82)91219-9",
    journal = "Phys. Lett. B",
    volume = "108",
    pages = "389--393",
    year = "1982"
}

@article{Albrecht:1982wi,
    author = "Albrecht, Andreas and Steinhardt, Paul J.",
    title = "{Cosmology for Grand Unified Theories with Radiatively Induced Symmetry Breaking}",
    doi = "10.1103/PhysRevLett.48.1220",
    journal = "Phys. Rev. Lett.",
    volume = "48",
    pages = "1220--1223",
    year = "1982"
}

@article{PlanckCollab,
  author  = {{Planck Collaboration}},
  title   = {Planck 2018 Results. VI. Cosmological Parameters},
  journal = {Astronomy \& Astrophysics},
  volume  = {641},
  pages   = {A6},
  year    = {2020},
  doi     = {10.1051/0004-6361/201833910},
  url     = {https://doi.org/10.1051/0004-6361/201833910}
}

@article{Suzuki,
  author  = {Suzuki, N. and Rubin, D. and Lidman, C. and Aldering, G.
             and Amanullah, R. and Barbary, K. and others},
  title   = {The Hubble Space Telescope Cluster Supernova Survey. V.
             Improving the Dark-Energy Constraints Above $z > 1$
             and Building an Early-Type-Hosted Supernova Sample},
  journal = {The Astrophysical Journal},
  volume  = {746},
  pages   = {85},
  year    = {2012},
  doi     = {10.1088/0004-637X/746/1/85},
  url     = {https://doi.org/10.1088/0004-637X/746/1/85}
}

@misc{SDSS,
  author       = {{Sloan Digital Sky Survey}},
  title        = {Sloan Digital Sky Survey (SDSS)},
  year         = {2024},
  howpublished = {\url{https://www.sdss.org}},
}

@article{Higuchi:1986wu,
    author = "Higuchi, Atsushi",
    title = "{Symmetric Tensor Spherical Harmonics on the $N$ Sphere and Their Application to the De Sitter Group SO($N$,1)}",
    reportNumber = "YTP-86-19",
    doi = "10.1063/1.527513",
    journal = "J. Math. Phys.",
    volume = "28",
    pages = "1553",
    year = "1987",
    note = "[Erratum: J. Math. Phys. 43, 6385 (2002)]"
}

@article{Higuchi:1991tn,
    author = "Higuchi, A.",
    title = "{Linearized Gravity in de Sitter Space-Time as a Representation of SO(4,1)}",
    doi = "10.1088/0264-9381/8/11/011",
    journal = "Class. Quant. Grav.",
    volume = "8",
    pages = "2005--2021",
    year = "1991"
}

@article{Higuchi:1991tk,
    author = "Higuchi, A.",
    title = "{Quantum Linearization Instabilities of de Sitter Space-Time. 1}",
    doi = "10.1088/0264-9381/8/11/009",
    journal = "Class. Quant. Grav.",
    volume = "8",
    pages = "1961--1981",
    year = "1991"
}

@article{Higuchi:1991tm,
    author = "Higuchi, A.",
    title = "{Quantum Linearization Instabilities of de Sitter Space-Time. 2}",
    doi = "10.1088/0264-9381/8/11/010",
    journal = "Class. Quant. Grav.",
    volume = "8",
    pages = "1983--2004",
    year = "1991"
}

@article{Letsios:2022tsq,
    author = "Letsios, Vasileios A.",
    title = "{(Non-)unitarity of strictly and partially massless fermions on de Sitter space II: an explanation based on the group-theoretic properties of the spin-3/2 and spin-5/2 eigenmodes}",
    eprint = "2206.09851",
    archivePrefix = "arXiv",
    primaryClass = "hep-th",
    doi = "10.1088/1751-8121/ad2c27",
    journal = "J. Phys. A",
    volume = "57",
    number = "13",
    pages = "135401",
    year = "2024"
}

@article{Letsios:2023qzq,
    author = "Letsios, Vasileios A.",
    title = "{(Non-)unitarity of strictly and partially massless fermions on de Sitter space}",
    eprint = "2303.00420",
    archivePrefix = "arXiv",
    primaryClass = "hep-th",
    doi = "10.1007/JHEP05(2023)015",
    journal = "JHEP",
    volume = "05",
    pages = "015",
    year = "2023"
}

@article{Hinterbichler:2026xqf,
    author = "Hinterbichler, Kurt",
    title = "{De Sitter Representations}",
    eprint = "2606.26221",
    archivePrefix = "arXiv",
    primaryClass = "hep-th",
    month = "6",
    year = "2026"
}

@article{Gorbenko:2019rza,
    author = "Gorbenko, Victor and Senatore, Leonardo",
    title = "{$\lambda \phi^4$ in dS}",
    eprint = "1911.00022",
    archivePrefix = "arXiv",
    primaryClass = "hep-th",
    month = "10",
    year = "2019"
}

@article{Smith:2026dae,
    author = "Smith, Joseph",
    title = "{A Note on the Perturbative Expansion of the Schwinger Model on $S^2$}",
    eprint = "2603.21938",
    archivePrefix = "arXiv",
    primaryClass = "hep-th",
    month = "3",
    year = "2026"
}

@article{Jayewardena:1988td,
    author = "Jayewardena, C.",
    title = "{Schwinger model on \ensuremath{S^2}}",
    journal = "Helv. Phys. Acta",
    volume = "61",
    pages = "636--711",
    year = "1988"
}

@article{Spindel:1976,
     author = {Schomblond, Christiane and Spindel, Philippe},
     title = {Conditions d{\textquoteright}unicité pour le propagateur $\Delta^1(x,y)$ du champ scalaire dans l{\textquoteright}univers de de {Sitter}},
     journal = {Annales de l'institut Henri Poincar\'e. Section A, Physique Th\'eorique},
     pages = {67--78},
     year = {1976},
     publisher = {Gauthier-Villars},
     volume = {25},
     number = {1},
     zbl = {0356.53010},
     language = {fr},
     url = {https://www.numdam.org/item/AIHPA_1976__25_1_67_0/}
}

@article{ThirringScwhinger,
  author = {Anous, Tarek and Gleeson, Johnny and Paul, Priyardashi and Rios Fukelman, Alan},
  journal = {to appear},
  number = {},
  title = {},
  volume = {},
  year = {}
}

@article{Green:2020dynamicalrg,
    author = "Green, Daniel and Premkumar, Akhil",
    title = "{Dynamical RG and Critical Phenomena in de Sitter Space}",
    eprint = "2001.05974",
    archivePrefix = "arXiv",
    primaryClass = "hep-th",
    doi = "10.1007/JHEP04(2020)064",
    journal = "JHEP",
    volume = "04",
    pages = "064",
    year = "2020"
}

@article{Akhmedov:2019cfd,
    author = "Akhmedov, E. T. and Moschella, U. and Popov, F. K.",
    title = "{Characters of different secular effects in various patches of de Sitter space}",
    eprint = "1901.07293",
    archivePrefix = "arXiv",
    primaryClass = "hep-th",
    month = "1",
    year = "2019"
}

@article{Kirsten:1993ug,
    author = "Kirsten, Klaus and Garriga, Jaume",
    title = "{Massless minimally coupled fields in de Sitter space: O(4) symmetric states versus de Sitter invariant vacuum}",
    eprint = "gr-qc/9305013",
    archivePrefix = "arXiv",
    reportNumber = "TUTP-92-1",
    doi = "10.1103/PhysRevD.48.567",
    journal = "Phys. Rev. D",
    volume = "48",
    pages = "567--577",
    year = "1993"
}

@inproceedings{Witten:2001kn,
    author = "Witten, Edward",
    title = "{Quantum gravity in de Sitter space}",
    booktitle = "{Strings 2001: International Conference}",
    eprint = "hep-th/0106109",
    archivePrefix = "arXiv",
    month = "6",
    year = "2001"
}

@article{Galante:2023uyf,
    author = "Galante, Damian A.",
    title = "{Modave lectures on de Sitter space {\&} holography}",
    eprint = "2306.10141",
    archivePrefix = "arXiv",
    primaryClass = "hep-th",
    doi = "10.22323/1.435.0003",
    journal = "PoS",
    volume = "Modave2022",
    pages = "003",
    year = "2023"
}

@inproceedings{Spradlin:2001pw,
    author = "Spradlin, Marcus and Strominger, Andrew and Volovich, Anastasia",
    title = "{Les Houches lectures on de Sitter space}",
    booktitle = "{Les Houches Summer School: Session 76: Euro Summer School on Unity of Fundamental Physics: Gravity, Gauge Theory and Strings}",
    eprint = "hep-th/0110007",
    archivePrefix = "arXiv",
    pages = "423--453",
    month = "10",
    year = "2001"
}

@article{Sun:2021thf,
    author = "Sun, Zimo",
    title = "{A note on the representations of SO(1,d + 1)}",
    eprint = "2111.04591",
    archivePrefix = "arXiv",
    primaryClass = "hep-th",
    doi = "10.1142/S0129055X24300073",
    journal = "Rev. Math. Phys.",
    volume = "37",
    number = "01",
    pages = "2430007",
    year = "2025"
}

@article{Sengor:2022kji,
    author = {{\c{S}}eng{\"o}r, Gizem},
    title = "{Particles of a de Sitter Universe}",
    eprint = "2212.10626",
    archivePrefix = "arXiv",
    primaryClass = "hep-th",
    doi = "10.3390/universe9020059",
    journal = "Universe",
    volume = "9",
    number = "2",
    pages = "59",
    year = "2023"
}

@article{Schaub:2024rnl,
    author = "Schaub, Vladimir",
    title = "{A Walk Through $Spin(1,d+1)$}",
    eprint = "2405.01659",
    archivePrefix = "arXiv",
    primaryClass = "hep-th",
    month = "5",
    year = "2024"
}

@book{Enayati:2022hed,
    author = "Enayati, Mohammad and Gazeau, Jean-Pierre and Pejhan, Hamed and Wang, Anzhong",
    title = "{The de Sitter (dS) Group and its Representations. An Introduction to Elementary Systems and Modeling the Dark Energy Universe}",
    eprint = "2201.11457",
    archivePrefix = "arXiv",
    primaryClass = "math-ph",
    doi = "10.1007/978-3-031-16045-5",
    isbn = "978-3-031-16047-9",
    publisher = "Springer",
    series = "Synthesis Lectures on Mathematics {\&} Statistics",
    year = "2023"
}

@article{Basile:2016aen,
    author = "Basile, Thomas and Bekaert, Xavier and Boulanger, Nicolas",
    title = "{Mixed-symmetry fields in de Sitter space: a group theoretical glance}",
    eprint = "1612.08166",
    archivePrefix = "arXiv",
    primaryClass = "hep-th",
    doi = "10.1007/JHEP05(2017)081",
    journal = "JHEP",
    volume = "05",
    pages = "081",
    year = "2017"
}

@article{Ottoson:1968,
  title = {A Classification of the Unitary Irreducible Representations of $SO_0(N, 1)$},
  author = {Ottoson, Ulf},
  date = {1968-09-01},
  journaltitle = {Communications in Mathematical Physics},
  shortjournal = {Commun.Math. Phys.},
  volume = {8},
  number = {3},
  pages = {228--244},
  issn = {1432-0916},
  doi = {10.1007/BF01645858},
}

@article{Schwarz:1971,
  title = {Unitary irreducible representations of the groups $SO(n,1)$},
  author = {Schwarz, Fritz},
  date = {1971-01-01},
  journaltitle = {Journal of Mathematical Physics},
  shortjournal = {J. Math. Phys.},
  volume = {12},
  number = {1},
  pages = {131--139},
  issn = {0022-2488},
  doi = {10.1063/1.1665471},
}

@article{Hirai:1962,
  title = {On irreducible representations of the Lorentz group of $n$-th order},
  author = {Hirai, Takeshi},
  date = {1962-01},
  journaltitle = {Proceedings of the Japan Academy},
  volume = {38},
  number = {6},
  pages = {258--262},
  publisher = {The Japan Academy},
  issn = {0021-4280},
  doi = {10.3792/pja/1195523378},
}

@article{Kitaev:2017hnr,
    author = "Kitaev, Alexei",
    title = "{Notes on $\widetilde{\mathrm{SL}}(2,\mathbb{R})$ representations}",
    eprint = "1711.08169",
    archivePrefix = "arXiv",
    primaryClass = "hep-th",
    month = "11",
    year = "2017"
}

@article{Higuchi:2010xt,
    author = "Higuchi, Atsushi and Marolf, Donald and Morrison, Ian A.",
    title = "{On the Equivalence between Euclidean and In-In Formalisms in de Sitter QFT}",
    eprint = "1012.3415",
    archivePrefix = "arXiv",
    primaryClass = "gr-qc",
    doi = "10.1103/PhysRevD.83.084029",
    journal = "Phys. Rev. D",
    volume = "83",
    pages = "084029",
    year = "2011"
}

@article{Allen:1985wd,
    author = "Allen, Bruce and Jacobson, Theodore",
    title = "{Vector Two Point Functions in Maximally Symmetric Spaces}",
    reportNumber = "UCSB-TH-4-1985",
    doi = "10.1007/BF01211169",
    journal = "Commun. Math. Phys.",
    volume = "103",
    pages = "669",
    year = "1986"
}

@article{SalehiVaziri:2024joi,
    author = "Salehi Vaziri, Kamran",
    title = "{A non-perturbative construction of the de Sitter late-time boundary}",
    eprint = "2412.00183",
    archivePrefix = "arXiv",
    primaryClass = "hep-th",
    month = "11",
    year = "2024"
}

@article{Sengor:2019mbz,
    author = {Seng{\"o}r, Gizem and Skordis, Constantinos},
    title = "{Unitarity at the Late time Boundary of de Sitter}",
    eprint = "1912.09885",
    archivePrefix = "arXiv",
    primaryClass = "hep-th",
    doi = "10.1007/JHEP06(2020)041",
    journal = "JHEP",
    volume = "06",
    pages = "041",
    year = "2020"
}

@article{Cohen:2024anu,
    author = "Cohen, Timothy and Green, Daniel and Huang, Yiwen",
    title = "{Operator origin of anomalous dimensions in de Sitter space}",
    eprint = "2407.08581",
    archivePrefix = "arXiv",
    primaryClass = "hep-th",
    reportNumber = "CERN-TH-2024-103",
    doi = "10.1103/PhysRevD.111.103513",
    journal = "Phys. Rev. D",
    volume = "111",
    number = "10",
    pages = "103513",
    year = "2025"
}

@article{VasilisGizem,
  title={A discrete series gauge field at the late-time boundary of $dS_4$},
  author={Botshekananfard, Manizheh and  Büşra Güraksın, Elif and  Letsios, Vasileios A. and Şengör, Gizem},
  journal={to appear},
  year={2026}
}

@article{Frohlich:2004ef,
    author = "Frohlich, Jurg and Fuchs, Jurgen and Runkel, Ingo and Schweigert, Christoph",
    title = "{Kramers-Wannier duality from conformal defects}",
    eprint = "cond-mat/0404051",
    archivePrefix = "arXiv",
    reportNumber = "HU-EP-04-19",
    doi = "10.1103/PhysRevLett.93.070601",
    journal = "Phys. Rev. Lett.",
    volume = "93",
    pages = "070601",
    year = "2004"
}

@article{Ambrosino:2025pjj,
    author = "Ambrosino, Federico and Runkel, Ingo and Watts, G{\'e}rard M. T.",
    title = "{Translation invariant defects as an extension of topological symmetries}",
    eprint = "2511.02007",
    archivePrefix = "arXiv",
    primaryClass = "hep-th",
    doi = "10.1142/S0217751X26480015",
    journal = "Int. J. Mod. Phys. A",
    volume = "41",
    number = "07",
    pages = "2648001",
    year = "2026"
}

@article{Witten:2015aba,
    author = "Witten, Edward",
    title = "{Fermion Path Integrals And Topological Phases}",
    eprint = "1508.04715",
    archivePrefix = "arXiv",
    primaryClass = "cond-mat.mes-hall",
    doi = "10.1103/RevModPhys.88.035001",
    journal = "Rev. Mod. Phys.",
    volume = "88",
    number = "3",
    pages = "035001",
    year = "2016"
}

@article{Tong:2019bbk,
    author = "Tong, David and Turner, Carl",
    title = "{Notes on 8 Majorana Fermions}",
    eprint = "1906.07199",
    archivePrefix = "arXiv",
    primaryClass = "hep-th",
    doi = "10.21468/SciPostPhysLectNotes.14",
    journal = "SciPost Phys. Lect. Notes",
    volume = "14",
    pages = "1",
    year = "2020"
}

@incollection{zagier2007dilogarithm,
  title={The dilogarithm function},
  author={Zagier, Don},
  booktitle={Frontiers in number theory, physics, and geometry II: on conformal field theories, discrete groups and renormalization},
  pages={3--65},
  year={2007},
  publisher={Springer}
}

@article{Neumann:1985yjj,
  title = {Volumes of Hyperbolic Three-Manifolds},
  author = {Neumann, Walter D. and Zagier, Don},
  date = {1985-01-01},
  journaltitle = {Topology},
  shortjournal = {Topology},
  volume = {24},
  number = {3},
  pages = {307--332},
  issn = {0040-9383},
  doi = {10.1016/0040-9383(85)90004-7},
}

@book{thurston2022geometry,
  title={The geometry and topology of three-manifolds: With a preface by Steven P. Kerckhoff},
  author={Thurston, William P},
  volume={27},
  year={2022},
  publisher={American Mathematical Society}
}

@article{LopezNacir:2016gzi,
    author = "L{\'o}pez Nacir, Diana and Mazzitelli, Francisco D. and Trombetta, Leonardo G.",
    title = "{$O(N)$ model in Euclidean de Sitter space: beyond the leading infrared approximation}",
    eprint = "1606.03481",
    archivePrefix = "arXiv",
    primaryClass = "hep-th",
    reportNumber = "CERN-TH-2016-126",
    doi = "10.1007/JHEP09(2016)117",
    journal = "JHEP",
    volume = "09",
    pages = "117",
    year = "2016"
}

@article{Namuduri:2023smx,
    author = "Namuduri, Manojna and Bilal, Adel",
    title = "{Effective gravitational action for 2D massive Majorana fermions on arbitrary genus Riemann surfaces}",
    eprint = "2308.05802",
    archivePrefix = "arXiv",
    primaryClass = "hep-th",
    doi = "10.1007/JHEP11(2023)194",
    journal = "JHEP",
    volume = "11",
    pages = "194",
    year = "2023"
}

@article{Anninos:2024iwf,
    author = {Anninos, Dionysios and Baracco, Chiara and M{\"u}hlmann, Beatrix},
    title = "{Remarks on 2D quantum cosmology}",
    eprint = "2406.15271",
    archivePrefix = "arXiv",
    primaryClass = "hep-th",
    doi = "10.1088/1475-7516/2024/10/031",
    journal = "JCAP",
    volume = "10",
    pages = "031",
    year = "2024"
}

@article{Anninos:2021ene,
    author = {Anninos, Dionysios and Bautista, Teresa and M{\"u}hlmann, Beatrix},
    title = "{The two-sphere partition function in two-dimensional quantum gravity}",
    eprint = "2106.01665",
    archivePrefix = "arXiv",
    primaryClass = "hep-th",
    doi = "10.1007/JHEP09(2021)116",
    journal = "JHEP",
    volume = "09",
    pages = "116",
    year = "2021"
}

@article{Brezin:1989db,
    author = "Brezin, Edouard and Douglas, Michael R. and Kazakov, Vladimir and Shenker, Stephen H.",
    title = "{The Ising Model Coupled to 2-$D$ Gravity: A Nonperturbative Analysis}",
    reportNumber = "RU-89-47",
    doi = "10.1016/0370-2693(90)90458-I",
    journal = "Phys. Lett. B",
    volume = "237",
    pages = "43--46",
    year = "1990"
}

@article{Anninos:2020ccj,
    author = {Anninos, Dionysios and M{\"u}hlmann, Beatrix},
    title = "{Notes on matrix models (matrix musings)}",
    eprint = "2004.01171",
    archivePrefix = "arXiv",
    primaryClass = "hep-th",
    doi = "10.1088/1742-5468/aba499",
    journal = "J. Stat. Mech.",
    volume = "2008",
    pages = "083109",
    year = "2020"
}

@article{Runkel:2020zgg,
    author = "Runkel, Ingo and Watts, G{\'e}rard M. T.",
    title = "{Fermionic CFTs and classifying algebras}",
    eprint = "2001.05055",
    archivePrefix = "arXiv",
    primaryClass = "hep-th",
    reportNumber = "kcl-mth-20-01",
    doi = "10.1007/JHEP06(2020)025",
    journal = "JHEP",
    volume = "06",
    pages = "025",
    year = "2020"
}

@article{Oshikawa:1996dj,
    author = "Oshikawa, Masaki and Affleck, Ian",
    title = "{Boundary conformal field theory approach to the critical two-dimensional Ising model with a defect line}",
    eprint = "cond-mat/9612187",
    archivePrefix = "arXiv",
    doi = "10.1016/S0550-3213(97)00219-8",
    journal = "Nucl. Phys. B",
    volume = "495",
    pages = "533--582",
    year = "1997"
}

@article{Oshikawa:1996ww,
    author = "Oshikawa, Masaki and Affleck, Ian",
    title = "{Defect lines in the Ising model and boundary states on orbifolds}",
    eprint = "hep-th/9606177",
    archivePrefix = "arXiv",
    doi = "10.1103/PhysRevLett.77.2604",
    journal = "Phys. Rev. Lett.",
    volume = "77",
    pages = "2604--2607",
    year = "1996"
}

@article{Quella:2006de,
    author = "Quella, Thomas and Runkel, Ingo and Watts, G{\'e}rard M. T.",
    title = "{Reflection and transmission for conformal defects}",
    eprint = "hep-th/0611296",
    archivePrefix = "arXiv",
    reportNumber = "KCL-MTH-06-12, NSF-KITP-06-110",
    doi = "10.1088/1126-6708/2007/04/095",
    journal = "JHEP",
    volume = "04",
    pages = "095",
    year = "2007"
}

@article{Bros:2010rku,
    author = "Bros, Jacques and Epstein, Henri and Moschella, Ugo",
    title = "{Particle decays and stability on the de Sitter universe}",
    eprint = "0812.3513",
    archivePrefix = "arXiv",
    primaryClass = "hep-th",
    doi = "10.1007/s00023-010-0042-7",
    journal = "Annales Henri Poincare",
    volume = "11",
    pages = "611--658",
    year = "2010"
}

@article{Nachtmann:1967nss,
    author = "Nachtmann, Otto",
    title = "{Quantum theory in de-Sitter space}",
    doi = "10.1007/BF01646319",
    journal = "Commun. Math. Phys.",
    volume = "6",
    number = "1",
    pages = "1--16",
    year = "1967"
}

@article{Guijosa:2003ze,
    author = "Guijosa, Alberto and Lowe, David A.",
    title = "{A New twist on dS / CFT}",
    eprint = "hep-th/0312282",
    archivePrefix = "arXiv",
    reportNumber = "BROWN-HET-1383, ICN-UNAM-03-15, NSF-KITP-03-107",
    doi = "10.1103/PhysRevD.69.106008",
    journal = "Phys. Rev. D",
    volume = "69",
    pages = "106008",
    year = "2004"
}

@article{Bonifacio:2023prb,
    author = "Bonifacio, James and Hinterbichler, Kurt",
    title = "{Fermionic shift symmetries in (anti) de Sitter space}",
    eprint = "2312.06743",
    archivePrefix = "arXiv",
    primaryClass = "hep-th",
    doi = "10.1007/JHEP04(2024)100",
    journal = "JHEP",
    volume = "04",
    pages = "100",
    year = "2024"
}

@incollection{Fisher:1965rna,
  author    = {Fisher, Michael E.},
  title     = {The Nature of Critical Points},
  booktitle = {Lectures in Theoretical Physics},
  editor    = {Brittin, Wesley E.},
  volume    = {VII C},
  pages     = {1--159},
  publisher = {University of Colorado Press},
  year      = {1965}
}

@article{Itzykson:1983gb,
    author = "Itzykson, C. and Pearson, R. B. and Zuber, J. B.",
    title = "{Distribution of Zeros in Ising and Gauge Models}",
    reportNumber = "SACLAY-SPHT-83-40",
    doi = "10.1016/0550-3213(83)90499-6",
    journal = "Nucl. Phys. B",
    volume = "220",
    pages = "415--433",
    year = "1983"
}

@article{Hoelbling:1996gv,
    author = "Hoelbling, Christian and Lang, C. B.",
    title = "{Universality of the Ising model on sphere-like lattices}",
    eprint = "hep-lat/9602025",
    archivePrefix = "arXiv",
    reportNumber = "UNIGRAZ-UTP-26-02-96",
    doi = "10.1103/PhysRevB.54.3434",
    journal = "Phys. Rev. B",
    volume = "54",
    pages = "3434",
    year = "1996"
}

@article{Holm:1995wa,
    author = "Holm, Christian and Janke, Wolfhard",
    title = "{Ising spins on a gravitating sphere}",
    eprint = "hep-lat/9512002",
    archivePrefix = "arXiv",
    reportNumber = "FUB-HEP-18-95, KOMA-95-81",
    doi = "10.1016/0370-2693(96)00199-2",
    journal = "Phys. Lett. B",
    volume = "375",
    pages = "69--74",
    year = "1996"
}

@article{Hoelbling:1995hi,
    author = "Hoelbling, Christian and Jakovac, A. and Jersak, J. and Lang, C. B. and Neuhaus, T.",
    editor = "Kieu, T. D. and McKellar, B. H. J. and Guttmann, A. J.",
    title = "{Spin and gauge systems on spherical lattices}",
    eprint = "hep-lat/9509009",
    archivePrefix = "arXiv",
    reportNumber = "UNIGRAZ-UTP-07-09-95",
    doi = "10.1016/0920-5632(96)00181-8",
    journal = "Nucl. Phys. B Proc. Suppl.",
    volume = "47",
    pages = "815--818",
    year = "1996"
}

@article{Diego:1993wq,
    author = "Diego, O. and Gonzalez, J. and Salas, J.",
    title = "{The Ising model on spherical lattices: Dimer versus Monte Carlo approach}",
    eprint = "hep-lat/9307018",
    archivePrefix = "arXiv",
    reportNumber = "PRINT-93-0562",
    doi = "10.1088/0305-4470/27/9/013",
    journal = "J. Phys. A",
    volume = "27",
    pages = "2965--2983",
    year = "1994"
}

@article{Brower:2024otr,
    author = "Brower, Richard C. and Owen, Evan K.",
    title = "{The Ising Model on $\mathbb S^2$}",
    eprint = "2407.00459",
    archivePrefix = "arXiv",
    primaryClass = "hep-lat",
    month = "6",
    year = "2024"
}
\end{document}